\documentclass[letterpaper,twocolumn,10pt]{article}
\usepackage{usenix-2020-09}

\usepackage{cite}
\usepackage{amsmath,amssymb,amsfonts}
\usepackage{algorithmic}
\usepackage{graphicx}
\usepackage{textcomp}
\usepackage[autostyle]{csquotes}
\usepackage{array}
\usepackage{url}
\usepackage{relsize}

\usepackage{tikz}
\usetikzlibrary{decorations.pathmorphing, arrows, arrows.meta}

\usepackage{siunitx}
\usepackage{multirow}

\usepackage{xspace}
\usepackage{listings}
\usepackage{paralist}
\usepackage{colortbl}
\usepackage{tabularx}
\usepackage{booktabs}
\usepackage{mdframed}
\usepackage{subcaption}

\usepackage[most]{tcolorbox}
\usepackage{wasysym}
\usepackage{threeparttable}
\usepackage{rotating}
\usepackage{tablefootnote}
\usepackage{ragged2e}

\usepackage{xstring}
\usepackage{pifont}
\newcommand{\resolve}[1]{%
    \IfEqCase{#1}{
        {yes}{\textcolor{blue}{\ding{52}}}%
        {no}{\textcolor{red}{\ding{56}}}%
        {orange}{\textcolor{#1}{\ding{74}}}%
        {LC}{\textcolor{orange}{\LEFTcircle}} %
        {RC}{\textcolor{orange}{\RIGHTcircle}} %
        {CC}{\textcolor{orange}{\Circle}} %
        {circle}{\textcolor{orange}{\CIRCLE}} %
    }[\PackageError{resolve}{Undefined option to resolve: #1}{}]%
}%

\usepackage[strict]{changepage}

\usepackage{framed}

\definecolor{formalshade}{rgb}{0.95,0.95,1}

\newenvironment{formal}{%
  \MakeFramed{\advance\hsize-\width\FrameRestore}%
  \noindent\hspace{-4.55pt}%
  \begin{adjustwidth}{}{7pt}%
}
{%
  \vspace{2pt}\end{adjustwidth}\endMakeFramed%
}

\newtcolorbox[auto counter]{summary}[1][]{title={\bfseries Summary~\thetcbcounter},enhanced,drop shadow={black!50!white},
  coltitle=black,
  top=0.3in,
  attach boxed title to top left=
  {xshift=1.5em,yshift=-\tcboxedtitleheight/2},
  boxed title style={size=small,colback=pink},#1
}

\mdfdefinestyle{lessonLearned}{roundcorner=20pt,backgroundcolor=lightgray}
\usepackage{soul}

\usepackage{hyperref}
\hypersetup{
    colorlinks,
    linkcolor={red},
    citecolor={blue},
    urlcolor={black}
}

\usetikzlibrary{calc,shadings,patterns,tikzmark}

\usepackage[nameinlink]{cleveref}
\crefformat{section}{\S#2#1#3} 
\crefformat{subsection}{\S#2#1#3}
\crefformat{subsubsection}{\S#2#1#3}
\crefname{secinapp}{appendix}{appendices}
\Crefname{secinapp}{Appendix}{Appendices}

\usepackage[inline]{enumitem}

\newcommand{\SO}{\textit{SO}\xspace} 
 
\newcommand{\DC}{\textsc{Dicos}\xspace}

\newenvironment{sinparaenum}%
{\begin{inparaenum}[\itshape(i)\upshape]}%
{\end{inparaenum}}

\definecolor{KWColor}{rgb}{0.37,0.08,0.25}
\definecolor{CommentColor}{rgb}{0.133,0.545,0.133}
\definecolor{StringColor}{rgb}{0,0.126,0.941}
\definecolor{Gray}{gray}{0.9}
\definecolor{lightgray}{gray}{0.93}
\definecolor{dkgreen}{rgb}{0,0.6,0}
\definecolor{gray}{rgb}{0.5,0.5,0.5}
\definecolor{mauve}{rgb}{0.58,0,0.82}
\definecolor{mahogany}{rgb}{0.75, 0.25, 0.0}
\definecolor{orange-red}{rgb}{1.0, 0.27, 0.0}
\definecolor{blue-violet}{rgb}{0.54, 0.17, 0.89}
\definecolor{darkviolet}{rgb}{0.58, 0.0, 0.83}
\definecolor{darkturquoise}{rgb}{0.0, 0.81, 0.82}
\definecolor{turquoiseblue}{rgb}{0.0, 1.0, 0.94}
\definecolor{turquoise}{rgb}{0.19, 0.84, 0.78}
\definecolor{brightturquoise}{rgb}{0.03, 0.91, 0.87}

\definecolor{hlbluecolour}{HTML}{B4B8FB}
\definecolor{hlgreencolour}{HTML}{B6FFBC}

\DeclareRobustCommand{\hlblue}[1]{{\sethlcolor{hlbluecolour}\hl{#1}}}
\DeclareRobustCommand{\hlgreen}[1]{{\sethlcolor{hlgreencolour}\hl{#1}}}

\usepackage{marginnote}
\colorlet{punct}{red!60!black}
\definecolor{background}{HTML}{EEEEEE}
\definecolor{delim}{RGB}{20,105,176}
\colorlet{numb}{magenta!60!black}
\lstdefinelanguage{json}{
    basicstyle=\scriptsize\ttfamily,
    showstringspaces=false,
    breaklines=true,
    frame=lines,
    literate=
     *{0}{{{\color{numb}0}}}{1}
      {1}{{{\color{numb}1}}}{1}
      {2}{{{\color{numb}2}}}{1}
      {3}{{{\color{numb}3}}}{1}
      {4}{{{\color{numb}4}}}{1}
      {5}{{{\color{numb}5}}}{1}
      {6}{{{\color{numb}6}}}{1}
      {7}{{{\color{numb}7}}}{1}
      {8}{{{\color{numb}8}}}{1}
      {9}{{{\color{numb}9}}}{1}
      {:}{{{\color{punct}{:}}}}{1}
      {,}{{{\color{punct}{,}}}}{1}
      {\{}{{{\color{delim}{\{}}}}{1}
      {\}}{{{\color{delim}{\}}}}}{1}
      {[}{{{\color{delim}{[}}}}{1}
      {]}{{{\color{delim}{]}}}}{1},
}

\newcolumntype{N}{>{\centering\arraybackslash}m{.41in}}
\newcolumntype{G}{>{\centering\arraybackslash}m{2in}}

\lstdefinelanguage{diff}{
  morecomment=[f][\color{blue}]{@@},     %
  morecomment=[f][\color{red}]-,         %
  morecomment=[f][\color{green}]+,       %
  morecomment=[f][\color{magenta}]{---}, %
  morecomment=[f][\color{magenta}]{+++},
}

\makeatletter
\newcommand\footnoteref[1]{\protected@xdef\@thefnmark{\ref{#1}}\@footnotemark}
\makeatother

\usepackage[multiple]{footmisc}

\begin{document}

\date{}

\author{
{\rm Alfusainey Jallow\textsuperscript{\textsection\textasteriskcentered}, Sven\ Bugiel\textsuperscript{\textsection}}\\
\small
\normalfont
\textsuperscript{\textsection} CISPA Helmholtz Center for Information Security, \textsuperscript{\textasteriskcentered} Saarland University
}

\title{\Large \bf Stack Overflow Meets Replication: Security Research Amid Evolving Code Snippets (Extended Version)}

\maketitle

\begin{abstract}

We study the impact of Stack Overflow code evolution on the stability of prior research findings derived from Stack Overflow data and provide recommendations for future studies.
We systematically reviewed papers published between 2005--2023 to identify key aspects of Stack Overflow that can affect study results, such as the language or context of code snippets.
Our analysis reveals that certain aspects are non-stationary over time, which could lead to different conclusions if experiments are repeated at different times.
We replicated six studies using a more recent dataset to demonstrate this risk.
Our findings show that four papers produced significantly different results than the original findings, preventing the same conclusions from being drawn with a newer dataset version.
Consequently, we recommend treating Stack Overflow as a time series data source to provide context for interpreting cross-sectional research conclusions.

\end{abstract}

\section{Introduction}\label{sec:introduction}

In recent years, several security-focused studies~\cite{acar2016YouGetWhereYouLook, fischer2017SOConsideredHarmful, FischerUsenix2019, dicos, zhang_code_weaknesses, mining_rule_violations, selvaraj_collaborative_editing, verdi19, rahman_snakes_in_paradies, ManesBaysal_MSR21, Meng_java_secure_coding_practices, toxicCodeSnippets, Abdalkareem, Chen_reliable_crowd_source_knowledge} have examined Stack Overflow to analyze the security of shared code on the platform, develop tools for secure code reuse, or use it as a proxy for studying developer behavior. This research is fostered by the quarterly releases of a dataset containing all content created on the Stack Exchange Inc.~platform since its launch in 2008. Right now, about 60 million posts (24 million questions and 35.8 million answers)\cite{questions_answers_query_results} containing over 91 million comments are publicly available for analysis.

However, Stack Overflow code and content evolves as the community adds snippets and updates existing ones~\cite{baltes2018sotorrent, zhang_code_weaknesses, dicos}.
This code evolution has already been shown to negatively affect developers who reuse a specific snippet version from Stack Overflow without tracking updates for security fixes~\cite{jallow_sp24}.

Like developers, security researchers may also study the content of the Stack Overflow data set only using the current version, i.e., \textit{cross-sectional} studies.
While cross-sectional studies provide valuable and novel insights into the underlying data, other disciplines~\cite{hamilton1994series, wilks-statistical_2011,Box-Steffensmeier_Freeman_Hitt_Pevehouse_2014} and also software engineering research~\cite{10.1145/3551349.3559517} suggest complementing such insights with longitudinal trend and time-series analysis provides a better context for interpreting findings.
Transferring these lessons to Stack Overflow code evolution raises questions about how this evolution affects cross-sectional research findings based on particular dataset versions and what lessons can be learned for future studies using Stack Overflow data.
While a shift in research results due to a shift in the data is intuitive, this phenomenon has not been systematically studied before for Stack Overflow-based data-driven research.
To offer new insights into this issue, this paper aims to address the following meta-research questions:

\begin{enumerate}[label=\textbf{MQ\arabic*:},leftmargin=*,labelwidth=2cm,itemsep=0.0em,noitemsep]
    \item \textit{Which aspects of Stack Overflow affect the results of prior research? }
    \item \textit{How much do Stack Overflow code snippets and surrounding context evolve? }
    \item \textit{How would the research results of prior work differ if replicated on a newer Stack Overflow dataset version?}
\end{enumerate}

To answer our research questions, we surveyed the literature for papers studying the security properties of code snippets on Stack Overflow.
This yielded 42 highly relevant papers, which we systematized according to the Stack Overflow aspects their methods relied on (e.g., programming language or the context of snippets).
This systematization shows that the targeted programming language may affect the stability of results over time and that most works leverage some form of code classification, whose results can be immediately affected by code revisions (\textit{MQ1}).
Further, we conducted a time series analysis to understand how code snippets and security- and privacy-related discussions on Stack Overflow evolve.
Our data shows that programming languages trend differently regarding their overall number of added snippets and their ratio of security-relevant edits.
Moreover, we found that the fraction of security-relevant comments on Stack Overflow steadily increased.
As a result, studies focusing on particular programming languages will likely find a different landscape when conducted at various points in time (\textit{MQ2}).
Together, these two insights provide an intuition about how prior research results may shift in light of the evolution of Stack Overflow content.
To provide concrete evidence for the impact of this evolution on research results (\textit{MQ3}), we conducted \textit{six} replication studies of prior work~\cite{zhang_code_weaknesses,dicos, mining_rule_violations, rahman_snakes_in_paradies, fischer2017SOConsideredHarmful, FischerUsenix2019} 
using a more recent data set version.
We find that the results of multiple works shift over time~\cite{zhang_code_weaknesses,dicos, mining_rule_violations, rahman_snakes_in_paradies}.
For example, the landscape of CWEs in C/C++ code snippets~\cite{zhang_code_weaknesses} has significantly shifted, and Stack Overflow now contains proportionally more vulnerable snippets with different ratios for CWE types.
Only the results of two papers studying crypto API misuse in Java snippets~\cite{fischer2017SOConsideredHarmful,FischerUsenix2019} remained stable.
We postulate that this may be due to the particular niche topic that requires domain experts to identify and fix such vulnerabilities.
Based on our replication studies, we offer advice for future research involving Stack Overflow data.
A shift in results does \textit{not} mean that the results are invalid but missing context.
We recommend that researchers consider data on Stack Overflow as time-series data and discuss their results as a trend model rather than a cross-sectional analysis---taking inspiration from methods in economics, environmental science, or medical studies.
This approach provides a more meaningful context for the results, allowing us to determine whether the observed issues are short-term trends or persistent systemic issues.

\section{Background \& Motivation}\label{sec:technical_background}

Developers seeking advice for a programming problem can create a question post on Stack Overflow, which other developers can answer.
The question and all the answers are called \textit{posts}, each with a unique \textit{identifier}.
Developers can also \textit{comment} on posted code snippets through Stack Overflow's commenting feature.
Comments made by other developers are known to induce updates to posts~\cite{commentInducedUpdates} or raise bug reports and point out security vulnerabilities~\cite{dicos, zhang_code_weaknesses, jallow_sp24}.
Stack Exchange Inc.~has a quarterly release cycle of all data created on the platform since its inception in 2008.
This recurring release allows researchers to tap into this rich data source (e.g.,~\cite{fischer2017SOConsideredHarmful, FischerUsenix2019, dicos, zhang_code_weaknesses, rahman_snakes_in_paradies, mining_rule_violations, toxicCodeSnippets, jallow_sp24}).
Unfortunately, the Stack Exchange dataset only provides versioning at the level of whole posts and not at the level of individual text and code snippets, which makes it non-trivial to track and analyze changes to individual code snippets contained in a post. 

The SOTorrent open dataset by Baltes et al.~\cite{baltes2018sotorrent} is based on the official Stack Exchange data and provides version control at the level of individual text and code snippets.
For this reason, it is popular among researchers studying Stack Overflow.

\begin{figure}[t]
 \centering
  \includegraphics[width=\linewidth]{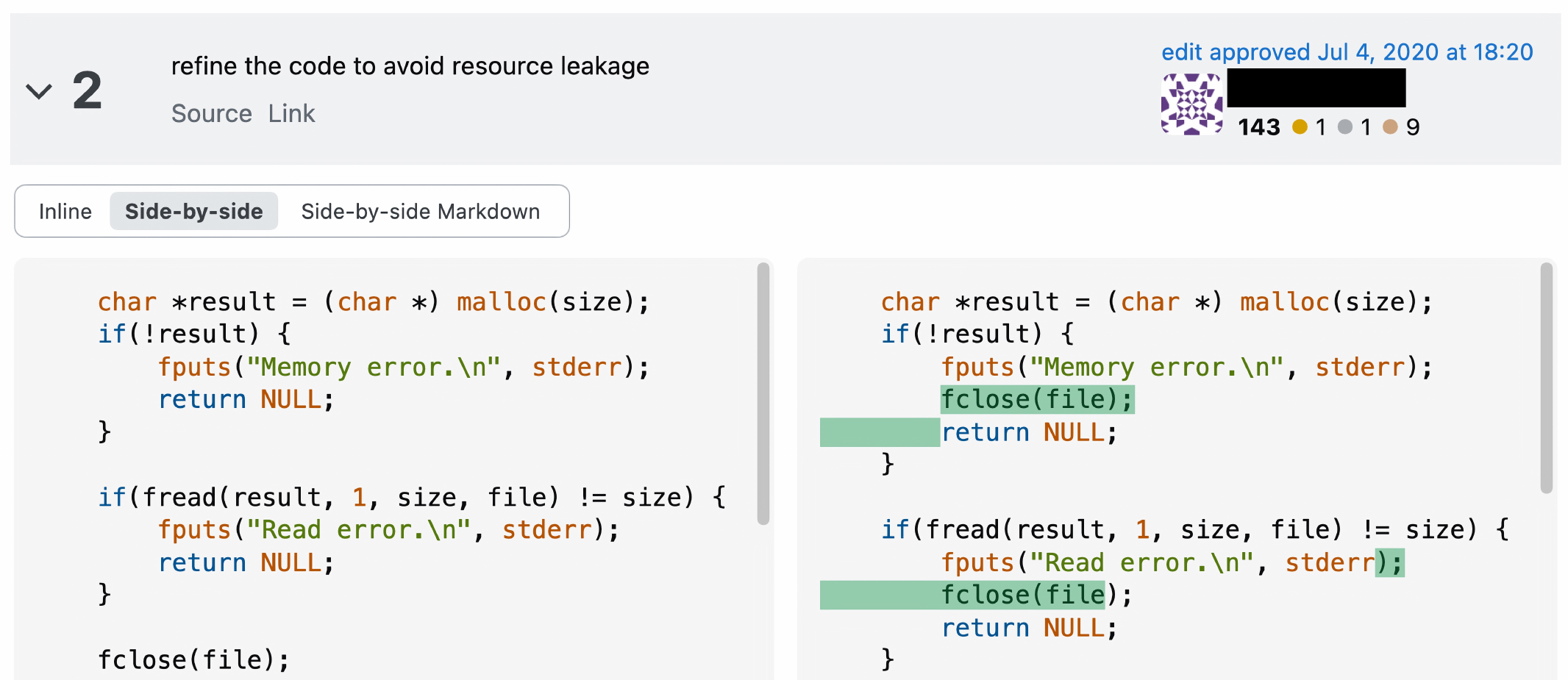}
  \caption{First version of the answer (left-hand side) labeled by the authors as insecure and unmodified with three CWE instances. The 2$^{nd}$ version of the answer on the right-hand side shows both instances of CWE-775 fixed on July 4$^{th}$, changing the snippet's status to \textit{improved}.
 }
\label{fig:background:answer_14424800}
\end{figure}

\paragraph*{Motivating Example} \Cref{fig:background:answer_14424800} depicts an example code snippet~\cite{post_14424800_revisions} from the 350+ examples of the results by Zhang et al.~\cite{zhang_code_weaknesses} that were affected by Stack Overflow evolution.
Zhang et al.\ used the December 2018 version of the SOTorrent dataset to study whether code revisions improve security.
They identified two CWEs in this code: two instances of CWE-775 and one instance of CWE-401, all introduced in the first version in January 2013.
By December 2018, this snippet had never been revised, leaving these security issues unresolved. 
The authors, therefore, labeled the snippet correctly as insecure and concluded that the snippet \textit{had never been revised} to address the three CWE instances.
By July 2020, the snippet had been revised to fix both instances of CWE-775.
This revision, posted \textit{after} the authors sampled their data, reduced the CWE instances from three to one, changing the snippet's status to \textit{improved}.

The example shows that a code snippet can undergo edits throughout its lifespan, causing its security status to change over time.
Consequently, code snippets might be insecure at a given time but secure at a future time or vice-versa.
Therefore, it seems prima facie intuitive for researchers studying Stack Overflow snippets to conduct measurements at multiple points in time using different dataset versions to account for these fluctuations.
Integrating these evolving trends and changes in snippets may enhance the robustness of research studies.

\section{Systematization of Relevant Works}\label{sec:code_based_studies}

To understand how prior work may be affected by Stack Overflow evolution (\textbf{MQ1}), we systematize studies that investigated the security properties of Stack Overflow code snippets.
First, we systematically surveyed the literature for relevant works (\Cref{sec:literature_search}).
Following this, we created a taxonomy of the methodologies by those studies for analyzing Stack Overflow datasets (\Cref{sec:criteria}).
Two researchers conducted this systematization and decided on the criteria for the taxonomy.

\subsection{Literature Search}\label{sec:literature_search}

We conducted a systematic literature review following the guidelines by Kitchenham and Charters~\cite{kitchenham_slr}.

\paragraph{Inclusion and Exclusion Criteria.}
The inclusion criteria are that the study:
\begin{enumerate*}[label=\textit{IC\arabic*.}]
    \item must focus on the Stack Overflow website.
    \item must examine code snippets on Stack Overflow.
    \item must analyze the security of code snippets or identify and address bugs or faults in snippets.
    \item must be published between 01/2005 and 12/2023.
    \item is a journal paper or in conference proceedings that are DBLP-indexed.
\end{enumerate*}
The exclusion criteria are:
\begin{enumerate*}[label=\textit{EC\arabic*.}]
    \item Studies focusing only on code reuse from Stack Overflow without studying the security of or identifying bugs or faults in snippets.
    \item Publications outside the date range in \textit{IC4}.
    \item Systematic reviews or meta-analyses, as only primary studies are considered.
    \item Non-peer-reviewed studies (e.g., technical reports).
\end{enumerate*}

\begin{figure}[t]
    \centering
    \includegraphics[width=\linewidth]{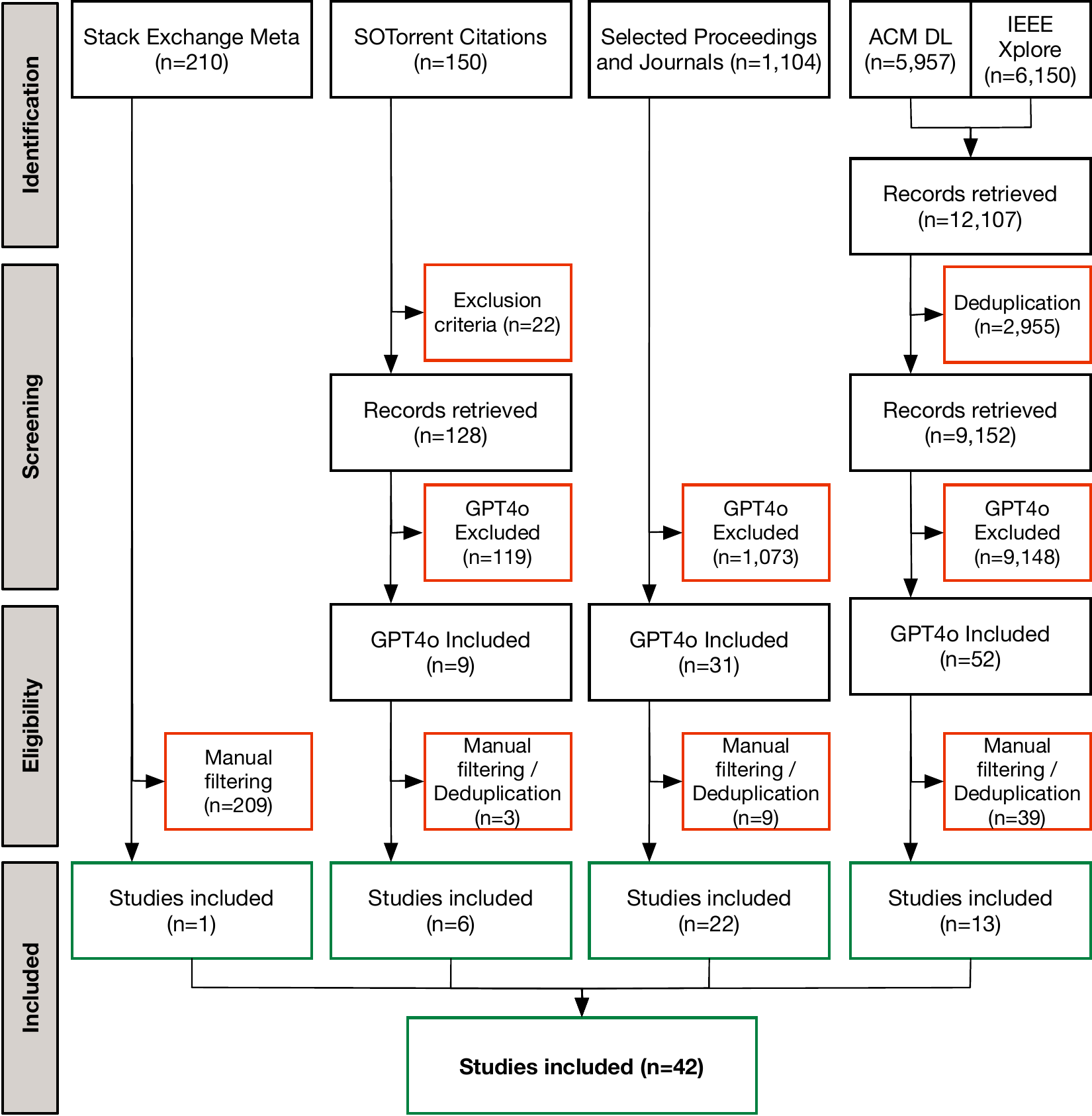}
    \caption{PRISMA diagram of our literature review}
    \label{fig:slr}
\end{figure}

\paragraph{Search Strategy.}
\Cref{fig:slr} depicts the flow of our systematic literature review.
We manually reviewed the titles and abstracts of studies listed on the dedicated Meta Exchange thread~\cite{academic_papers} listing academic papers that use Stack Exchange datasets.
This yielded one relevant paper.
Next, we retrieved all studies that cited the SOTorrent dataset~\cite{baltes2018sotorrent}, a widely used dataset for studying Stack Overflow.
Of these 150 studies, 128 fit the inclusion criteria.
Following the recommendations by Kitchenham and Charters~\cite{kitchenham_slr}, we automated part of our review.
We employed OpenAI's GPT4o model to scan abstracts and identify studies focusing on the security of Stack Overflow code snippets.
A preliminary evaluation of GPT4o showed that it performs well in classifying studies based on our inclusion and exclusion criteria (details see \Cref{sec:appendix:llm}).
We use GPT4o to efficiently screen non-relevant studies based on a better contextual understanding of security concepts than keyword-based filtering.
Papers identified as relevant by GPT4o are manually verified through full-text analysis.
With the help of GPT4o, we identified six additional relevant studies among the 128 candidates.
Based on the reviewed studies, we created a list of search terms to identify \textit{all} research studies related to Stack Overflow, irrespective of whether they analyzed code snippets or investigated their security.
The search terms are listed in \Cref{sec:appendix:systematization:keywords}.
We used these terms to search the proceedings of 29 conferences and the volumes of 3 journals (full list in \Cref{sec:appendix:systematization:venues}). 
We identified 1,104 matching studies across all venues.
Using GPT-4o, we analyzed their abstracts, resulting in 22 confirmed relevant studies.
We conducted additional searches in IEEE Xplore and ACM Digital Library using the same search terms.
After deduplication, this resulted in 9,152 unique publications, which we retrieved from the two systems.
Among these, using GPT-4o, we confirmed an additional 13 studies that met our criteria, totaling 42 relevant studies.

\begin{table}[!htbp]
    
    \caption{Comparison of security-focused studies on Stack Overflow. We replicated the \hlblue{highlighted} studies.}
    \label{tab:comparision_table}
\centering
\fontsize{6}{7}\selectfont
 \setlength{\tabcolsep}{2pt}
\def\arraystretch{1.5}
\setlength{\aboverulesep}{0pt}
\setlength{\belowrulesep}{0pt}
 \begin{threeparttable}
    \begin{tabularx}{\linewidth}{Xr|c|>{\columncolor{lightgray}}cc>{\columncolor{lightgray}}cc>{\columncolor{lightgray}}c|c>{\columncolor{lightgray}}cc>{\columncolor{lightgray}}cc}
    &
    &
    \tikz \node[rotate=90] {\textbf{Dataset Snapshot}};
    &
    \tikz \node[rotate=90] {\textbf{D1. Prog. Languages}};
    &
    \tikz \node[rotate=90] {\textbf{D2. Code Scanners}};
    &
    \tikz \node[rotate=90] {\textbf{D3. Code Evolution}};
    &
    \tikz \node[rotate=90] {\textbf{D4. Surrounding Context}};
    &
    \tikz \node[rotate=90] {\textbf{D5. Sample Size Filter}};
    &
    \tikz \node[rotate=90] {\textbf{R1. Artifact Availability}};
    &
    \tikz \node[rotate=90] {\textbf{R2. Language Detection}};
    &
    \tikz \node[rotate=90] {\textbf{R3. Code Reuse Detect.}};
    &
    \tikz \node[rotate=90] {\textbf{R4. Human-Centered}};
    \tabularnewline
    \midrule
    \cellcolor{blue!25}Zhang et al.      & \cellcolor{blue!25}\cite{zhang_code_weaknesses}          & 12/2018 & 
    C & 
    \resolve{yes} & 
    \resolve{yes} & 
    \resolve{no}  & 
    \resolve{yes} & 
    \resolve{no}  & 
    \resolve{orange}, M & 
    \resolve{no}  & 
    \resolve{no}  & 
    \tabularnewline
    \midrule
    
    \cellcolor{blue!25}Hong et al.       & \cellcolor{blue!25}\cite{dicos}                          & 12/2020 & 
    C  &  
    \resolve{orange}, N & 
    \resolve{yes} & 
    \resolve{yes} & 
    \resolve{yes} &
    \resolve{CC} & 
    \resolve{yes} & 
    \resolve{orange}, S & 
    \resolve{no}  & 
    \tabularnewline
    \midrule
    
    \cellcolor{blue!25}Fischer et al.    & \cellcolor{blue!25}\cite{FischerUsenix2019}              & 03/2018 & 
    J  & 
    \resolve{orange}, M & 
    \resolve{no} & 
    \resolve{no} & 
    \resolve{yes} & 
    \resolve{LC} & 
    N/A & 
    \resolve{no}  & 
    \resolve{no}  & 
    \tabularnewline
    \midrule
    
    \cellcolor{blue!25}Fischer et al.    & \cellcolor{blue!25}\cite{fischer2017SOConsideredHarmful} & 03/2016 & 
    J   &
    \resolve{orange}, M &
    \resolve{no} & 
    \resolve{no} & 
    \resolve{yes} &
    \resolve{LC} & 
    N/A &
    \resolve{orange}, P & 
    \resolve{no}  & 
    \tabularnewline
    \midrule

    \cellcolor{blue!25}Rahman et al.                         & \cellcolor{blue!25}\cite{rahman_snakes_in_paradies}     & 12/2018 & 
    P   &
    \resolve{orange}, SM & 
    \resolve{no} & 
    \resolve{no} & 
    \resolve{yes} &
    \resolve{RC} & 
    \resolve{yes} &
    \resolve{no} & 
    \resolve{no} & 
    \tabularnewline
    \midrule

    \cellcolor{blue!25}Campos et al.                       & \cellcolor{blue!25}\cite{mining_rule_violations}         & 12/2018 & 
    JS   &  
    \resolve{yes} & 
    \resolve{no} & 
    \resolve{no} & 
    \resolve{yes} &
    \resolve{orange} & 
    \resolve{yes} & 
    \resolve{yes} & 
    \resolve{no}  & 
    \tabularnewline    
    \midrule

    Verdi et al.                        & \cite{verdi19}                        & 09/2018 & 
    C   & 
    \resolve{no} & 
    \resolve{no} & 
    \resolve{no} & 
    \resolve{yes} &
    \resolve{LC} & 
    \resolve{yes} & 
    \resolve{yes}    & 
    \resolve{no}  & 
    \tabularnewline
    \midrule
    
    Selvaraj et al.                  & \cite{selvaraj_collaborative_editing} & 01/2022 & 
        C   &  
    \resolve{yes} & 
    \resolve{yes} & 
    \resolve{no} & 
    \resolve{yes} &
    \resolve{LC} & 
    \resolve{orange}, M & 
    \resolve{no}  & 
    \resolve{no}  & 
    \tabularnewline
    \midrule

    Acar et al.                         & \cite{acar2016YouGetWhereYouLook}     & 10/2015 & 
    J   &  
    \resolve{no} & 
    \resolve{no} & 
    \resolve{no} & 
    \resolve{no} &
    \resolve{no} & 
    N/A & 
    \resolve{no} & 
    \resolve{yes} & 
    \tabularnewline
    \midrule

    Chen et al.                         & \cite{Chen_reliable_crowd_source_knowledge}     & \textbf{?}/2018 & 
    J   &  
    \resolve{no} & 
    \resolve{no} & 
    \resolve{no} & 
    \resolve{yes} &
    \resolve{no} & 
    N/A & 
    \resolve{no} & 
    \resolve{no} & 
    \tabularnewline
    \midrule

    Meng et al.                         & \cite{Meng_java_secure_coding_practices}     & 08/2017 & 
    J   &  
    \resolve{no} &
    \resolve{no} & 
    \resolve{yes} & 
    \resolve{yes} &
    \resolve{LC} & 
    \resolve{yes} & 
    \resolve{no} & 
    \resolve{no} &
    \tabularnewline
    \midrule

    Ragkhitwets                         & \cite{toxicCodeSnippets}     & 01/2016 & 
    J   &  
    \resolve{no}  & 
    \resolve{yes}  & 
    \resolve{no}  & 
    \resolve{yes}  & 
    \resolve{LC} & 
    \resolve{yes} &
    \resolve{orange},SI,CC & 
    \resolve{yes} & 

    \tabularnewline
    \midrule

    Bai et al.                         & \cite{bai_insecure_code_propagration}     & N/A & 
    J   &  
    \resolve{no}  &
    \resolve{no}  & 
    \resolve{no}  & 
    \resolve{yes}  & 
    \resolve{no} & 
    N/A & 
    \resolve{orange},M & 
    \resolve{yes} &

    \tabularnewline
    \midrule

     Bagherzadeh et al.                         & \cite{Bagherzadeh_akka_actor_bugs}     & N/A & 
    J,S   &  
    \resolve{no}  & 
    \resolve{no}  & 
    \resolve{yes}  & 
    \resolve{yes}  & 
    \resolve{LC} & 
    \resolve{yes} &
    \resolve{no} & 
    \resolve{no} & 

    \tabularnewline
    \midrule
    
    Chen et al.                         & \cite{chen_intelligent_detection_system}     & ?/2018 & 
    J   &  
    \resolve{orange},M  & 
    \resolve{no}  & 
    \resolve{no}  & 
    \resolve{yes}  &  
    \resolve{no} & 
    N/A & 
    \resolve{no} & 
    \resolve{no} & 

    \tabularnewline
    \midrule

     Zhang et al.                         & \cite{ZhangSoApiMisuse}     & 10/2016 & 
    J   &  
    \resolve{orange},P  & 
    \resolve{no}  & 
    \resolve{no}  & 
    \resolve{yes}  & 
    \resolve{no} & 
    \resolve{yes} & 
    \resolve{yes},CC & 
    \resolve{no} & 

    \tabularnewline
    \midrule

     Rahman et al.                         & \cite{rahman_reusability_insight}     & 08/2021 & 
    J   &  
    \resolve{no}  & 
    \resolve{no}  & 
    \resolve{no}  & 
    \resolve{yes}  & 
    \resolve{no} & 
    \resolve{yes} & 
    \resolve{yes},CC & 
    \resolve{yes} & 

    \tabularnewline
    \midrule

     Reinhardt et al.                         & \cite{reinhardt_augmenting_stack_overflow}     & N/A & 
    J   &  
    \resolve{yes}  & 
    \resolve{no}  & 
    \resolve{no}  & 
    \resolve{yes}  & 
    \resolve{no} & 
    \resolve{yes} & 
    \resolve{yes},CC & 
    \resolve{no} & 

    \tabularnewline
    \midrule

     Licorish et al.                         & \cite{Licorish_contextual_profiling}     & ?/2016 & 
    J   &  
    \resolve{yes}  & 
    \resolve{no}  & 
    \resolve{yes}  & 
    \resolve{no}  & 
    \resolve{no} & 
    \resolve{yes} & 
    \resolve{no} &
    \resolve{no} & 

    \tabularnewline
    \midrule

     Schmidt et al.                         & \cite{Schmidt_CopypastaVulGuard}     & 03/2022 & 
    JS,P   &  
    \resolve{no}  & 
    \resolve{no}  & 
    \resolve{yes}  & 
    \resolve{no}  & 
    \resolve{RC} & 
    \resolve{yes} & 
    \resolve{no} & 
    \resolve{no} & 

    \tabularnewline
    \midrule

    Yi Liu et al.                         & \cite{Liu_deepanna}     & N/A & 
    J   &  
    \resolve{orange},M  & 
    \resolve{no}  & 
    \resolve{yes}  & 
    \resolve{no}  & 
    \resolve{no} & 
    \resolve{yes} & 
    \resolve{no} & 
    \resolve{no} & 

    \tabularnewline
    \midrule

     Ren et al.                         & \cite{Ren_Demystify_API_Usage_Directives}     & 03/2019 & 
    J   &  
    \resolve{orange},M  & 
    \resolve{no}  & 
    \resolve{yes}  & 
    \resolve{yes}  &
    \resolve{no} & 
    \resolve{yes} & 
    \resolve{no} & 
    \resolve{yes} &

    \tabularnewline
    \midrule

     Licorish et al.                         & \cite{Licorish_dissecting_copy_delete_replace_swap_mutations}     & ?/2016 & 
    J   &  
    \resolve{yes}  & 
    \resolve{no}  & 
    \resolve{no}  & 
    \resolve{yes}  & 
    \resolve{no} & 
    \resolve{yes} & 
    \resolve{no} & 
    \resolve{no} & 

    \tabularnewline
    \midrule

    Rangeet Pan                         & \cite{Pan_deep_learning_bug_fixing}     & N/A & 
    P   &  
    \resolve{no}  & 
    \resolve{no}  & 
    \resolve{yes}  & 
    \resolve{yes}  & 
    \resolve{no} & 
    \resolve{yes} & 
    \resolve{no} &
    \resolve{no} &

    \tabularnewline
    \midrule

     Ye et al.                         & \cite{Ye_Insecure_Code_Snippet_Detection}     & N/A & 
    J   &  
    \resolve{orange},M  &
    \resolve{no}  & 
    \resolve{no}  & 
    \resolve{yes}  & 
    \resolve{no} & 
    \resolve{yes} & 
    \resolve{no} & 
    \resolve{no} & 

    \tabularnewline
    \midrule

    Chen et al.                         & \cite{Chen_crowd_debugging}     & N/A & 
    J   &  
    \resolve{orange},M  & 
    \resolve{no}  & 
    \resolve{yes}  & 
    \resolve{yes}  & 
    \resolve{no} & 
    \resolve{yes} & 
    \resolve{yes},CX & 
    \resolve{yes} & 

    \tabularnewline
    \midrule

    Zhang et al.                         & \cite{Zhang_TensorFlow_program_bugs}     & N/A & 
    P   &  
    \resolve{no}  & 
    \resolve{no}  & 
    \resolve{no}  & 
    \resolve{yes}  & 
    \resolve{RC} & 
    N/A & 
    \resolve{no} & 
    \resolve{no} & 

    \tabularnewline
    \midrule

    Alhanahnah et al.                         & \cite{Alhanahnah_best_secure_coding_practice}     & N/A & 
    J   &  
    \resolve{yes}  & 
    \resolve{no}  & 
    \resolve{yes}  & 
    \resolve{yes}  & 
    \resolve{RC} & 
    N/A & 
    \resolve{no} & 
    \resolve{no} & 

    \tabularnewline
    \midrule

    Imai et al.                    & \cite{Imai_Time_Series_Analysis_Copy_and_Paste_Impact}     & N/A & 
    J   &  
    \resolve{yes}  & 
    \resolve{no}  & 
    \resolve{no}  & 
    \resolve{no}  & 
    \resolve{no} & 
    \resolve{orange},R & 
    \resolve{orange},Se & 
    \resolve{no} & 

    \tabularnewline
    \midrule

    Fischer et al.                    & \cite{Fischer_Effect_of_Google_Search_on_Software_Security}     & 03/2018 & 
    J   &  
    \resolve{orange},M  & 
    \resolve{no}  & 
    \resolve{no}  & 
    \resolve{yes}  & 
    \resolve{no} & 
    N/A & 
    \resolve{no} &
    \resolve{yes} & 

    \tabularnewline
    \midrule

    Almeida et al.                    & \cite{Almeida_Rexstepper}     & N/A & 
    JS   &  
    \resolve{yes}  & 
    \resolve{no}  & 
    \resolve{no}  &
    \resolve{yes}  & 
    \resolve{no} & 
    N/A & 
    \resolve{no} & 
    \resolve{no} & 
    \tabularnewline
    \midrule

    Islam et al.                    & \cite{Islam_repairing_deep_neural_networks}     & N/A & 
    P   &  
    \resolve{no}  & 
    \resolve{no}  & 
    \resolve{no}  & 
    \resolve{yes}  & 
    \resolve{RC} & 
    N/A & 
    \resolve{no} & 
    \resolve{no} & 
    \tabularnewline
    \midrule

    Mahajan et al.                    & \cite{Mahajan_Recommending_stack_overflow_posts}     & 03/2019 & 
    J   &  
    \resolve{orange},A  & 
    \resolve{no}  & 
    \resolve{no}  & 
    \resolve{yes}  & 
    \resolve{RC} & 
    \resolve{yes} & 
    \resolve{no} & 
    \resolve{yes} & 
    \tabularnewline
    \midrule

    Mahajan et al.                    & \cite{Mahajan_repairing_runtime_exceptions}     & 03/2019 & 
    J   &  
    \resolve{yes}  & 
    \resolve{no}  & 
    \resolve{no}  & 
    \resolve{yes}  & 
    \resolve{RC} & 
    N/A & 
    \resolve{no} &
    \resolve{yes} & 
    \tabularnewline
    \midrule

    Yadavally et al.                    & \cite{Yadavally_program_dependence_learning}     & N/A & 
    J,C   &  
    \resolve{no}  & 
    \resolve{no}  & 
    \resolve{no}  & 
    \resolve{yes}  & 
    \resolve{RC} & 
    \resolve{orange},S & 
    \resolve{no} & 
    \resolve{no} & 
    \tabularnewline
    \midrule

    Firouzi et al.                    & \cite{Firouzi_unsafe_code_context}     & 09/2018 & 
    C\#   &  
    \resolve{no}  & 
    \resolve{no}  & 
    \resolve{no}  & 
    \resolve{yes}  & 
    \resolve{no} & 
    \resolve{yes} & 
    \resolve{no} & 
    \resolve{no} & 
    \tabularnewline
    \midrule

    Ghanbari et al.                    & \cite{Ghanbari_fault_localization}     & 09/2018 & 
    P   &  
    \resolve{orange},F  & 
    \resolve{no}  &
    \resolve{no}  & 
    \resolve{yes}  &
    \resolve{no} & 
    \resolve{yes} & 
    \resolve{no} & 
    \resolve{no} &
    \tabularnewline
    \midrule

    Gao et al.                    & \cite{Gao_fixing_recurring_crash_bugs}     & N/A & 
    J   &  
    \resolve{yes}  & 
    \resolve{no}  & 
    \resolve{no}  & 
    \resolve{yes}  & 
    \resolve{no} & 
    N/A & 
    \resolve{no} & 
    \resolve{no} & 
    \tabularnewline
    \midrule

    M. Chakraborty                    & \cite{Chakraborty_bug_propagation}     & 01/2021 & 
    P   &  
    \resolve{no}  & 
    \resolve{no}  & 
    \resolve{no}  & 
    \resolve{yes}  & 
    \resolve{no} & 
    N/A & 
    \resolve{no} & 
    \resolve{no} & 
    \tabularnewline
    \midrule    

    Moghadam et al.                    & \cite{Moghadam_mutation_testing}     & N/A & 
    J   &  
    \resolve{no}  &
    \resolve{no}  & 
    \resolve{no}  & 
    \resolve{yes}  & 
    \resolve{no} & 
    \resolve{yes} & 
    \resolve{no} & 
    \resolve{no} &
    \tabularnewline
    \midrule

    Madsen et al.                    & \cite{Madsen_javaScript_promises}     & N/A & 
        JS   &  
        \resolve{orange},S  & 
        \resolve{no}  & 
        \resolve{yes}  & 
        \resolve{yes}  & 
        \resolve{no} & 
        \resolve{yes} & 
        \resolve{no} & 
        \resolve{no} & 
        \tabularnewline
        \midrule

    Jhoo et al.                    & \cite{Jhoo_static_analyzer}     & N/A & 
    P   &  
    \resolve{orange},S  & 
    \resolve{no}  & 
    \resolve{no}  &
    \resolve{yes}  & 
    \resolve{no} & 
    \resolve{yes} &
    \resolve{no} &
    \resolve{no} &
    \tabularnewline
    \midrule 

    \bottomrule
    \end{tabularx}

    \begin{tablenotes}[flushleft]
        \item \textbf{D1}: J $=$ Java, C $=$ C/C++, JS $=$ JavaScript, P $=$ Python, S $=$ Scala, P $=$ PHP
        \item \textbf{D2}: SM $=$ String matching, M $=$ Machine learning, N $=$ NLP, R $=$ Regular Expression, A $=$ Abstract Program Graph, P $=$ Pattern Recognition, F $=$ Fault Localization, S $=$ Static analysis
        \item \textbf{R2}: M $=$ Machine Learning, R $=$ Regular Expression, S $=$ Static analysis
        \item \textbf{R3}: S $=$ SourcererCC, SI $=$ Simian, CC $=$ CCFinder, CX $=$ CCFinderX, P = PDG, Se $=$ SeByte, M $=$ MOSS
    \end{tablenotes}

    \end{threeparttable}
\end{table}

\subsection{Comparison Criteria}\label{sec:criteria}
Our literature search yielded 42 relevant studies.
We reviewed the full text of each study and its artifacts to develop criteria for comparison of different works.
We found nine criteria, explained in the following, and \Cref{tab:comparision_table} compares the 43 considered studies based on these criteria.
Criteria \textbf{D1}--\textbf{D5} define dependencies on Stack Overflow data (used in \Cref{sec:knowledge_evolution}), where criteria \textbf{R1}--\textbf{R4} are relevant for our replication studies (in \Cref{sec:case_studies}).
For presentation, we adopt the pictogram-based visualization style common in \textit{Systematization of Knowledge} papers~\cite{soklib}.

    \noindent
    \textbf{D1.~Programming Languages}:
    Lists the programming language(s) of code snippets investigated in a study. If a study selects snippets independent of language, we assign \resolve{orange}. 

    \noindent
    \textbf{D2.~Code Scanners}:
    \textit{Off-the-shelve} scanners (\resolve{yes}) or \textit{custom} code classifiers (\resolve{orange}) can be used to find security weaknesses or general code quality issues in code snippets.
    For custom solutions, we list the techniques for finding security flaws, e.g., static analysis, machine learning, or graph query analysis.
    We assign \resolve{no} for studies detecting security issues manually.

    \noindent
    \textbf{D3.~Code Evolution}:
    Indicates whether a study considered~(\resolve{yes}) code evolution in their methodology or not (\resolve{no}).

    \noindent
    \textbf{D4.~Surrounding Context:}
    Comments and post descriptions are natural language text that provides context around code snippets.
    We assign \resolve{yes} if a paper utilizes surrounding context to classify code snippets (e.g., using NLP); otherwise, \resolve{no}.

    \noindent
    \textbf{D5. Sample Size Filter}:
    Studies may exclude code snippets based on specific criteria.
    For example, a study examining cryptographic API usage in Java might filter out snippets that do not use those APIs, or a study focusing on the effects of code revisions might exclude single-version snippets.
    We assign \resolve{no} if a study includes all snippets without filtering; otherwise, we assign \resolve{yes}.

    \noindent
    \textbf{R1.~Artifact Availability}:
    Code and data artifacts are important for revisiting and replicating prior findings; here, specifically, they are used to compare results on different versions of Stack Overflow datasets without undergoing the tedious (and error-prone) effort of re-implementing prior approaches.
    If paper artifacts (either code, data, or both) are unavailable, we assign \resolve{no}.
    We distinguish between artifacts that are \textit{fully} or \textit{partially} available.
    An artifact is \textit{fully functional available}~(\resolve{orange}) if both data and code are available and, most importantly, the code is free of bugs and can be used directly without changes to the code base.
    We assign \resolve{circle} if the code of a fully available artifact has bugs that require significant effort to fix.
    An artifact is \textit{partially} available if only code (\resolve{RC}) or only data (\resolve{LC}) is available.
    If the code of a partially available artifact requires significant bug fixing, we assign \resolve{CC}.

    \noindent
    \textbf{R2.~Language Detection}:
    The Stack Exchange and SOTorrent datasets do not state the programming languages of code snippets, requiring researchers to determine these.
    Researchers can rely on other data (\resolve{yes}) like \textit{tags} of posts, or they can employ code analysis tools (\resolve{orange}).
    When a tool is used, we indicate the underlying technique (e.g., machine learning or static analysis) used for language detection.
    If the authors did not mention their detection technique, we assign \textbf{N/A}.

    \noindent
    \textbf{R3.~Code Reuse Detection:}
Several studies indicate that developers reuse code from Stack Overflow, sometimes attributing the copied code snippets with their URL~\cite{Baltes2018Attribution}. 
This criterion differentiates between studies using attribution (\resolve{yes}) and studies using tools (\resolve{orange}), e.g., clone detection, to identify code reuse. If a tool is used, we give the name of the tool or technique. We assign \resolve{no} if code reuse is not considered.

        \noindent
    \textbf{R4.~Human-Centered Research}:
Differentiates human-centered studies of Stack Overflow data involving software developers as participants.
If the study’s methodology includes developer participation, we assign \resolve{yes}; otherwise, we assign a \resolve{no} for non-user-driven studies.
\medskip

\section{Relevance of Stack Overflow Evolution}\label{sec:knowledge_evolution}

\Cref{tab:comparision_table} lists the relevant works from our literature review.
In the following, we elaborate on the criteria for these papers (\textbf{D1}--\textbf{D5}) and how these can be affected by content evolution on Stack Overflow.
For space reasons, we restrict our explanations to the six highlighted studies in \Cref{tab:comparision_table}, with explanations of the remaining studies in \Cref{sec:appendix:systematization}.

The work by Zhang et al.~\cite{zhang_code_weaknesses} investigated whether code revisions on Stack Overflow are associated with improving or worsening the security of code snippets (\textbf{D3}: \resolve{yes}).
The study focused on \textit{C} and \textit{C++} code snippets (\textbf{D1}: C/C++) with at least 5 LoC (\textbf{D5}: \resolve{yes}). 
The \textit{CppCheck} static analysis tool was used to detect security weaknesses (\textbf{D2}: \resolve{yes}, \textbf{D4}: \resolve{no}).

The \textit{DICOS} tool by Hong et al.~\cite{dicos} analyzes the change history of C/C++ code snippets (\textbf{D1}: C/C++; \textbf{D3}: \resolve{yes}) to discover insecure code snippets by looking for changes in security-sensitive APIs, control flow information, and/or security-related keywords in the surrounding context (\textbf{D2}:~\resolve{orange},~N; \textbf{D4}: \resolve{yes}). However, the authors only considered the first and last revision of each snippet (\textbf{D5}: \resolve{yes}).

Fischer et al.~\cite{fischer2017SOConsideredHarmful} investigated the reuse of insecure Java code snippets in Android apps (\textbf{D1}: Java).
They employed a machine learning classifier to identify snippets with insecure usage of crypto APIs (\textbf{D2}: \resolve{orange},~M; \textbf{D5}: \resolve{yes}).
The authors considered neither code evolution (\textbf{D3}: \resolve{no}) nor surrounding context in their methodology (\textbf{D4}: \resolve{no}).

Follow-up work by Fischer et al.~\cite{FischerUsenix2019} relied on flagging Java snippets with crypto API misuse and suggesting more secure snippets (\textbf{D1}: Java; \textbf{D2}: \resolve{orange},~D; \textbf{D5}: \resolve{yes}). As in their preceding work, the authors considered neither code evolution (\textbf{D3}: \resolve{no}) nor context in their methodology (\textbf{D4}:~\resolve{no}).

Rahman et al.~\cite{rahman_snakes_in_paradies} analyzed Python code snippets (\textbf{D1}: Python) to identify insecure coding practices using string matching (\textbf{D2}: \resolve{orange}, SM). They focused on snippets from answers attributed in GitHub project (\textbf{D5}: \resolve{yes}) but did not consider code evolution or context (\textbf{D3:} \resolve{no}, \textbf{D4}: \resolve{no}).

Campos et al.~\cite{mining_rule_violations} studied JavaScript code snippets (\textbf{D1}: JavaScript) to identify rule violations using ESLint, a JavaScript linter (\textbf{D2: \resolve{yes}}). They relied exclusively on ESLint to detect security or code quality issues without considering the surrounding context (\textbf{D4}: \resolve{no}) or the evolution of snippets (\textbf{D3}: \resolve{no}). Further, they focused only on snippets with a minimum of 10 lines of code (\textbf{D5}: \resolve{yes}).

\begin{framed}
\textbf{MQ1: What aspects of Stack Overflow affect the results of prior research?} Criteria \textbf{D1}--\textbf{D5} in \Cref{tab:comparision_table} show that programming language-specific trends and evolution may affect the stability of results over time.
Further, most of these works leverage some form of code classification, making their results susceptible to changes as the code evolves.
Only four works rely upon code evolution in their methodology, showing that ongoing evolution beyond the study's timeframe can influence their results.
Additionally, twelve studies focus on the context surrounding code snippets, so any changes in context, like the addition of security-relevant comments, could also impact their findings.
\end{framed}

\section{Evolution of Stack Overflow}\label{sec:statistics}

Based on the identified aspects of Stack Overflow that \textit{may} affect prior research (MQ1), we now look at the global evolution of the programming languages and security-relevant contexts on Stack Overflow.
Measuring the security of code snippets directly, however, is a challenging task and requires dedicated methodologies (e.g., \cite{zhang_code_weaknesses,dicos,FischerUsenix2019,fischer2017SOConsideredHarmful}).
We address the evolution of code snippet security through case studies in \Cref{sec:case_studies} and focus here on programming languages and security-relevant edits and comments. 
\begin{figure}[t]
    \centering
    \includegraphics[width=1\linewidth]{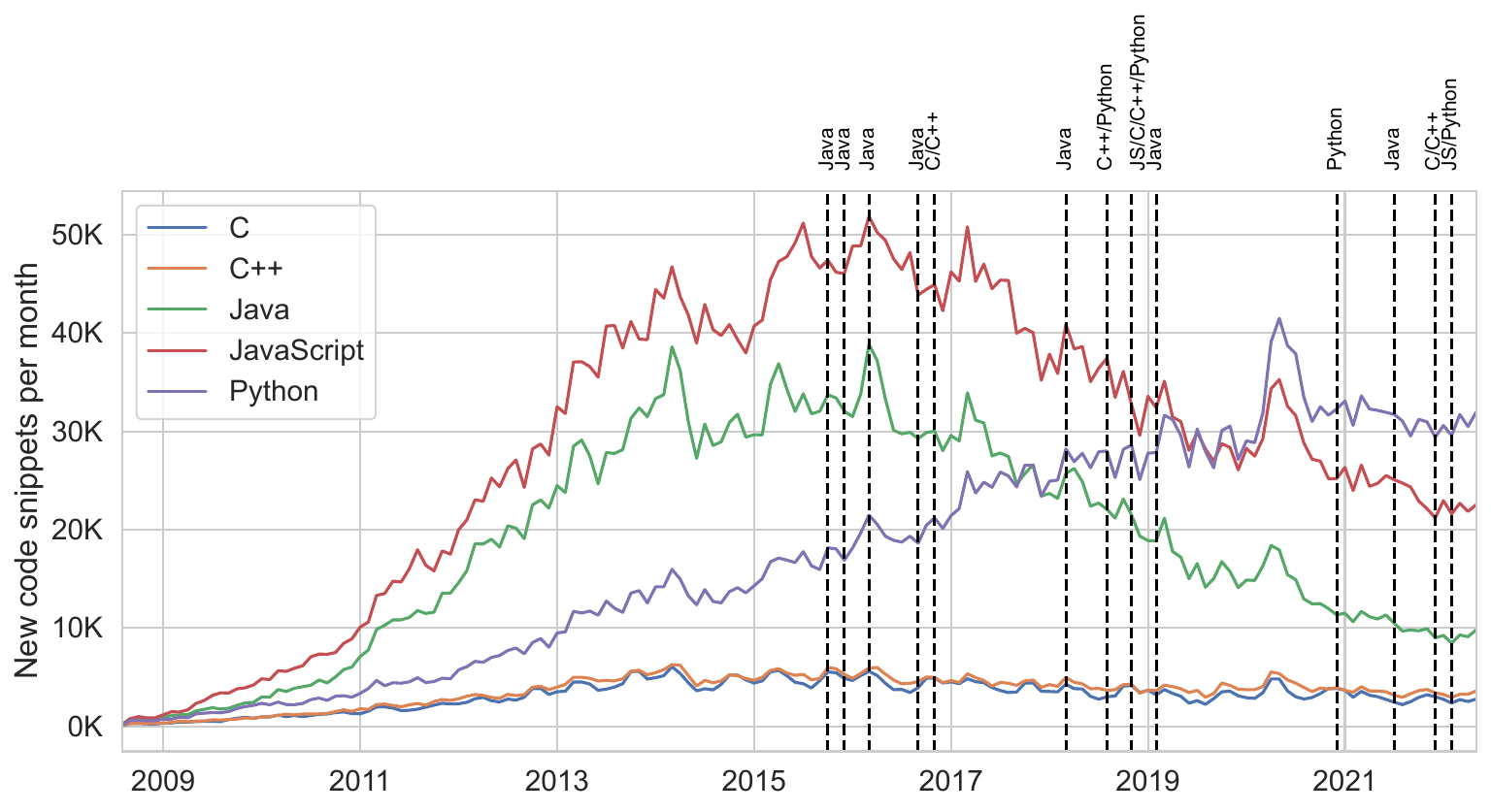}
    \caption{Added code snippets on Stack Overflow per month. Dashed lines indicate data collection points of snippets by the works in our systematization (if known; see \Cref{tab:comparision_table}).}
    \label{fig:prog_language_trend}
\end{figure}

\paragraph{Programming Languages}

\Cref{fig:prog_language_trend} depicts the monthly addition of new code snippets in the most considered programming languages in the works from \Cref{tab:comparision_table}.
The data shows that while C/C++ is relatively stable over time but at a comparatively low rate, Java and JavaScript peaked between 2013 and 2018 and have been shrinking on Stack Overflow since 2018.
We indicate with dashed lines when the papers from our systematization sampled their data (if known).

\begin{figure}[t]
    \centering
    \includegraphics[width=1\linewidth]{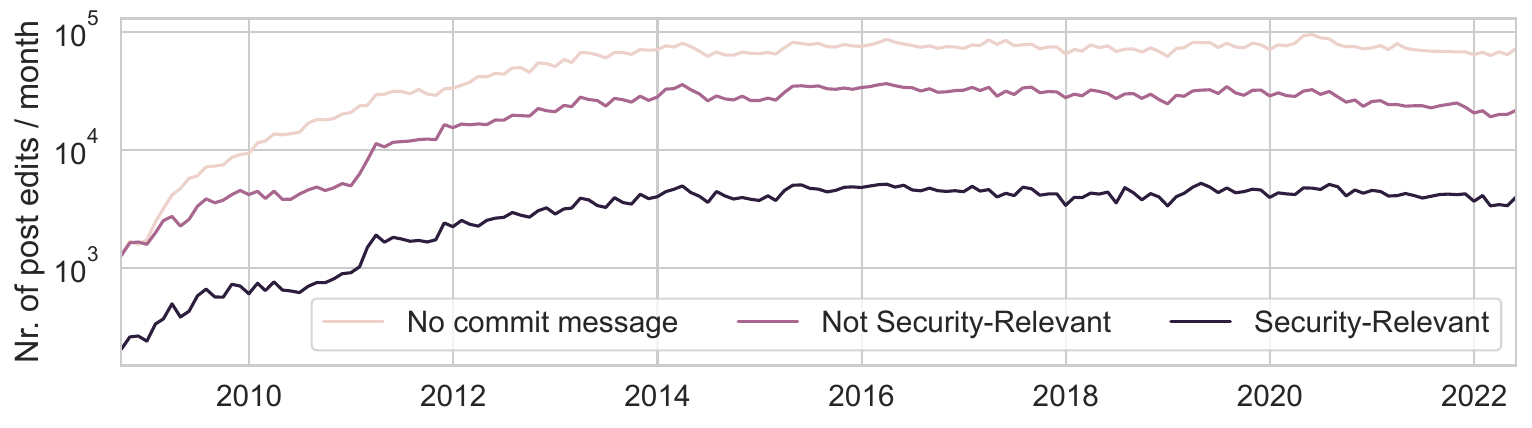}
    \caption{Number of monthly (30-day interval) post edits categorized by their security relevance.}
    \label{fig:commits_global}
\end{figure}

\paragraph{Security-Relevant Edits and Comments}
We found that the average code snippet on Stack Overflow receives 1.70 edits (max = 754, P$_{75}$ = 2, P$_{99}$ = 6).
A breakdown of average code snippet edits per language is provided in \Cref{tab:appendix:editslang} in \Cref{sec:appendix:systematization}.
\Cref{fig:commits_global} shows the number of monthly \textit{post} edits on Stack Overflow between 09/2008--06/2022, broken down by their security relevance and commit message.
To identify security-relevant edits, we apply the publicly shared NLP-based classifier by Jallow et al.~\cite{jallow_sp24, jallow24_osf_artifacts} on the commit messages of the edits.
We found 9,443,509 code edits without a commit message, 3,731,935 non-security-relevant commit messages, and 549,863 commit messages indicating security relevance.
Our data shows that 514,666 answer posts have received at least one security-relevant commit (max = 27).
We calculate the percentage of security-relevant commits (PSC):
\[
PSC=\frac{\text{Number of security-relevant commits} *100}{\text{Total number of commits}}
\]

\begin{figure}[t]
\centering
\begin{subfigure}[b]{\linewidth}
   \includegraphics[width=1\linewidth]{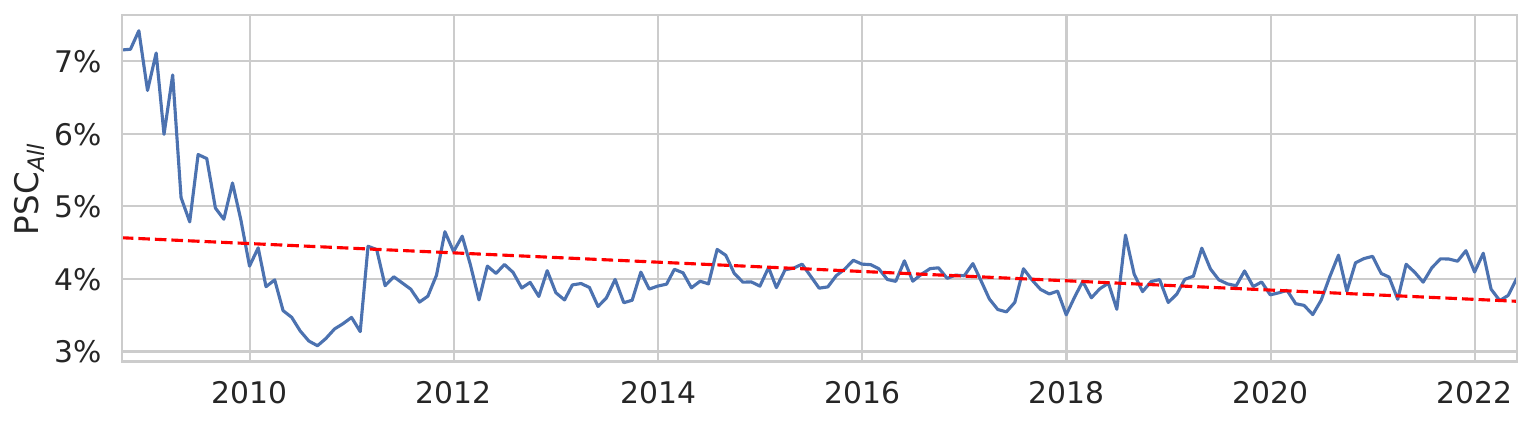}
   \caption{Including empty commit messages.}
   \label{fig:pcs_all_all} 
\end{subfigure}

\begin{subfigure}[b]{\linewidth}
   \includegraphics[width=1\linewidth]{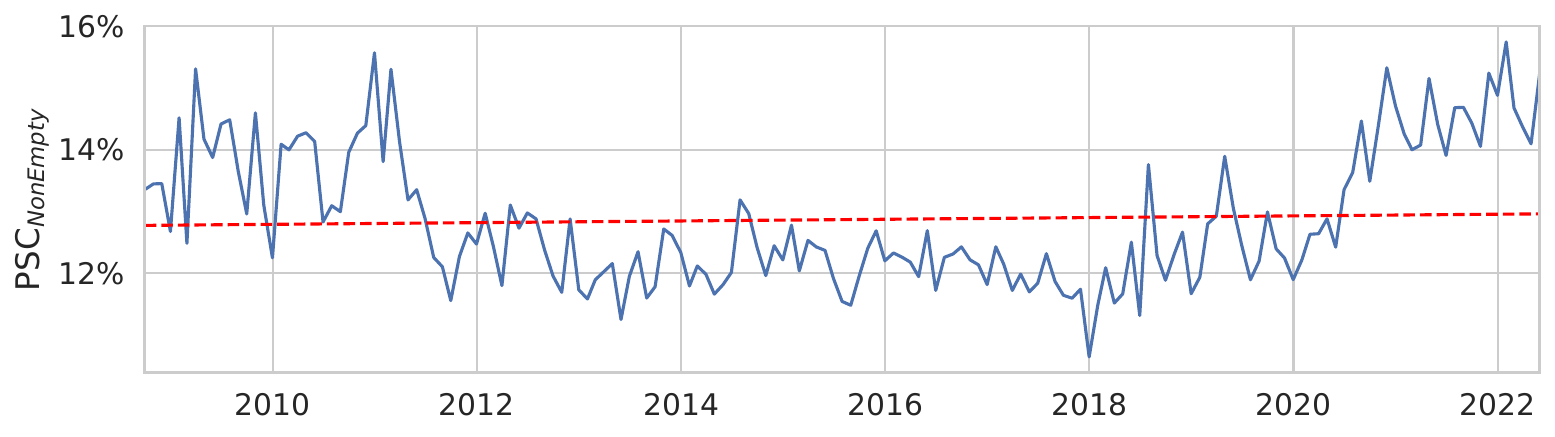}
   \caption{Excluding empty commit messages.}
   \label{fig:pcs_nonempty_all}
\end{subfigure}

\caption{Percentage of security-relevant commits (PSC) in monthly intervals. Dashed lines are fitted linear regressions.}
\label{fig:pcs_all}
\end{figure}

On June 20, 2020, Stack Overflow changed to CommonMark~\cite{so_commonmark} and adjusted all 338,622 nonconforming posts with automated edits.
For data sanity, we exclude these script-generated edits from the PSC calculation.

\Cref{fig:pcs_all} depicts the PSC for all code edits on Stack Overflow, where \Cref{fig:pcs_all_all} counts empty commit messages to the total number of commits and \Cref{fig:pcs_nonempty_all} excludes them.
The monthly average PSC is PSC$_{\textrm{All}}$$=4.2\%\pm0.1$ and PSC$_{\textrm{NonEmpty}}$$=13.1\%\pm0.2$ ($CI=95\%$).
We use the Kwiatkowski–Phillips–Schmidt–Shin~(KPSS) test~\cite{kpss} to test for stationarity of the PSC over the long term.
Since KPSS is known to exhibit a high rate of Type-I errors, indicating non-stationary too often, we additionally test non-stationary of the PSC with the Augmented Dickey Fuller~(ADF) test~\cite{adf}.
\Cref{tab:appendix:kpssadf} in \Cref{sec:appendix:systematization} details the results and illustrates the corresponding PSC.
In summary, when focusing on non-empty commit messages, we find that the overall PSC is trend stationary (KPSS $p>0.05$; ADF $p>0.05$), the PSC for C code snippets is difference stationary (KPSS $p>0.05$; ADF $p<0.001$) with an increasing PSC (i.e., the difference between data points is stationary and the PSC has a linear upward trend over time), and the PSC for JavaScript snippets is non-stationary (KPSS $p<0.05$; ADF $p>0.05$).
All but C++ and Python edits show a stationary PSC when considering non-empty commit messages.
We additionally fitted linear regressions to the PSC to indicate their long-term developments (see \Cref{sec:appendix:systematization}).

Regarding comments, we found that 3,991,533 (8.1\%) of 49,087,103 comments were security-relevant.
These relevant comments are for 3,128,208 posts, of which 98,139 also received a security-relevant commit message.
The mean percentage of security-relevant comments PSC$_{\textrm{Comments}}$ is 7.78$\pm$0.143, where posts with C/C++ snippets have a significantly higher mean PSC$_{\textrm{Comments}}$ between 11--12\% (see \Cref{tab:appendix:psc_comments} in \Cref{sec:appendix:systematization}).
Further, KPSS and ADF tests show that the overall PSC$_{\textrm{Comments}}$ is difference stationary and that the PSC$_{\textrm{Comments}}$ for C/C++, Java, JavaScript, and Python posts is non-stationary (see \Cref{tab:appendix:kpssadf_comments} in \Cref{sec:appendix:systematization}).
A fitted linear regression confirms this increasing trend of PSC$_{\textrm{Comments}}$ ($R^2=0.93$, $p<0.001$ for all comments).
A potential explanation for this development could be the decreasing number of new code snippets (e.g., because simple questions are now posted to GenAI tools) and a continued community effort to curate the Stack Overflow content.

\begin{framed}
\textbf{MQ2: How much do Stack Overflow code snippets and surrounding context evolve?}
Our data shows that programming languages trend differently in their overall number of added snippets and their ratio of security-relevant edits.
Thus, studies focusing on particular languages will likely find a different landscape when conducted at different times.
Further, many comments raised security-relevant issues but were largely not on posts that received a security-relevant edit.
Over time, the ratio of security-relevant comments steadily increased, indicating that the community strives to improve content quality.
\end{framed}

\section{Replication Case Studies}\label{sec:case_studies}
\begingroup
    \sisetup{detect-weight,
            output-decimal-marker={,},
            group-minimum-digits=2,
            group-separator={.}
            }
\begin{table*}[t]
    \caption{\hlblue{Results by Zhang et al.} (cf.~Table 1 in \cite{zhang_code_weaknesses}) versus our \hlgreen{replication study} using SOTorrent22 and Cppcheck v2.13.}
    \label{tab:zhang_dataset_comparison}
    \centering
    \footnotesize

    \begin{tabular}{lrrr|lrrr}\toprule
    \multicolumn{4}{c|}{\textbf{Based on SOTorrent18 (Original)}} & \multicolumn{4}{c}{\textbf{Based on SOTorrent22 (Replication)}}\\
    \midrule
     & \textit{Answer \#} & \textit{Code Snippet \#} & \textit{Code Version \#} &  & \textit{Answer \#} & \textit{Code Snippet \#} & \textit{Code Version \#}\\\midrule
    SOTorrent & 867,734 & 1,561,550 & 1,833,449 & SOTorrent & 1,096,380 & 1,944,378 & 2,340,975 \\
    \rowcolor{gray!15}LOC $>=$ 5 & 527,932 & 724,784 & 919,947 & LOC $>=$ 5 & 695,326 & 938,643 & 1,234,443 \\
    Guesslang & 490,778 & 646,716 & 826,520 & Cppcheck 2.13 & 323,321 & 388,749 & 507,997 \\
    \rowcolor{gray!15}$Code_w$ & \cellcolor{blue!25} \textbf{11,235} & \cellcolor{blue!25} \textbf{11,748} & \cellcolor{blue!25} \textbf{14,934} & $Code_w$ & \cellcolor{green!25} \textbf{28,521} & \cellcolor{green!25} \textbf{30,254} & \cellcolor{green!25} \textbf{38,248} \\
    \bottomrule
    \end{tabular}
\end{table*}
\endgroup

We present replication case studies to answer \textbf{MQ3} as to whether the findings of prior research that relied on specific versions of the Stack Overflow dataset change due to StackOverflow evolution.
In contrast to \Cref{sec:knowledge_evolution} and \Cref{sec:statistics}, we aim to find concrete evidence for the impact of Stack Overflow evolution on research results.
In this section, we detail six replication case studies~\cite{zhang_code_weaknesses,dicos,fischer2017SOConsideredHarmful,FischerUsenix2019,rahman_snakes_in_paradies,mining_rule_violations}.

\paragraph*{Excluded papers.} As shown in \Cref{tab:comparision_table}, we focused on six papers for replication.
In 16 studies~\cite{bai_insecure_code_propagration, Bagherzadeh_akka_actor_bugs, rahman_reusability_insight, Schmidt_CopypastaVulGuard, Pan_deep_learning_bug_fixing, Zhang_TensorFlow_program_bugs, Islam_repairing_deep_neural_networks, Yadavally_program_dependence_learning, Firouzi_unsafe_code_context, Chakraborty_bug_propagation, Moghadam_mutation_testing}, the detection of security weaknesses and bugs in code snippets was not automated; instead, these studies relied on manual processes to label and identify security issues and/or bugs in code snippets.
This approach typically involved human annotators reviewing and classifying the code for potential weaknesses, which is hard to compare in a replication study with different human evaluators. 
On the other hand, three studies\cite{Ren_Demystify_API_Usage_Directives, Chen_crowd_debugging, Fischer_Effect_of_Google_Search_on_Software_Security} leveraged machine learning techniques to automate the detection of security vulnerabilities in code snippets.
While these studies employed advanced algorithms to facilitate automated analysis, they failed to release the underlying source code and datasets, particularly the training data used for training their models, making replication hard.
Additionally, nine studies~\cite{reinhardt_augmenting_stack_overflow, Liu_deepanna, Ye_Insecure_Code_Snippet_Detection, chen_intelligent_detection_system, Alhanahnah_best_secure_coding_practice, Almeida_Rexstepper, Gao_fixing_recurring_crash_bugs, Madsen_javaScript_promises, Jhoo_static_analyzer} did not specify which version of the dataset they used to collect code snippets from Stack Overflow.
Lastly, nine studies~\cite{acar2016YouGetWhereYouLook,toxicCodeSnippets,bai_insecure_code_propagration,Ren_Demystify_API_Usage_Directives,Fischer_Effect_of_Google_Search_on_Software_Security,Mahajan_Recommending_stack_overflow_posts,Chen_crowd_debugging,Mahajan_repairing_runtime_exceptions,rahman_reusability_insight} were excluded in favor of non-user-driven studies.

\paragraph*{Datasets} The selected papers investigated two kinds of Stack Overflow datasets. 
First, Fischer et al.~\cite{fischer2017SOConsideredHarmful, FischerUsenix2019} used the official data dump provided by Stack Exchange, Inc~\cite{SOOfficialDump}.
Here, we perform a replication study using the September 2023 version (denoted \textit{StackExchange23}).
Second, Zhang et al.~\cite{zhang_code_weaknesses}, Hong et al.~\cite{dicos}, Campos et al.~\cite{mining_rule_violations} and Rahman et al.~\cite{rahman_snakes_in_paradies} used the \textit{SOTorrent} dataset by Baltes et al.~\cite{baltes2018sotorrent}.
Here, we perform a replication study using a newer dataset version.
The authors of the SOTorrent dataset stopped providing new releases in December 2020.
Fortunately, the tool used to make new releases of the dataset is open-source\cite{sotorrent_post_history_extractor}, allowing us to create a new release of SOTorrent (denoted \textit{SOTorrent22}) based on the \textit{June 2022} version of the Stack Exchange dataset.

\paragraph{Notation} Colored text is used to distinguish between results from the \hlblue{original} studies and those from our \hlgreen{replication}.

\subsection{Case Study 1: C/C++ Code Weaknesses}\label{sec:zhang}
Zhang et al.~\cite{zhang_code_weaknesses} studied whether revisions to C/C++ snippets increase or decrease the snippets' security.
Their work addressed the following questions:
\textbf{RQ1}: \textit{What are the types of code weaknesses that are detected in C/C++ code snippets on Stack Overflow?}
\textbf{RQ2}: \textit{How do code with weaknesses evolve through revisions?}
\textbf{RQ3}: \textit{What are the characteristics of the users who contributed code with weaknesses?}

\subsubsection{Original Methodology}
The authors employed a data-driven approach (\textbf{R4}: \resolve{no}), focusing exclusively on answer posts from the \textit{SOTorrent18} dataset (released in 12/2018), without considering code reused from Stack Overflow (\textbf{R3}: \resolve{no}).
The left-hand side of \Cref{tab:zhang_dataset_comparison} shows the originally collected data.
The authors extracted 867,734 answers containing 1,561,550 code snippets with C/C++ tags with 1,833,449 versions.
From this data set, they filtered all snippets with less than five LoC.
This resulted in 724,784 code snippets (with 919,947 versions) from 527,932 answers.
However, the authors noticed that using \textit{tags} alone is insufficient to determine the language of code snippets, i.e., not all snippets contain valid C/C++ code.
Using the \textit{Guesslang} machine learning classifier~\cite{guesslang} they filtered non-C/C++ code snippets (\textbf{R2}: \resolve{orange}, \textbf{M}).
This resulted in 646,716 code snippets (having 826,520 versions) from 490,778 answers.
In a final step, the authors leveraged the \textit{CppCheck} static analysis tool to scan all 826,520 versions to detect security weaknesses. 
We reuse the terms by the authors to denote code snippets, snippet versions, and answers with security weaknesses as $Code_w$, $Version_w$, and $Answer_w$, respectively.The authors' final dataset to answer their research questions are $Version_w=14,934$ from $Code_w=11,748$ in $Answer_w=11,235$. 

\subsubsection{Re-Implementation} 

The authors did not publicly release their source and data artifacts (\textbf{R1}: \resolve{no}) but provided us on request with a CSV file of the $Answer_w=11,235$.
The unavailability of the code artifact forced us to re-implement their methodology for replication.

\begin{figure}[t]
 \centering
  \includegraphics[width=\linewidth]{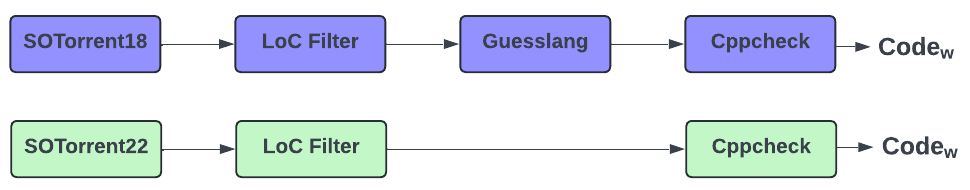}
  \caption{ Comparison of \hlblue{Zhang et al.'s}~\cite{zhang_code_weaknesses} methodology with the approach used in our \hlgreen{replication} study.
 }
\label{fig:zhang:methodology_flowchart}
\end{figure}

An exact re-implementation of Zhang et al.'s~\cite{zhang_code_weaknesses} methodology was impossible.
The Guesslang version used by the authors is outdated and unavailable.
Replacing the old version with the newest version, v2.2.1, yielded significantly different numbers for the same \textit{SOTorrent18} data set (490,778 vs.249,451).
Additionally, we noticed that Guesslang v2.2.1 misclassified 4,901 C/C++ answers (out of $Answer_w=11,235$) from the authors' results as Java.
Given the substantial differences of Guesslang v2.2.1, we devised an alternative methodology that skips Guesslang (see \Cref{fig:zhang:methodology_flowchart}).
We relied solely on Cppcheck to identify valid C/C++ snippets and detect security weaknesses.
Cppcheck attempts to compile code snippets, failing if the snippets are not valid C/C++ code.
Thus, Cppcheck formed an additional language detection step in the original methodology, which is now the only such step.
The intuition is that any invalid C/C++ code detected by Guesslang would be rejected by Cppcheck.

\begin{table*}[ht]
    \caption{Side-by-side summary of the \hlblue{main claims in Zhang et al.'s paper}~\cite{zhang_code_weaknesses} for \textbf{RQ1} and \textbf{RQ2} and \hlgreen{claims based on the results of our replication} using \textit{SOTorrent22}.}
    \label{tab:zhang_rq1_rq2_results_comparison}

    \footnotesize
    \centering
    \begin{tabularx}{\linewidth}{@{} >{\RaggedRight\hsize=1.0\hsize}X |
                                >{\RaggedRight\hsize=1.0\hsize}X
                                }
        \toprule
      \multicolumn{1}{c|}{\textbf{Original Results Based on SOTorrent18}}   & \multicolumn{1}{c}{\textbf{Replication Results Based on SOTorrent22}}   \\  \midrule
      \rowcolor{gray!18}\multicolumn{2}{c}{\textbf{RQ1} \textit{What are the types of code weaknesses that are detected in C/C++ code snippets on Stack Overflow?}} \\  \midrule

      The authors found 36\% (i.e., \hlblue{32 out of 89}) of all the C/C++ CWE types in C/C++ code snippets on Stack Overflow.  &
      We found 37\% (i.e., \hlgreen{33 out of 89}) of all the C/C++ CWE types in C/C++ code snippets on \SO.
      CWE-476 is a newly introduced type for snippets after December 2018 and has 1,159 instances. \\  \midrule

      The authors identified \hlblue{12,998} CWE instances within the latest versions of the \hlblue{7,481} answers. &
      We identified \hlgreen{7,679} CWE instances within the latest versions of the \hlgreen{5,721} answers. \\ \midrule
      
      The authors found CWE-758 to be the \hlblue{sixth} most prevalent CWE type in C/C++ code snippets with \hlblue{482} (3.7\%) instances. & We found CWE-758 to be the \hlgreen{second} most prevalent CWE type in C/C++ code snippets with \hlgreen{10,911} instances. \\ \midrule 

      The authors found \hlblue{10,533} CWE instances in the TOP-6 most prevalent CWE types affecting C/C++ code snippets on \SO. & While we found \hlgreen{42,984} CWE instances in the TOP-6 most prevalent CWE types affecting C/C++ code snippets on \SO. \\ \midrule

      \rowcolor{gray!15}\multicolumn{2}{c}{\textbf{RQ2} \textit{How does code with weaknesses evolve through revisions?}} \\ \midrule

      As the number of revisions increases from one to $\geq 3$, the proportion of improved $Code_w$ increases from \hlblue{30.1\% to 41.8\%}. & As the number of revisions increases from one to $\geq 3$, the proportion of improved $Code_w$ increases from \hlgreen{3.1\% to 7.4\%}. \\  \midrule

      In $Code_w$ with different rounds of revisions, \hlblue{a larger proportion} of code snippets have reduced rather than increased the number of security weaknesses. & We observed \hlgreen{a smaller proportion} of code snippets whose associated security weaknesses reduced with different rounds of code revisions. \\  \midrule

      The authors found \hlblue{92.6\%} (i.e., 10,884) of the 11,748 $Code_w$ had weaknesses introduced when their code snippets were initially created on Stack Overflow. They found \hlblue{10,884} $Code_w$ introduced in the snippets' first version. & We found \hlgreen{93.1\%} (i.e., 28,155) of the 30,254 $Code_w$ had weaknesses introduced in their first version.  However, we discovered significantly more $Code_w$ introduced when code snippets were initially created: \hlgreen{28,155}.\\  \midrule
      
      \hlblue{69\%} (i.e., 8,103 out of 11,748) of the $Code_w$ has never been revised. & \hlgreen{80.6\%} (i.e., 24,388 out of 30,254)  of $Code_w$ has never been revised. \\

    \bottomrule
    \end{tabularx}
\end{table*}

A second challenge for re-implementation was that the authors did not specify the exact Cppcheck version used in their study and did not respond to multiple inquiries.
Ultimately, we resorted to brute force by testing 15 different versions of Cppcheck on the provided $Answer_w$ list against the reported results in the paper. 
We found version v1.86 to be closest to the original results.
This version detected $Answer_w=15,724$, of which 11,142 (99.2\% of 11,235) were also identified by the authors.
We surmise the additional 4,489 answers result from erroneous filtering with Guesslang in the original approach.

We evaluated our re-implementation with the \textit{SOTorrent18} data set that was also used by the authors, aiming to verify that our approach is within an acceptable error margin from the authors' results and allowing us to replicate their study with a newer data set faithfully.
\Cref{tab:appendix_zhang_evaluation} in \Cref{sec:appendix:casestudy_1} compares with the original findings and shows that our approach resembles the original methodology closely enough.

\subsubsection{Replication}
We replicate their findings using \textit{SOTorrent22}, released four years after the original dataset.
We use Cppcheck v2.13 to show how the results would differ if the study were conceived years later.
\Cref{tab:appendix:zhang} in \Cref{sec:appendix:casestudy_1} shows results for different combinations of Cppcheck and SOTorrent versions.

\Cref{tab:zhang_dataset_comparison} compares the \hlblue{original results} with the \hlgreen{replication} based on \textit{SOTorrent22} and Cppcheck v2.13.
We found that the number of code snippets with $LoC >= 5$ increased between 2018 and 2022 (\hlblue{724,784} $\nearrow$ \hlgreen{938,643}), a growth rate of 29.5\%.
Among these, the $Code_w$ also increased (\hlblue{11,748} $\nearrow$ \hlgreen{30,254}), a growth rate of 157.5\%.

\Cref{tab:zhang_rq1_rq2_results_comparison} presents a side-by-side comparison of the authors' claims and our findings based on the newer dataset for the authors' \textbf{RQ1} and \textbf{RQ2}. 
Below, we summarize the main points.

\textbf{\textit{Revisiting RQ1 Findings}}: We found that an additional CWE type (CWE-476) has appeared since December 2018, which is now the sixth most prevalent CWE type.
Similarly, the authors identified \hlblue{12,998 CWE instances} in the latest versions of \hlblue{7,481 answers}, whereas we found \hlgreen{7,679 instances in 5,721 answers}. 
Further, we noticed a shift in the ranking of CWE types: CWE-758 climbed \hlblue{6th} $\nearrow$ \hlgreen{2nd}; CWE-401 dropped \hlblue{2nd} $\searrow$ \hlgreen{3rd}; CWE-775 fell \hlblue{3rd} $\searrow$ \hlgreen{7th}.

\begin{formal}
We found that several of Zhang et al.'s~\cite{zhang_code_weaknesses} original conclusions regarding the types of code weaknesses are no longer valid.
\textit{SOTorrent22} contains proportionally more vulnerable snippets with different ratios for CWE types and the emergence of a new CWE type.
\end{formal}

\textbf{\textit{Revisiting RQ2 Findings}}: 
The authors noted that as the number of revisions to $Code_w$ increased from 1 to 3+, the proportion of \textit{improved} $Code_w$ rose from \hlblue{30.1\% to 41.8\%} (see \Cref{tab:zhang_proportion_comparison} in \Cref{sec:appendix:casestudy_1}).
As a result, the authors concluded that a larger proportion of $Code_w$ has reduced rather than increased, indicating that revisions improve code security.
In contrast, our results showed significantly lower improvement rates.
As revisions increased from 1 to 3 or more, the proportion of improved $Code_w$ only rose from \hlgreen{3.1\% to 7.4\%}.

\begin{formal}
Our replication shows that if the authors had conducted their study 4 years later, they would have observed only a slight increase in the proportion of \textit{improved} $Code_w$.
\end{formal}

\textbf{\textit{Revisiting RQ3 Findings}}:

\begin{table*}[ht]
    \caption{Side-by-side summary of the \hlblue{main claims in the original paper} by Zhang et al.~\cite{zhang_code_weaknesses} for their \textbf{RQ3} and \hlgreen{claims based on the results of our replication} using \textit{SOTorrent22}.}
    \label{tab:zhang_rq3_results_comparison}

    \small
    \centering
    \begin{tabularx}{\linewidth}{@{} >{\RaggedRight\hsize=1.0\hsize}X |
                                >{\RaggedRight\hsize=1.0\hsize}X
                                }
        \toprule
      \multicolumn{1}{c}{\textbf{Original Results Based on SOTorrent18}}   & \multicolumn{1}{c}{\textbf{Replication Results Based on SOTorrent22/Cppcheck v2.13}}   \\  \midrule
    
      \rowcolor{gray!15}\multicolumn{2}{c}{\textbf{RQ3} \textit{What are the Characteristics of the Users who Contributed to Code With Weaknesses?}} \\ \midrule

      The majority of the C/C++ $Version_w$ was contributed by a small number of users. \hlblue{72.4 percent (i.e., 10,652)} of $Version_w$ were posted by 36 percent (i.e., \hlblue{2,292}) of users. \hlblue{64.0 percent (i.e., 4,070)} of the users who contribute $Version_w$ have contributed only one $Version_w$.&
      The majority of the C/C++ $Version_w$ were contributed by a small number of users. \hlgreen{79.5 percent (i.e., 38,481)} of $Version_w$ were posted by 36 percent (i.e., \hlgreen{4,625}) of users. \hlgreen{86.2 percent (i.e., 11,077)} of the users who contribute $Version_w$ have contributed only one $Version_w$. (See \Cref{fig:zhang_rq3_accumuluative})\\\midrule

      Among all the \hlblue{85,165} users who posted C/C++ code snippets, only \hlblue{7.5} percent (i.e., \hlblue{6,361}) of them posted code snippets that have weaknesses. &
      Among all the \hlgreen{75,779} users who posted C/C++ code snippets, \hlgreen{17.0} percent (i.e., \hlgreen{12,845}) of them posted code snippets that have weaknesses. \\\midrule

      More active users are less likely to introduce $Code_w$. The weakness density of a user’s code drops when the number of contributed code revisions by the user increases. Users with higher reputations tend to introduce fewer CWE instances in their contributed code versions. &
      A Pearson correlation analysis and a linear regression indicate only a statistically significant but \hlgreen{weak} negative relationship between the number of revisions and weakness density. A Pearson correlation analysis and a linear regression show an \hlgreen{extremely small} effect size for reputation as a predictor for the number of CWE instances. \\\midrule

      \hlblue{78.0} percent of users contribute code with only one CWE type. Furthermore, \hlblue{42.2} percent (i.e., \hlblue{2,686}) of the users contribute only one CWE instance in all their $Version_w$. \hlblue{81.8} percent (i.e., \hlblue{5,206}) of the users contribute less than ﬁve CWE instances in all their $Version_w$. &
      \hlgreen{74.6} percent of users contribute code with only one CWE type. Furthermore, \hlgreen{50.9} percent (i.e., \hlgreen{6,533}) of the users contribute only one CWE instance in all their $Version_w$. \hlgreen{85.3} percent (i.e., \hlgreen{10,938}) of the users contribute less than ﬁve CWE instances in all their $Version_w$.\\\midrule

      Users tend to commit \hlblue{the same types} of CWE instances repeatedly. We observe that \hlblue{37.7} percent of users are likely to introduce a single type of CWE instances in their posted code versions. &
      Users tend to commit \hlgreen{different types} of CWE instances with similar probabilities. We observe that \hlgreen{15.6} percent of users are likely to introduce a single type of CWE instances in their posted code versions, but \hlgreen{32.7} percent of users introduce several types of CWE. Further, there is a \hlgreen{shift in the CWE types}: CWE 457, 476, 758, and 788 are significantly more frequent, while CWE 908 disappeared.\\
      
    \bottomrule
    \end{tabularx}
\end{table*}

\Cref{tab:zhang_rq3_results_comparison} compares the results related to RQ3 and we elaborate on these in the following.

\begin{figure}[t]
    \centering
    \includegraphics[width=\linewidth]{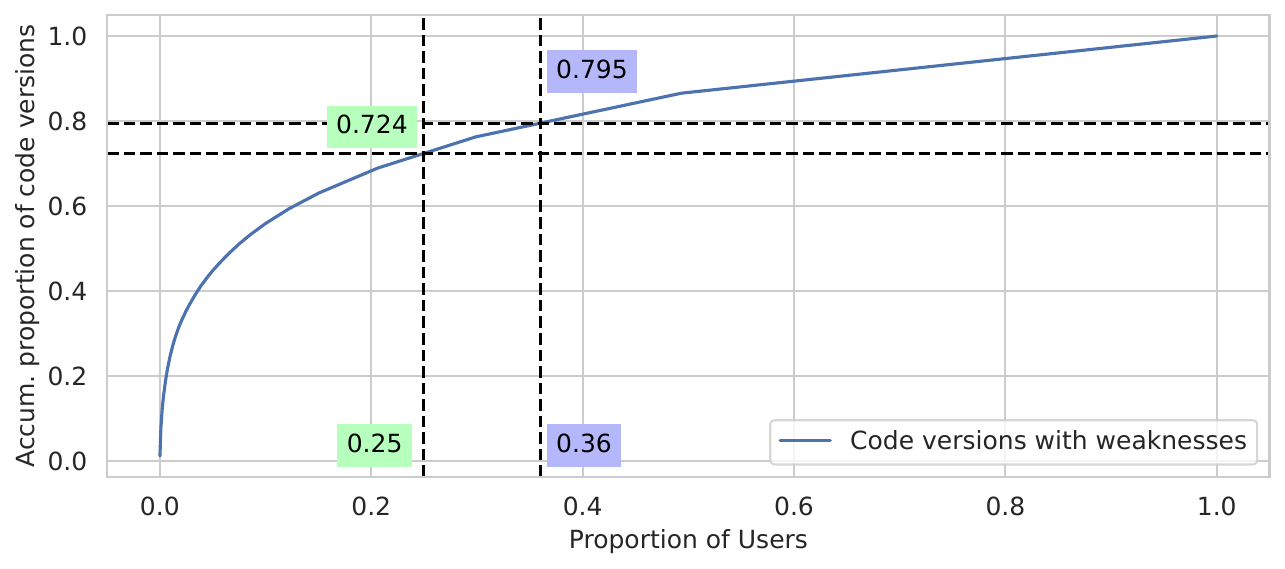}
    \caption{The accumulative proportion of $Version_w$ posted by the proportion of users. Annotations indicate the \hlblue{original} (cf.~Figure~9 in \cite{zhang_code_weaknesses}) and \hlgreen{replicated} significant data points. }
    \label{fig:zhang_rq3_accumuluative}
\end{figure}

The authors found that \textit{``the majority of the C/C++ $Version_w$ were contributed by a small number of users''} and that \textit{``72.4 percent (i.e., 10,652) of $Version_w$ were posted by 36 percent (i.e., 2,292) of users''}.
In our replication, we found a shift where an even smaller number of users contributed $Version_w$, see \Cref{fig:zhang_rq3_accumuluative}.
We found that 72.4 percent (i.e., 35,034) of $Version_w$ were posted by 25 percent (i.e., 3,205) of users.
In our data set, 36 percent (i.e., 4,625) of users contributed 79.5 percent (i.e., 38,481) of the $Version_w$.
Moreover, Zhang et al. reported that \textit{``64.0 percent (i.e., 4,070) of the users who contribute $Version_w$ have contributed only one $Version_w$''}.
We found that 86.2 percent (i.e., 11,077) of users contributed only one $Version_w$.
Further, they reported that \textit{``among all the 85,165 users who posted C/C++ code snippets, only 7.5 percent (i.e., 6,361) of them posted code snippets that have weaknesses''}.
In contrast, in our replication study, 17.0 percent (i.e., 12,845) of 75,779 users contributed code snippets with weaknesses.

\begin{figure}[t]
    \centering
    \includegraphics[width=\linewidth]{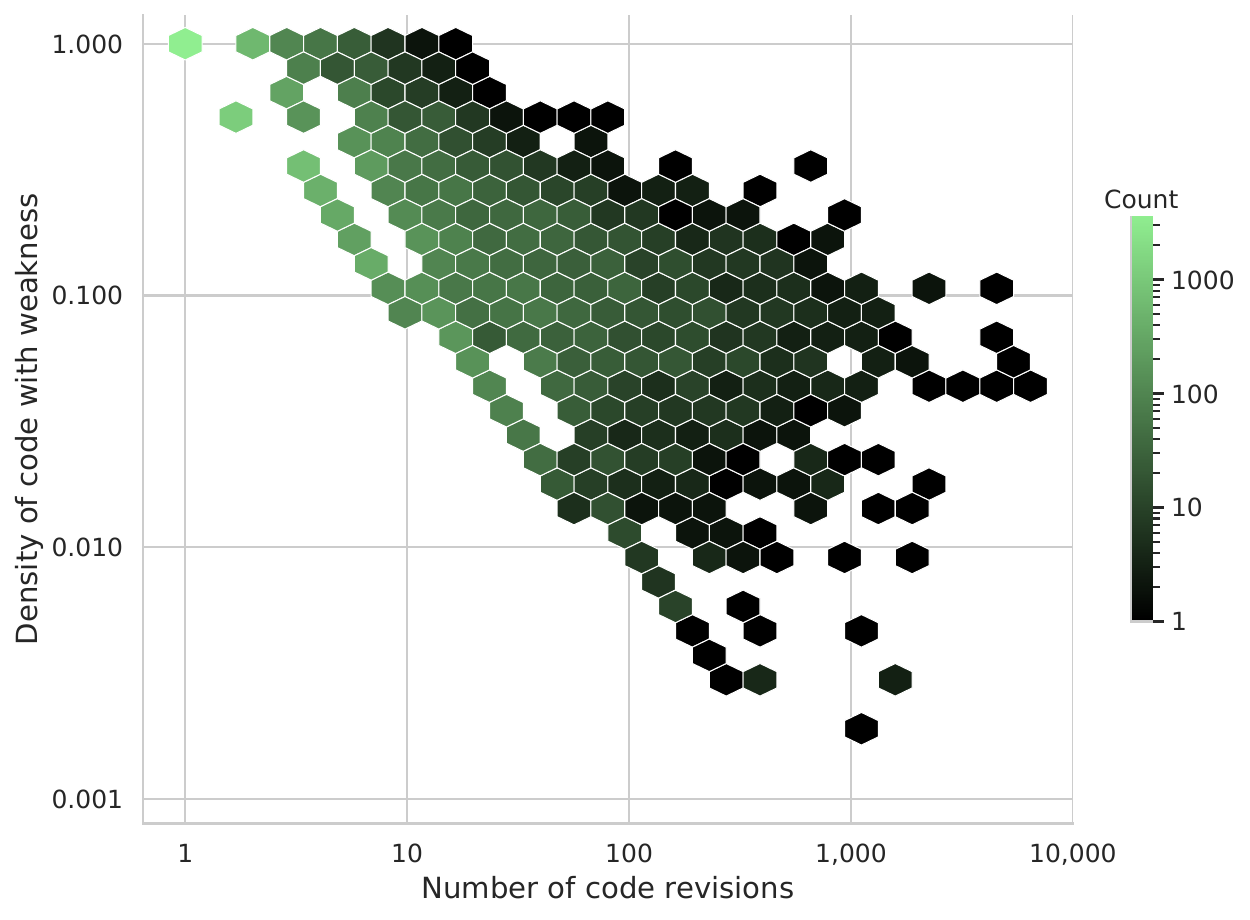}
    \caption{As the number of code revisions increases for a user, the density of contributed $Version_w$ by that user drops. Based on data from \hlgreen{replication study}. (cf.~Figure~10 in \cite{zhang_code_weaknesses})}
    \label{fig:zhang_rq3_weaknessdensity}
\end{figure}

Next, Zhang et al.~explored the connection between user activity and code weaknesses.
They found that ``more active users are less likely to introduce $Code_w$.''
We depict the same connection in \Cref{fig:zhang_rq3_weaknessdensity}, adopting the authors' plot style.
The authors concluded that \textit{``the weakness density of a user’s code drops when the number of contributed code revisions by the user increases.''}
We explored the relation between the number of code revisions and the density of contributed $Version_w$ by users with statistical testing.
A Pearson correlation analysis revealed a weak correlation, $r(4) = -0.190, p < 0.001 $, suggesting that as the number of revisions increases, the density of code with weaknesses decreases slightly.
A linear regression analysis was conducted to examine further the relationship between the number of revisions (independent variable) and the weakness density (dependent variable). The results indicated that the model was statistically significant $F(1, 12843) = 478.4$, $p < 0.001$, suggesting a significant inverse association of code revision count with weakness density. However, the model explained only a small proportion of the variance in weakness density $R^2 = 0.036$.
These findings suggest that the weakness density is expected to decrease by $0.0005$ for each additional revision. However, the low $R^2$ value indicates that the number of revisions explains only 3.6\% of the variability in weakness density, suggesting that other factors may also play a significant role.
Overall, the correlation and regression analyses support the conclusion that there is a statistically significant but weak inverse relationship between the number of revisions and weakness density. Further research is needed to explore additional variables and potential non-linear relationships that might better explain the variability in weakness density.

More details on the replication findings for \textit{RQ3} are provided in \Cref{sec:appendix:casestudy_1_rq3} and  \Cref{sec:appendix:casestudy_1_rq3_mistake}.

\begin{formal}
We found that the fraction of users with vulnerable C/C++ snippets more than doubled compared to the original findings.
Moreover, the number of users that contributed just one vulnerable snippet also increased.
Further, the authors reported that users who contributed multiple vulnerable snippet versions repeatedly contributed the same CWE type.
We found that these users contribute different types of CWE with the same likelihood.
\end{formal}

\subsection{Case Study 2: Discovering Insecure Code}\label{sec:dicos}
Hong et al.~\cite{dicos} built \DC to discover insecure code snippets by examining snippet revisions for changes in security-sensitive APIs, security-related keywords, and control flows.
A code snippet is classified as insecure if at least two types of changes (see §3 in~\cite{dicos}) occurred between its initial and most recent version.
The authors used tags to identify the programming language of code snippets (\textbf{R2}: \resolve{yes}).
Although the \DC source code is available on GitHub\cite{dicos_github}, it contains several bugs that required fixing.
Further, their dataset of labeled snippets is not available, even on request (\textbf{R1}:~\resolve{CC}).

The authors followed a data-driven approach (\textbf{R4}: \resolve{no}) to answer the following  questions:
\begin{itemize*}
    \item[\textbf{RQ1}] \textit{Are older posts more likely to provide insecure code snippets?}
    \item[\textbf{RQ2}] \textit{Are accepted answer posts more secure than non-accepted posts?}
    \item[\textbf{RQ3}] \textit{What types of insecure code snippets were discovered?}
    \item[\textbf{RQ4}] \textit{What is the status of reusing insecure code snippets in popular open-source software?}
\end{itemize*}

We replicate the study for \textbf{RQ2} and \textbf{RQ3} to determine if their results still hold today.
We excluded \textbf{RQ1} because code evolution does not affect its findings, i.e., code evolution cannot retroactively \textit{add} old posts, only evolve them.
However, we used the results of \textbf{RQ1} to evaluate the released \DC tool, which we will discuss later in this section.
We excluded \textbf{RQ4} because it deals with code reuse from Stack Overflow in open-source projects on GitHub (\textbf{R3}: \resolve{orange}, \textbf{S}).
Though RQ4 is an interesting question for replication, our focus is on code evolution within Stack Overflow.

\subsubsection{Original Methodology}

The authors reported 93\% precision, 94\% recall, and 90\% accuracy in discovering insecure \textbf{C/C++} code snippets with \DC while for \textbf{Android} code snippets, it has 86\% precision, 89\% recall, and 86\% accuracy.
The evaluation was done using the \textit{SOTorrent20} dataset, released in 12/2020, from which they extracted 668,520 posts containing 1,514,547 code snippets.
For replication, we need to re-evaluate \DC's precision, recall, and accuracy using the more recent SOTorrent22 dataset.
The authors evaluated the accuracy of \DC against the results by Fischer et al.~\cite{fischer2017SOConsideredHarmful} and Verdi et al.~\cite{verdi19}.
However, this methodology creates a barrier to replication: the authors only used the datasets provided by Fischer et al.~and Verdi et al.~rather than directly running \DC and these tools on the same input.
Specifically, they compared the insecure snippets identified by \DC against the labeled code snippets from Fischer et al.~and the insecure C++ snippets found by Verdi et al.
To replicate this experiment, we would need to follow the same approach, using the labeled examples from related work to compare with \DC's findings on a newer version of SOTorrent.
Unfortunately, since these tools are unavailable and the labeled data from Fischer et al.~and Verdi et al.~only represent the old SOTorrent data set, this evaluation approach is infeasible when using a newer SOTorrent version.

\subsubsection{Implementation} 

We used the \DC code base to replicate the authors' findings on a newer dataset. 
We discovered several bugs, which we fixed for future use with \DC, and noticed that some methods were only partially implemented.
For example, Joern\cite{joern_io} was not fully used for control flow analysis, and their Jaccard index-based pairing technique produced many false positives.
We plan to release a refactored \DC version with the option to use clone detection as a pairing technique.

\begin{table*}[t]
\caption{
Comparison of precision, recall, and accuracy \hlblue{measured by authors} (Table~5 in \cite{dicos}) vs.~our \hlgreen{replication (SOTorrent22)}.
}
\label{tab:dicos_accuracy_measurement}

\centering
\footnotesize
\noindent
\begin{tabular}{N N N N N N | N N N N N }\toprule
    \multicolumn{1}{N }{\textbf{}} & \multicolumn{5}{c |}{\textbf{\hlblue{Original} accuracy measurement results for}} & \multicolumn{5}{c }{\textbf{\hlgreen{Replicated} accuracy measurement results for C,}} \\  
    \multicolumn{1}{c }{\textbf{}} & \multicolumn{5}{c }{\textbf{C, C++, and Android posts based on \textit{SOTorrent20}}} & \multicolumn{5}{c }{\textbf{C++, and Android posts based on \textit{SOTorrent22}}} \\ 
    \cmidrule(lr){1-1}
    \cmidrule(lr){2-6}
    \cmidrule(ll){7-11}
    \multicolumn{1}{ N }{\textbf{ID}} & \textbf{\#Posts} & \textbf{\#TP} & \textbf{\#FP} & \textbf{\#TN} & \textbf{\#FN} & \textbf{\#Posts} & \textbf{\#TP} & \textbf{\#FP} & \textbf{\#TN} & \textbf{\#FN} \\ 
    \cmidrule(lr){1-1}
    \cmidrule(lr){2-6}
    \cmidrule(ll){7-11}
    \multicolumn{1}{ N }{\textbf{G1}} & 788 & 757 & 31 & \textcolor{gray!25}{N/A} & \textcolor{gray!25}{N/A} & 788 & 95 & 693 & \textcolor{gray!25}{N/A} & \textcolor{gray!25}{N/A} \\
    \multicolumn{1}{ c }{\textbf{G2}} & 400 & 346 & 54 & \textcolor{gray!25}{N/A} & \textcolor{gray!25}{N/A} & 400 & 33 & 367 & \textcolor{gray!22}{N/A} & \textcolor{gray!25}{N/A} \\
    \multicolumn{1}{ c }{\textbf{G3}} & 200 & 162 & 38 & \textcolor{gray!25}{N/A} & \textcolor{gray!25}{N/A} & 600 & 66 & 534 & \textcolor{gray!25}{N/A} & \textcolor{gray!25}{N/A} \\
    \multicolumn{1}{ c }{\textbf{G4}} & 400 & \textcolor{gray!25}{N/A} & \textcolor{gray!25}{N/A} & 318 & 82 & 400 & \textcolor{gray!25}{N/A} & \textcolor{gray!25}{N/A} & 379 & 17 \\ 
    \multicolumn{1}{ c }{\textbf{G5}} & 200 & \textcolor{gray!25}{N/A} & \textcolor{gray!25}{N/A} & 185 & 15 & 200 & \textcolor{gray!25}{N/A} & \textcolor{gray!25}{N/A} & 188 & 12 \\ \midrule
    \multicolumn{1}{ c }{\textbf{Total}} & 1,988 & 1,265 & 123 & 503 & 97 & 2,388 & 194 & 1,594 & 567 & 29 \\ \midrule
    \multicolumn{2}{ l }{\textbf{Precision}}  &  &  &  & \textcolor{red!80}{\textbf{0.91}} &  &  &  &  & \textcolor{red!80}{\textbf{0.11}} \\
    \multicolumn{2}{ l }{\textbf{Recall}}  &  &  &  & \textcolor{red!80}{\textbf{0.93}} &  &  &  &  & \textcolor{red!80}{\textbf{0.87}} \\
    \multicolumn{2}{ l }{\textbf{Accuracy}}  &  &  &  & \textcolor{red!80}{\textbf{0.89}} &  &  &  &  & \textcolor{red!80}{\textbf{0.32}} \\
    \bottomrule
\end{tabular}

\end{table*}

We evaluated the authors' \DC implementation using the same dataset version to verify that we have a reproducible methodology as the basis for replication.
However, we could not reproduce their reported Stack Overflow post counts using the same SQL queries~\cite{dicos_github_sql_file}.
While they reported 987,367 C/C++ and 970,916 Android posts, we found 867,962 C/C++ and 986,900 Android posts.
Additionally, we identified 26,550 insecure posts, 14,092 more than the 12,458 reported.
Unfortunately, we received no response from the authors for clarification.
We did replicate their \textit{\textbf{RQ1}} findings, observing a similar trend but with different yearly secure/insecure post counts.
Details are in \Cref{sec:appendix:casestudy_2}.
Although we could not reproduce the exact insecure post numbers, we replicated their study on a newer \textit{SOTorrent} dataset version, using their code.

\subsubsection{Replication}
We used \textit{SOTorrent22}, which was released \textit{two} years after the \textit{SOTorrent20} dataset used by the authors. 
We followed the authors' approach to collect posts from \textit{SOTorrent22}. 
The authors extracted 1,958,283 posts (987,367 C/C++ and 970,916 Android), whereas we extracted 2,858,003 posts (1,489,148 C/C++ and 1,368,855 Android posts). 
This indicates a 51\% increase in C/C++ posts and a 41\% increase in Android posts since December 2020.
The authors filtered all single-version posts, resulting in 668,520 (34\% of 1,958,283) multi-version posts.
After filtering all single-version posts from the extracted \textit{SOTorrent22} posts, we obtained 1,046,052 posts (36.6\% of 2,858,003) with at least two versions.

Finally, the authors applied \DC on their dataset of \hlblue{668,520 posts} and discovered \hlblue{12,458 (1.9\%) insecure posts} (8,941 C/C++ and 3,517 Android). 
Of these, \hlblue{788 (6.3\%) insecure posts contained all three types of changes} while the remaining 11,670 insecure posts contained two types of changes.
Using the same approach on our dataset of \hlgreen{1,046,052 posts}, we discovered \hlgreen{30,343 (2.9\%) insecure posts}, i.e., an \textbf{increase of 52\%}.
Among these, \hlgreen{4,887 (16.1\%) insecure posts contained all three features}, i.e., an \textbf{increase of 155\%}, while the remaining 25,472 posts contained two features.

\textit{\textbf{Accuracy of \DC}}: 
To compute the accuracy, the authors manually verified a subset of the 12,458 discovered insecure posts using the following groups:

\begin{itemize}[itemindent=0em,labelsep=0pt,labelwidth=0em,leftmargin=0pt,noitemsep,parsep=0pt,partopsep=0pt]
    \item[] \textbf{G1.} All posts with three changes
    \item[] \textbf{G2.} Top 200 posts with two changes
    \item[] \textbf{G3.} Randomly selected 100 posts with two changes
    \item[] \textbf{G4.} Top 200 posts with only one change
    \item[] \textbf{G5.} Top 100 posts without changes
\end{itemize}

Groups \textbf{G1--G3} were used to measure true and false positive rates for detecting insecure posts. 
Groups \textbf{G4} and \textbf{G5} were used to measure the true and false negative rates.
Two researchers manually verified the posts for each group, recording the positive and negative rates for C/C++ and Android.
We replicated this approach to calculate the precision, recall, and accuracy of \DC using \textit{SOTorrent22}.
We adhered to the authors' original method, verifying 788 insecure posts for \textbf{G1} by randomly selecting 788 posts for manual verification by two researchers.
For \textbf{G3}, we faced the challenge that the authors conducted a one-time random selection of 100 posts with two features.
In contrast, we performed three random selections of 100 posts and averaged the results.

\Cref{tab:dicos_accuracy_measurement} compares the original accuracy of \DC reported by the authors with our findings using a newer dataset. 
We found that \DC had an \hlgreen{11\% precision} (compared to the authors' \hlblue{91\%}), \hlgreen{32\% accuracy} (versus \hlblue{89\%}), and an \hlgreen{87\% recall} (versus \hlblue{93\%}).
These results indicate that the performance of \DC has significantly decreased due to the code evolution on Stack Overflow.
The replicated accuracy measurements for C/C++ and Android are in \Cref{sec:appendix:casestudy_2}.

\begin{formal}
We found that code evolution has adversely affected the precision and accuracy of \DC's approach to detecting vulnerable snippets.
Considering the stable recall, \DC on newer Stack Overflow versions is better suited for detecting \textit{secure} snippets.
\end{formal}

\begin{table*}[ht]
    \caption{Comparison of the results by \hlblue{Hong et al.}~\cite{dicos} and our \hlgreen{replication} using \textit{SOTorrent22}.}
    \label{tab:dicos_side-by-side-summary}
    \footnotesize
    \centering
    \begin{tabularx}{\linewidth}{@{} >{\RaggedRight\hsize=1.0\hsize}X |
                                >{\RaggedRight\hsize=1.0\hsize}X
                                }
        \toprule
      \multicolumn{1}{c}{\textbf{\DC with SOTorrent20}}   & \multicolumn{1}{c}{\textbf{\DC with SOTorrent22}}   \\  \midrule
     \rowcolor{gray!18}\multicolumn{2}{c}{\textbf{Data Collection}} \\  \midrule
      The authors extracted 987,367 C/C++ posts and 970,916 Android posts, totaling \hlblue{1,958,283 answer posts}. & We extracted 1,460,627 C/C++ posts and 1,339,692 Android posts, totaling \hlgreen{2,800,319 answer posts}.
      This is a 43\% increase in the number of C/C++ and Android answer posts created on Stack Overflow after December 2020. \\ \midrule
      After filtering out single-version posts, the authors collected \hlblue{668,520 multi-version answers}, which they used to evaluate \DC. & We collected \hlgreen{1,046,052 multi-version answers} after filtering out single-version posts. \\ \midrule
      The authors found \hlblue{12,458 insecure posts}; 8,941 insecure C/C++ posts,
      and 3,517 insecure Android posts. & In contrast, we found \hlgreen{30,359 insecure posts}; 22,167 insecure C/C++ posts and 8,192 insecure Android posts. \\ \midrule
      \DC has \hlblue{91\% precision, 93\% recall, and 89\% accuracy}. & We observed \hlgreen{11\% precision, 87\% recall and 32\% accuracy}. \\ \midrule

      \rowcolor{gray!18}\multicolumn{2}{c}{\textbf{RQ2} \textit{Are accepted answer posts more secure than non-accepted posts?}} \\  \midrule
      The ratio of insecure posts was almost the same between accepted (\hlblue{1.67\%}) and non-accepted (\hlblue{1.99\%}) posts. & The ratio of insecure posts is very similar between accepted (\hlgreen{7.72\%}) and non-accepted (\hlgreen{6.61\%}) posts, and the overall ratio of insecure posts increased. \\ \midrule
      
      \rowcolor{gray!18}\multicolumn{2}{c}{\textbf{RQ3} \textit{What types of insecure code snippets were discovered?}} \\  \midrule 
       The most prevalent type of insecure code snippets was \hlblue{\textit{undefined behavior}}, accounting for 42\% of the total; & The most prevalent type of insecure code snippets was \hlgreen{\textit{memory leak}}, accounting for 39.25\% of the total; \\
       \midrule 
       Authors observed \hlblue{\textit{Null-terminated string issue}} as the second most prevalent security weakness. & In contrast, we found three security weaknesses as the second most prevalent weaknesses: \hlgreen{\textit{Undefined behavior}}, \hlgreen{\textit{Out-of-bounds error}}, and \hlgreen{\textit{Others}}. \\

    \bottomrule
    \end{tabularx}
\end{table*}
\begin{figure}[t]
\centering
\includegraphics[width=\linewidth]{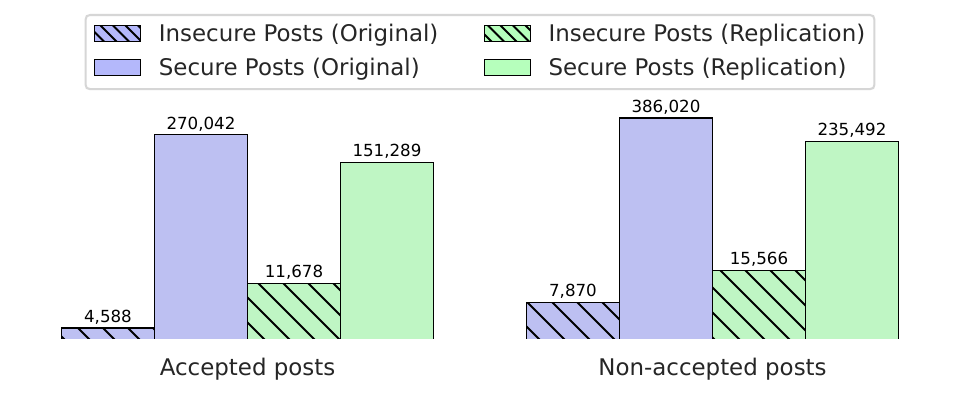}
\caption{Ratio of insecure posts between accepted and non-accepted posts discovered by \DC (logarithmic scale) as reported by the \hlblue{authors} (Figure 7 in \cite{dicos}) and found in our \hlgreen{replication study}.}
\label{fig:appendix:dicos_rq2_comparison}
\end{figure}

\begin{figure}[t]
\centering
\includegraphics[width=\linewidth]{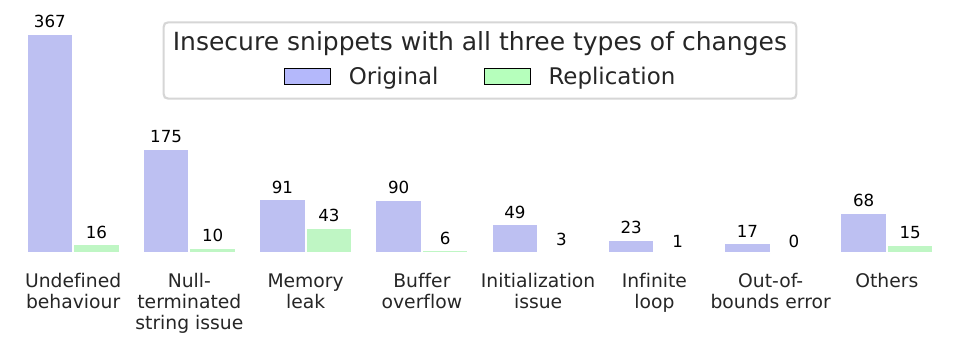}
\caption{Types of discovered insecure code snippets with three types of changes discovered by \DC as reported by the \hlblue{authors} (Figure 8 in \cite{dicos}) and found in our \hlgreen{replication study}.}
\label{fig:appendix:dicos_rq3_comparison}
\end{figure}
\textit{\textbf{Revisiting RQ2 Findings}}: The authors investigated the relationship between the security weaknesses of code snippets in accepted and non-accepted answers and found no difference between the ratios of insecure posts for \hlblue{accepted (1.67\%) and non-accepted (1.99\%) answers}.
\Cref{fig:appendix:dicos_rq2_comparison} compares the original results (refer to Figure 7 in \cite{dicos}) and the replication findings
Using the newer version of the SOTorrent dataset, we observed a higher ratio of insecure posts between \hlgreen{accepted (7.72\%) and non-accepted (6.61\%) answers}.
A two-sample z-test for proportions of the insecure to secure posts ratio between the original and our replication results shows a significant difference ($Z=-126.888$, $p<0.001$).

\textit{\textbf{Revisiting RQ3 Findings}}:  
The authors manually categorized the 788 insecure posts containing all three types of changes by their weakness types. \Cref{fig:appendix:dicos_rq3_comparison} 
compares the original and replication results.
The authors reported eight types of insecure code snippets, with \hlblue{undefined behavior (42\%)} being the most common.
Our findings indicate that \hlgreen{memory leaks (39.3\%)} are now the most prevalent security weakness in C/C++ code snippets.
Furthermore, while the authors identified \hlblue{null-terminated strings} as the second most common issue, our analysis found \hlgreen{undefined behavior, out-of-bound errors, and others} instead.

\begin{formal}
For the newer data set version, the types of weaknesses and their frequencies have shifted.
\end{formal}

\subsection{Case Study 3: Snakes in Paradies}\label{sec:case_study_5}
Rahman et al.~\cite{rahman_snakes_in_paradies} investigated Python code snippets posted on Stack Overflow, aiming to characterize the prevalence of insecure Python-related coding practices. The authors used the SOTorrent18 dataset, released in 09/2018, to empirically answer the following research questions:
\textbf{RQ1}: \textit{How frequently do insecure coding practices appear in Python-related Stack Overflow answers?}
\textbf{RQ2}: \textit{How does user reputation relate to the frequency of insecure Python-related coding practices?}
\textbf{RQ3}: \textit{What are the characteristics of Python-related questions that include answers with insecure code practices?}

The authors did not investigate the reuse of Python snippets from Stack Overflow (\textbf{R3}: \resolve{no}) and followed a purely data-driven approach to answer their research questions (\textbf{R4}: \resolve{no}).

\subsubsection{Original Methodology}
The authors focused exclusively on Python code snippets in answer posts in the SOTorrent18 dataset. If a question post, viewed more than once and with a score $>$ 0, is attributed inside Python source files on GitHub, 
then all answers of the question are considered. Attributing a question post in Python projects on GitHub is the information the authors used to determine the language of snippets (\textbf{R2}: \resolve{yes}). This resulted in 10,861 questions with 44,966 answers, from which they extracted 529,054 code snippets, which formed their final dataset.
To detect insecure coding patterns in code snippets, the authors used string matching to determine if a standard library or third-party Python API, known to be used in an insecure way, is found in a code snippet. The authors used six categories to group the insecure Python APIs they considered in their study. Table II of the original paper describes the six categories and their corresponding insecure coding pattern.

\subsubsection{Implementation}
The authors published the code artifacts of their study but not their data artifacts (\textbf{R1}: \resolve{RC}).
We used the published code~\cite{snakes_in_paradies_artifacts} for replication. However, we needed to collect the dataset from SOTorrent22.
This implementation was straightforward since the authors extensively described their data collection approach. We tested our implementation using SOTorrent18, the same version the authors used, and came to the same numbers as in the original paper.

\subsubsection{Replication}
We replicated their findings using \textit{SOTorrent22}, released four years after the original dataset version. 
After applying the authors' filtering criteria, we obtained \hlgreen{12,095} questions containing \hlgreen{72,202} answers, of which \hlgreen{10,140} were accepted answers. This means the number of code snippets matching the authors' filtering criteria has dropped since 2018: \hlblue{529,054} $\searrow$ \hlgreen{239,575}.

\textbf{\textit{Revisiting RQ1 Findings}}:
Our findings regarding the number of questions with at least one insecure answer differ significantly: \hlblue{18.1\% (out of 10,861)} dropped to \hlgreen{4.9\% (out of 12,095)}. Similarly, the percentage of accepted answers containing at least one insecure snippet also decreased: \hlblue{9.8\% (out of 7,444)} $\searrow$ \hlgreen{2.2\% (out of 10,139)}.
Although the vulnerability rankings from the original study remain unchanged, we observed a shift in the number of affected snippets: \textit{code injection} increased (\hlblue{2,319} $\nearrow$ \hlgreen{5,734}), while \textit{insecure cipher} (\hlblue{564} $\searrow$ \hlgreen{356}), \textit{insecure connection} (\hlblue{624} $\searrow$ \hlgreen{276}), and \textit{data serialization} (\hlblue{153} $\searrow$ \hlgreen{140}) all dropped. Like the original study, no snippets were impacted by XSS vulnerabilities.

\begin{formal}
    We found fewer insecure and accepted answers than the original, with percentages dropping significantly. Vulnerability rankings remained consistent, but there were notable shifts in affected snippets, including an increase in \textit{code injection} cases and decreases in others. No XSS vulnerabilities were found, as in the original study.
\end{formal}

\textbf{\textit{Revisiting RQ2 Findings}}: The authors answered this question by computing the \textit{normalized} reputation score of users that contributed at least one insecure code snippet with those that contributed answers with no insecure code snippets. Normalization was required to reduce bias since long-time Stack Overflow users tend to have higher reputations. 
Using the Mann-Whitney U and Cliff’s Delta non-parametric tests, the authors found \hlblue{no significant difference} between answer providers with high and low reputation, suggesting that both are equally likely to introduce insecure code snippets.
We also found \hlgreen{no significant difference} between the two user groups. However, we observed different $p-$value (\hlblue{$0.9$} $\nearrow$ \hlgreen{$6.2$}) and Cliff's delta (\hlblue{$0.01$} $\nearrow$ \hlgreen{$0.03$}) values.

\textbf{\textit{Revisiting RQ3 Findings}}:
The authors employed Latent Dirichlet Allocation (LDA)-based topic modeling to group questions associated with an insecure answer. They found that answers to questions \hlblue{associated with \textbf{web}, \textbf{string}, and \textbf{RNG}} topics contain at least one insecure code snippet. In contrast, we found that \hlgreen{answers to questions related to \textbf{web} topics are no longer associated with insecure code snippets} but questions associated with \textbf{string} and \textbf{RNG} topics still have answers containing at least one insecure code snippet.

\begin{formal}
We found that user reputation does not influence the likelihood of posting insecure code snippets, confirming the original observation. However, we discovered that the association between \textit{web} topics and insecure code snippets is no longer valid.
\end{formal}

\subsection{Case Study 4: Mining Rule Violations in JavaScript}\label{sec:case_study_6}
Campos et al.~\cite{mining_rule_violations} investigated the prevalence of violations in JavaScript code snippets on Stack Overflow, aiming to answer the following research questions:
\textbf{RQ1}: \textit{How commonplace are rule violations in JavaScript code snippets?}
\textbf{RQ2}: \textit{What are the most common rules violated in JavaScript code snippets?}
\textbf{RQ3}: \textit{Are JavaScript code snippets flagged with possible errors being reused in GitHub projects?}

We replicated the study for \textbf{RQ1} and \textbf{RQ2} to assess if the findings remain valid today. We omitted RQ3, as it concerns code reuse from Stack Overflow, while our research focuses on the evolution of code within Stack Overflow.

\subsubsection{Original Methodology}
The authors used a data-driven approach to answer their research questions (\textbf{R4}: \resolve{no}) and focused on accepted answers to questions with JavaScript tags (\textbf{R2}: \resolve{yes}) in the SOTorrent18 dataset.
If a code snippet in an accepted answer has multiple versions, the latest version of the snippet is selected, resulting in 336,643 code snippets. Using the ESLint static analysis tool, the authors found that \textbf{all code snippets} in their data set contains rule violations, with stylistic issues being the most prevalent violation, accounting for 82.9\% of the total. Relying on attribution to detect code reuse (\textbf{R3}: \resolve{yes}), they found 36 of the JavaScript code snippets containing violations to be reused in 845 GitHub projects.

\subsubsection{Implementation}
The authors made their source code and data publicly available (\textbf{R1}: \resolve{orange}), enabling us to replicate their findings on a newer dataset. However, we found a discrepancy between their paper's data collection method and the released dataset. Although they claimed to discard snippets with fewer than 10 LoC, their dataset includes \textbf{42,158} snippets with nine LoC. Since the script used for data collection wasn't shared, 
reproducing the exact number of code snippets from SOTorrent18 using the approach discussed in the paper was impossible. 
Similarly, when we attempted to reproduce the author's findings for their \textbf{RQ1} using their own source code and dataset of code snippets, we could not come to their conclusion that \textit{no code snippet in their dataset were free of violations}. Instead, we found 153,159 (45.5\% of 336,643) code snippets that only contain parse errors without any rule violations.

We contacted the authors for clarification and learned that they chose a minimum LoC of nine and included parse errors as violations. Since our replication study follows the original methodology, we adhered to their clarification by setting the LoC threshold to nine and counting parse errors as violations.

\subsubsection{Replication}
We replicate \textbf{RQ1} and \textbf{RQ2} using \textit{SOTorrent22}, released four years after the dataset version used in their study.

\textbf{\textit{Revisiting RQ1 Findings:}} The original study reported that \hlblue{no JavaScript code snippet was free of violations}, but our replication found \hlgreen{nine snippets without any violations}. Additionally, the number of violations in JavaScript code snippets increased from \hlblue{5,587,357} to \hlgreen{7,385,044}, with the average violations per snippet rising from \hlblue{11.94} to \hlgreen{28.8}.

\begin{formal}
    The observation that no JavaScript code snippet on Stack Overflow is violation-free no longer holds, and we found more violations in snippets than originally reported.
\end{formal}

\textbf{\textit{Revisiting RQ2 Findings:}} The original study grouped violations into six categories and reported the top three most common rule violations for each category.
Below, we compare the authors' \hlblue{original findings} with our \hlgreen{replication results}.

\medskip \noindent \textbf{Stylistic Issues:} 
Violations increased by 28.4\% (from \hlblue{4,632,348} to \hlgreen{5,946,283}), with the three most common violations in this category being: 
\begin{sinparaenum} 
    \item[(1)] \textit{semi}: \hlblue{1,477,808} $\nearrow$ \hlgreen{1,990,461}, 
    \item[(2)] \textit{quotes}: \hlblue{700,770} $\nearrow$ \hlgreen{1,072,468}, and 
    \item[(3)] \textit{no-trailing-spaces}: \hlblue{374,012} to \hlgreen{473,294}. 
\end{sinparaenum}
In the original study, the authors found and removed \hlblue{3,461,739} violations related to indentation rule violations from this category, reasoning that multiple snippets were merged into a single file. We followed the same approach and removed \hlgreen{4,910,529} indentation violations.

\medskip
\noindent
\textbf{Variable:} Violations increased by 28.7\% (from \hlblue{787,824} to \hlgreen{1,013,612}), with the top issues being:
\begin{sinparaenum}
    \item [(1)] \textit{no-undef}: \hlblue{719,679} $\nearrow$ \hlgreen{913,103},
    \item [(2)] \textit{no-unused-vars}: \hlblue{67,816} $\nearrow$ \hlgreen{100,124}, and
    \item [(3)] \textit{no-undef-init}: \hlblue{150} $\nearrow$ \hlgreen{189}.
\end{sinparaenum}

\medskip
\noindent
\textbf{Best Practices (BP):} Violations increased by 216.5\% (\hlblue{57,578} $\nearrow$ \hlgreen{182,232}), with the most common being:
\begin{sinparaenum}
    \item [(1)] \textit{eqeqeq}: \hlblue{53,321} $\nearrow$ \hlgreen{66,664},
    \item [(2)] \textit{no-multi-spaces}: \hlblue{54,768} $\nearrow$ \hlgreen{50,352},
    \item[(3)] \textit{no-redeclare}: \hlgreen{18624}, and
    \item[(4)] \textit{curly}: \hlblue{14,989} $\nearrow$ \hlgreen{18,292}.
\end{sinparaenum}
The \textit{eqeqeq} violation is now the most common, while \textit{curly} fell to the fourth position.

\medskip
\noindent
\textbf{Possible Errors (PE):} Violations increased by 39\% (\hlblue{6,303} $\nearrow$ \hlgreen{8,762}), with the most common being:
\begin{sinparaenum}
    \item [(1)] \textit{no-irregular-whitespace}: \hlblue{2,037} $\nearrow$ \hlgreen{2,196},
    \item[(2)] \textit{no-cond-assign}: \hlblue{910} $\nearrow$ \hlgreen{1,196}, and
    \item[(3)] \textit{no-dupe-keys}: \hlgreen{874}, which replaced \textit{no-unreachable} as the third most common violation.
\end{sinparaenum}

\medskip
\noindent
\textbf{Node.js/Common.js:} Violations increased by 27.5\% (from \hlblue{3,304} to \hlgreen{4,211}), with the most common violations being:
\begin{sinparaenum}
    \item[(1)] \textit{handle-callback-err}: \hlblue{2,855} $\nearrow$ \hlgreen{3,650},
    \item[(2)] \textit{no-path-concat}: \hlblue{444} $\nearrow$ \hlgreen{555}, and
    \item[(3)] \textit{no-new-require}: \hlblue{5} $\nearrow$ \hlgreen{6}.
\end{sinparaenum}

\medskip
\noindent
\textbf{ECMAscript 6:} Violations increased by 139.8\% (from \hlblue{548} to \hlgreen{1,314}), with the most common being:
\begin{sinparaenum}
    \item [(1)] \textit{template-curly-spacing}: \hlblue{164} $\nearrow$ \hlgreen{516},
    \item[(2)] \textit{no-useless-constructor}: \hlblue{154} $\nearrow$ \hlgreen{238}, and
    \item[(3)] \textit{no-const-assign}: \hlgreen{129}, replacing \textit{no-this-before-super} as the third most common violation.
\end{sinparaenum}

\medskip
\noindent
\textbf{Parse Errors:} We found \hlgreen{267,795} code snippets with \textit{parse errors} but no violations. While the original study included parse errors in the total violation count, we introduced a seventh category to separately list snippets with only parse errors.

\begin{formal}
Due to Stack Overflow's evolution, the number of violations in each category has risen. Further, the ranking of specific violations within certain categories has changed.
\end{formal}

\section{Case Study 5\&6: Stack Overflow Considered Harmful and Helpful}\label{sec:appendix:casestudy_34}
Fischer et al.~\cite{fischer2017SOConsideredHarmful} conducted a data-driven study (\textbf{R4}: \resolve{no}) of crypto API misuse in Java snippets.
They classified snippets of the \textit{March 2016} Stack Exchange data set into \textit{secure} or \textit{insecure} using a custom-built machine learning classifier.
Through static analysis, they detected insecure snippets reused in Android apps (\textbf{R3}: \resolve{orange}, \textbf{P}).

Unfortunately, the authors could only make their set of labeled snippets publicly available (\textbf{R1}: \resolve{LC}) and did not mention how they detected the language of snippets (\textbf{R2}: \textbf{N/A}).
Re-implementing their approach is challenging because their pipeline involves outdated and custom-built components.
For instance, JavaBaker~\cite{java_baker_github} is no longer functional and maintained, and we were unable to install the Partial Project Analysis tool~\cite{eclipse_ppa_website}. 
In a follow-up study, Fischer et al.~\cite{FischerUsenix2019} proposed solutions to help developers use more secure code from Stack Overflow.
The authors again only published their dataset of labeled code snippets (\textbf{R1}: \resolve{LC}) and did not indicate how they detected the language of code snippets (\textbf{R2}: \textbf{N/A}).
The study neither considered the participation of developers (\textbf{R4}: \resolve{no}) nor code reuse from Stack Overflow (\textbf{R3}: \resolve{no}).
The authors used the \textit{March 2018} Stack Exchange data set and a machine learning classifier to identify secure and insecure snippets.
The training dataset and classifier are also unavailable, and the pipeline also includes outdated components, some identical to the previous study.
As for their previous study, these challenges made re-implementing their approach infeasible.

\begin{table}[t]
    \centering
    \footnotesize
        \caption{Composition of the $DS_{2016}$ and $DS_{2018}$ datasets}
    \begin{tabular}{lrrr}
    \toprule
            & \multicolumn{1}{c}{\textbf{All}} & \multicolumn{1}{c}{\textbf{Secure}} & \multicolumn{1}{c}{\textbf{Insecure}} \\\midrule
            \rowcolor{lightgray} & \multicolumn{3}{c}{$DS_{2016}$ ($N=2,370$)}   \\
            \midrule
          Questions & 1,147 & 813 (70.9\%) & 334 (29.1\%) \\
          Answers   & 1,223 & 870 (71.1\%) & 353 (28.9\%) \\
          \midrule
          \rowcolor{lightgray} & \multicolumn{3}{c}{$DS_{2018}$ ($N=16,343$)}   \\
          \midrule
          Questions & 1,0294 & 4,189 (40.7\%) & 6,105 (59.3\%) \\
          Answers   & 6,049 & 2,057 (34.0\%) & 3,992 (66.0\%)\\
    \bottomrule
    \end{tabular}
    \label{tab:fischer_datasets_compared}
\end{table}

Instead of re-implementation, we relied on the labeled datasets the authors made available to study \textbf{if} code evolution in those datasets could affect their findings.
However, without their machine learning classifier, we cannot fully assess \textbf{how} their findings might be affected by code evolution since we cannot classify new snippets added on Stack Overflow after their study in a scalable, reasonable manner.
Consequently, this section cannot provide the same level of confidence about the impact of StackOverflow evolution on the research results as the previous sections.

The first labeled dataset from the 2016 study includes 4,019 security-related snippets. 
The second dataset from the 2018 study includes 16,343 security-related snippets.
We refer to these as $DS_{2016}$ and $DS_{2018}$ respectively (see \Cref{tab:fischer_datasets_compared}).
Unfortunately, $DS_{2016}$ lacks the post IDs of the snippets, making it impossible to track snippet evolution.
To try resolving this, we indexed all Android posts and their snippet versions in Apache Solr~\cite{apacheSolr}, using the snippets as search strings to find matching posts.
This process identified post IDs for 2,370 (59\%) snippets of the original data set.
The $DS_{2018}$ dataset includes post IDs for the tracking of snippet evolution.

\subsection{Evolution of Labeled Code Snippets}\label{sec:fischer:evolution}
We used the \textit{StackExchange23} data set to study the evolution of snippets in the labeled datasets of both studies.
We collected all snippets from the $DS_{2016}$ and $DS_{2018}$ datasets that were revised in \textit{StackExchange23} and manually labeled each edit.
We tracked the evolution of the $DS_{2016}$ snippets from 03/2016 to 09/2023 (denoted as $I_{16-23}$), i.e., seven years of changes.
\Cref{fig:fischerDatasetEvolutionExample} in \Cref{sec:appendix:casestudy_34} exemplifies an answer post with two snippets that evolved twice in $I_{16-23}$.
For the $DS_{2018}$ dataset, we examined a period of one year earlier ($U_{17-18}$) and five years later ($U_{18-23}$).

\begin{table}[t]
    \centering
    \footnotesize
    \caption{Labels assigned during manual classification}
    \rowcolors{2}{white}{lightgray}
    \begin{tabularx}{\linewidth}{p{2.2cm}X}
    \toprule
         \textbf{Label} & \textbf{Meaning}  \\
         \midrule
         1. Cosmetic & Does not affect the snippet functionality, e.g.~identifier renaming, comments, code formatting\\
         2. Functional & Affect functionality, but not security-relevant, e.g.~adding test cases or refactoring\\
         3.~Security-relevant & Changes to code segments handling security-relevant data or interacting with a security API, e.g.,~changing security API, adding encryption, changing key size. Fixes to bugs and vulnerabilities also fall under this category. \\
         4. Security label change & Changes to snippets that change the security classification from \emph{insecure} to \emph{secure} or vice versa according to the definition by Fischer et al.~\cite{fischer2017SOConsideredHarmful}. \\
         \midrule
         5. Bad source classification & Coders disagree with the original security classification in the dataset by Fischer et al.\\
         6. Post deleted & Post has been deleted from Stack Overflow \\
         \hline
    \end{tabularx}
    \label{tbl:label_categories}
\end{table}

\paragraph{Labeling}
Two researchers independently assigned each edit one of four labels (see \Cref{tbl:label_categories}).
They then discussed and resolved conflicts.
We excluded posts assigned labels 5 or 6 for data sanitization.
The researchers reached a high level of agreement (Krippendorff's Alpha~\cite{krippendorff2011computing} of $\alpha=0.96$).
\begin{table}[t]
    \centering
    \small
    \caption{Categories of posts containing code snippets that evolved in different periods in the two datasets.}
    \label{tab:fischer_classification_results}
    \begin{tabular}{lrrr}
         Label & $U_{17-18}$ & $U_{18-23}$ & $I_{16-23}$ \\
         \midrule
         1. Cosmetic            & 107 & 443 &  256  \\
         2. Functionality       & 25  & 25  &  22 \\
         3. Security-relevant   & 35  & 24  &  44 \\
         \rowcolor{hlgreencolour} 4. Security-label      & 3  &  8   &   9 \\
         \midrule
        \phantom{1234} $\Sigma$ & \emph{170}&  \emph{500} &   \emph{331}\\
         \midrule
         5. Wrong label         & 2  &  2 &  6\\
         6. Post deleted        & 0  &  0 &  1\\
         \bottomrule
         \multicolumn{4}{c}{\scriptsize $U_{17-18}$ and $U_{18-23}$ for $DS_{2018}$; $I_{16-23}$ for $DS_{2016}$}
    \end{tabular}
    
\end{table}

\subsection{Results and Findings}
\Cref{tab:fischer_classification_results} shows the results of classifying the posts containing code snippets that evolved within the selected time frames.
Our analysis reveals that only \textbf{nine} code snippets from the $DS_{2016}$ dataset labeled by Fischer et al.~evolved between 2016 and 2023 with a security-relevant label change (see highlighted row in \Cref{tab:fischer_classification_results}). 
Initially labeled insecure by Fischer et al., the vulnerabilities in these nine snippets were fixed during the seven years following Fischer et al.'s data collection.
Similarly, if Fischer et al.~\cite{FischerUsenix2019} would have performed their data collection a year earlier ($U_{17-18}$), \textbf{three} snippet labels would have changed.
If they would have crawled Stack Overflow five years later ($U_{18-23}$), \textbf{eight} snippet labels would have been different.
The limited number of fixes for insecure cryptographic API usage in these snippets suggests that such corrections are rare, potentially because of the particular niche expert topic of crypto API misuse in Java/Android code.

\begin{formal}
Snippets identified as vulnerable by Fischer et al.~\cite{fischer2017SOConsideredHarmful,FischerUsenix2019} remained vulnerable despite code evolution.
We postulate that this may be due to the particular niche topic of crypto API usage that requires domain experts to identify and fix such vulnerabilities.
\end{formal}

\begin{figure}[t]
    \centering
    \includegraphics[width=\linewidth]{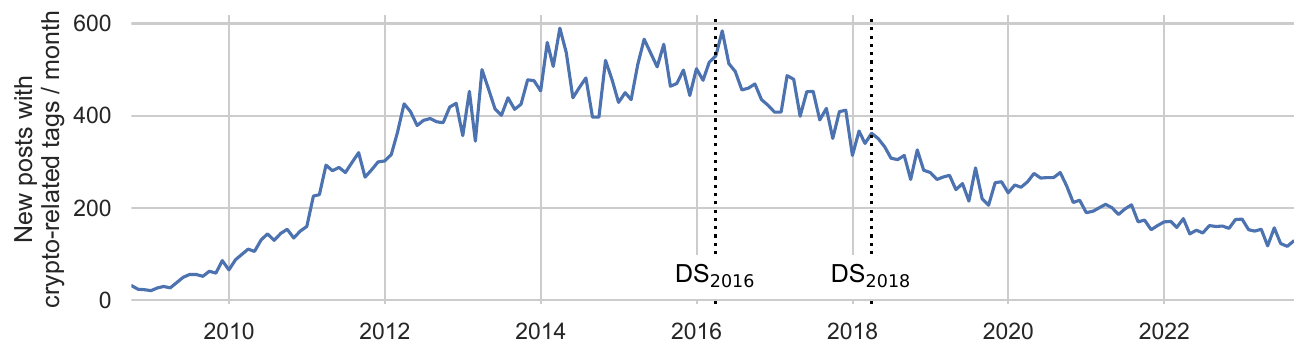}
    \caption{Nr.~of monthly created posts with \textit{Java} and/or \textit{Android} tag plus at least one top-50 crypto-related tag. Vertical lines indicate data collection points by Fischer et al.~\cite{fischer2017SOConsideredHarmful,FischerUsenix2019}}
    \label{fig:crypto_tag_ts}
\end{figure}

\begin{figure}[!htbp]
    \centering
    \includegraphics[width=.7\linewidth]{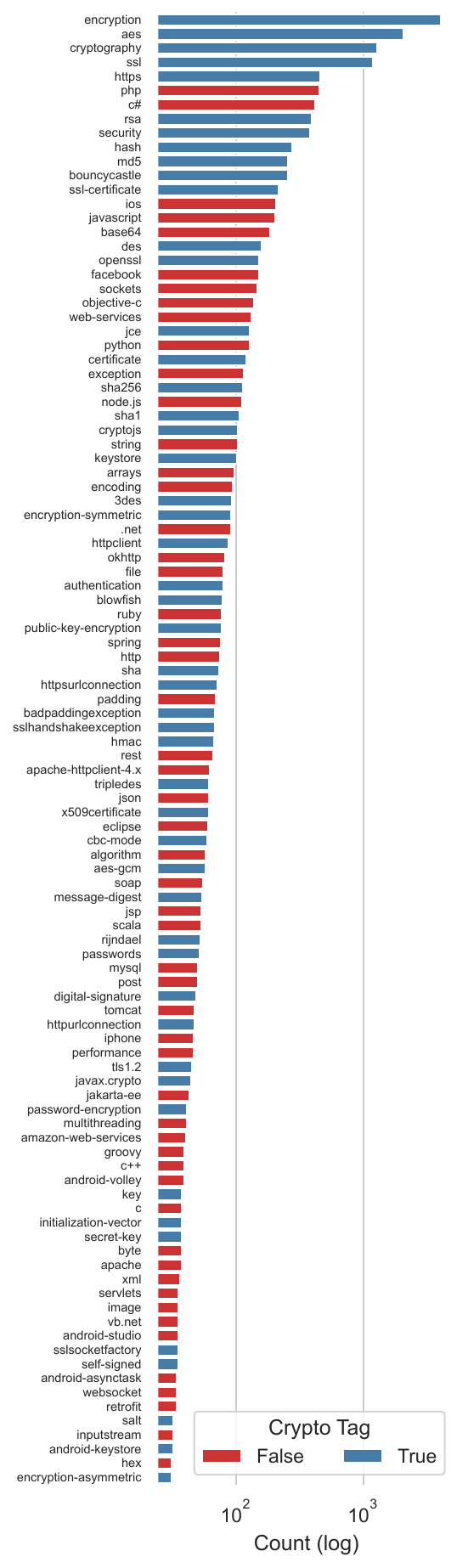}
    \caption{Top tags co-located with \textit{Java} or \textit{Android} tag in the data set by Fischer et al. Top-50 crypto-related tags are indicated.}
    \label{fig:tags_top100}
\end{figure}

\begin{figure}[!htbp]
    \centering
    \includegraphics[width=0.7\linewidth]{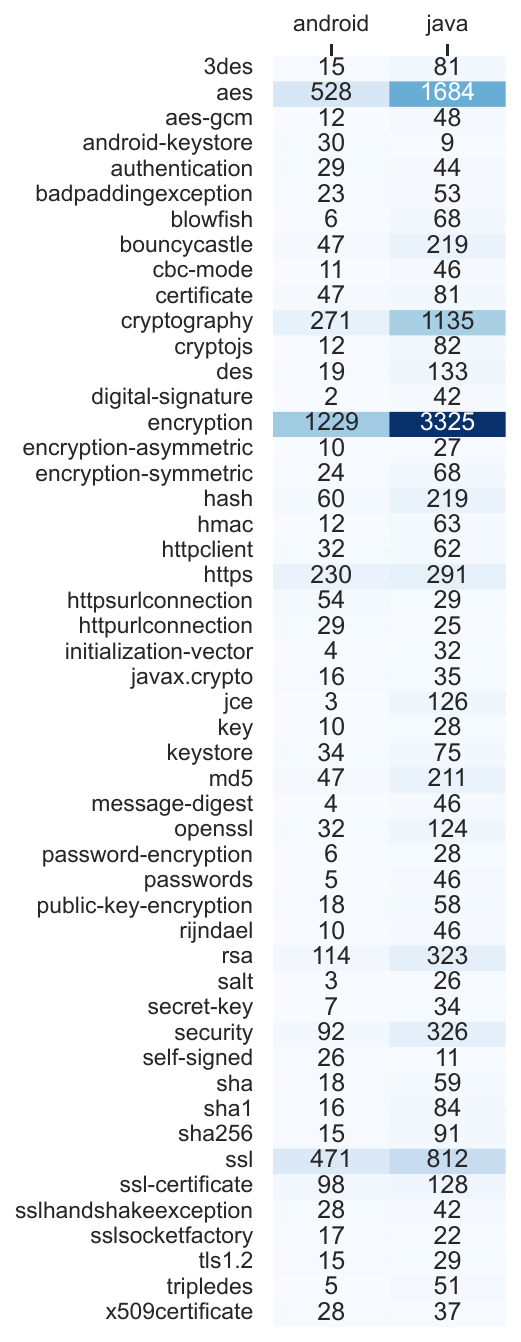}
    \caption{Frequency of co-location between top-50 crypto-related tags and \textit{Java} and \textit{Android} tags in the data set by Fischer et al.}
    \label{fig:heatmap_top50}
\end{figure}

\begin{figure*}[t]
 \center
  \includegraphics[width=.95\textwidth]{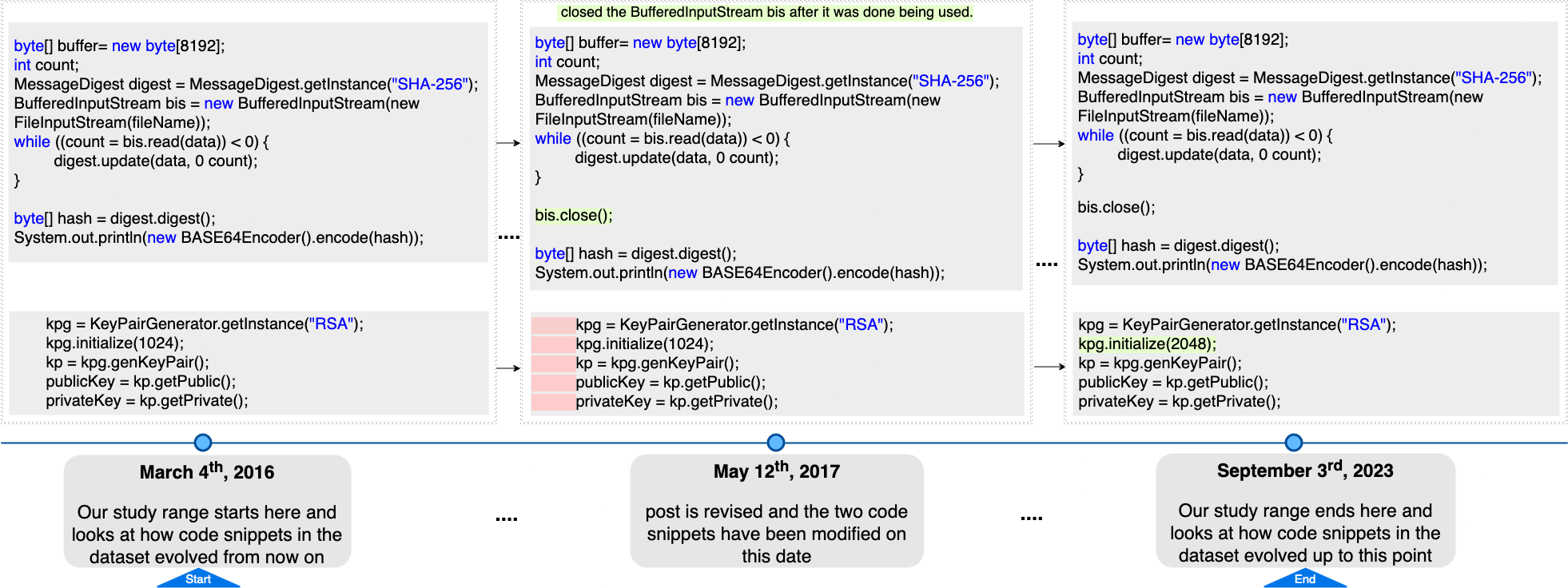}
  \captionsetup{font=small,labelfont=bf}
  \caption{The evolution of an answer post in $DS_{2016}$ between March $4^{th}$, 2016 and September $3^{rd}$, 2023. Two additional versions have been created during this time interval. The first version shows the post on March $4^{th}$, 2016, labeled by Fischer et al.~as secure (top snippet) and insecure (bottom snippet). The post was revised on May $12^{th}$, 2017, and the two code snippets were edited. The top code snippet undergoes a \textbf{security-relevant} edit because a resource leak bug was fixed. The edit to the bottom code snippet is merely \textbf{cosmetic}. In the final revision of the post, only the bottom code snippet is edited. This edit is a \textbf{label change} because the code snippet was originally labeled as insecure by Fischer et al.\ because the key size is considered weak, and the edit changes to a stronger key size in the final revision. In our classification, we label the top code snippet as a security-relevant change and the bottom code snippet as a security \textbf{label change} (see also \Cref{tbl:label_categories}).}
  \label{fig:fischerDatasetEvolutionExample}
\end{figure*}

However, since we do not have access to their code classifier, we cannot measure the impact of \textit{new} snippets \textit{added} to Stack Overflow after the data sets were collected.
Therefore, while our analysis supports the stability of their results within the original datasets, the potential impact of new code snippets on the broader conclusions remains unknown, e.g., if the fraction of vulnerable snippets remains stable.
To estimate the potential impact of new code snippets, we counted all posts added after the original studies that shared meaningful tags with those identified by Fischer et al.
Since Fischer et al.~studied the security of Java or Android code snippets, we counted the frequency of tags co-located with these two tags within the data set by Fischer et al.~and two researchers identified the top 50 crypto-related co-located tags.
These tags include \textit{encryption}, \textit{aes}, \textit{ssl}, or \textit{cryptography}.
\Cref{fig:tags_top100} and \Cref{fig:heatmap_top50} in \Cref{sec:appendix:casestudy_34} provide more details.
\Cref{fig:crypto_tag_ts} depicts the number of posts created each month that are tagged with \textit{Android} and/or \textit{Java} plus at least one of the top 50 crypto-related tags.
We found 24,767 new answers containing those tags added after $DS_{2016}$, corresponding to a growth of 87.6\% to the 28,267 posts in this dataset version.
Using the crypto-tags as a proxy for new potentially vulnerable snippets, together with the stability of Fischer et al.'s original results, this growth indicates that Fischer et al.'s results likely form a lower bound for the number of vulnerable Java and Android snippets about crypto APIs in today's Stack Overflow.
In our opinion, this is the most viable and fair approach to understanding the impact of code evolution on their findings.

\begin{formal}
Without code artifacts, we cannot measure whether the fraction of vulnerable snippets remains stable.
Using tags as a proxy, we estimate that the number of vulnerable snippets has increased since Fischer et al.'s study.
\end{formal}

\Cref{tab:appendix:tagfilterlist} lists all tags (see \Cref{fig:tags_top100}) that are not crypto-related and were filtered for computing the frequency of co-location with the tags \textit{android} and \textit{java} (see \Cref{fig:heatmap_top50}).

\begin{table}[t]
    \caption{List of filtered tags}
    \label{tab:appendix:tagfilterlist}
    \centering
    \begin{tabularx}{\linewidth}{X}
c\#, .net, facebook, sockets, base64, javascript, php, web-services, string, exception, node.js, ios, python, encoding, spring, objective-c, file, padding, mysql, apache-httpclient-4.x, ruby, soap, rest, bytearray, algorithm, eclipse, android-asynctask, android-volley, arrays, json, performance, okhttp, hex, byte, android-studio, facebook-graph-api, http, jsp, xml, servlets, tomcat, hibernate, post, multithreading, apache, okhttp3, c, scala, jakarta-ee, amazon-web-services, groovy, websocket, iphone, amazon-s3, c++, vb.net, retrofit, image, inputstream
    \end{tabularx}
\end{table}

\Cref{tab:appendix:rankedcryptotags} shows the top-20 crypto-related tags co-located with \textit{android} or \textit{java} among all 23,876,743 questions on Stack Overflow.
\Cref{tab:appendix:rankedtags} shows the top 10 co-located tags.
This shows that crypto-related questions are a narrow topic among developers.

\begin{table}[h]
    \caption{Top-20 crypto-related tags colocated with \textit{Java} and \textit{Android} among all questions on Stack Overflow. Rank among all co-located tags.}
    \label{tab:appendix:rankedcryptotags}
    \centering
    \small
    \begin{tabular}{rllr}
    \toprule
    \textbf{Count} & \textbf{Tag1} & \textbf{Tag2} & \textbf{Rank} \\
    \midrule
    7,554 & java & ssl & 143 \\
    6,936 & java & encryption & 159 \\
    6,180 & java & security & 189 \\
    4,477 & java & authentication & 263 \\
    3,382 & android & authentication & 330 \\
    2,767 & java & cryptography & 399 \\
    2,600 & android & security & 426 \\
    2,514 & android & encryption & 446 \\
    2,388 & java & https & 472 \\
    2,333 & android & ssl & 488 \\
    1,958 & java & httpurlconnection & 592 \\
    1,911 & java & hash & 609 \\
    1,854 & java & httpclient & 629 \\
    1,833 & java & aes & 634 \\
    1,788 & android & httpurlconnection & 651 \\
    1,755 & java & bouncycastle & 666 \\
    1,516 & java & rsa & 759 \\
    1,342 & java & keystore & 850 \\
    1,286 & java & ssl-certificate & 872 \\
    1,257 & java & certificate & 903 \\
    \bottomrule
    \end{tabular}
\end{table}

\begin{table}[h]
    \caption{Top-10 tags co-located with \textit{Java} and \textit{Android} among all questions on Stack Overflow. Rank among all co-located tags.}
    \label{tab:appendix:rankedtags}
    \centering
    \small
    \begin{tabular}{rllr}
    \toprule
    \textbf{Count} & \textbf{Tag1} & \textbf{Tag2} & \textbf{Rank} \\
    \midrule
    128,083 & java & spring & 1 \\
    79,498 & java & swing & 2 \\
    74,193 & java & spring-boot & 3 \\
    67,465 & android & android-studio & 4 \\
    59,401 & java & eclipse & 5 \\
    58,764 & java & hibernate & 6 \\
    57,068 & android & android-layout & 7 \\
    55,218 & android & kotlin & 8 \\
    52,237 & java & arrays & 9 \\
    47,781 & java & maven & 10 \\
    \bottomrule
    \end{tabular}
\end{table}

\section{Limitations and Challenges}\label{sec:discussions}
Before concluding with an overview of related works (\Cref{sec:related_work}) and recommendations for future works (\Cref{sec:conclusion}), we briefly discuss the limitations of our study and discuss basic challenges for studying the security of Stack Overflow snippets.

\paragraph{Quality of original studies.} This study replicates prior research using the original methods on a newer dataset.
Consequently, any flaws or biases in the original studies are also reflected in our replication.

\paragraph{Generalizability of results.} Our study specifically targeted research investigating the security properties of Stack Overflow code snippets. Thus, our findings may not apply to Stack Overflow-based studies that explore other aspects, such as user behavior, limiting the generalizability of our conclusions.

\paragraph{Dependence on available artifacts.} The quality and availability of the artifacts by the original studies posed another limitation. If a published artifact inaccurately implemented a study's methodology, these inaccuracies carried over into our replication. Despite our best efforts to reuse the same source artifacts, any errors or bugs in the original implementation may have impacted our results. Therefore, the reliability of our replication study is inherently linked to the accuracy and completeness of the original research artifacts.

However, the biggest challenge we faced in our study was the lack of code and data artifacts from prior research (see \Cref{tab:comparision_table}), which forced us to re-implement the original methodology in various case studies.
This was not always possible (e.g., training a machine learning-based classifier without access to the original data) and can be error-prone (e.g., replacing deprecated toolchains).

\paragraph{Language detection.}
A general challenge for studies on Stack Overflow is determining the programming language of code snippets.
Related work relied on post tags or the Guesslang tool (see \textbf{R2} in \Cref{tab:comparision_table}).
We evaluated these approaches briefly and found neither approach is reliable (see \Cref{sec:appendix:languagedetection} for details).
In comparison, ChatGPT outperformed either tool, but it is neither scalable in data set size nor economically reasonable for our replication studies to create a ground truth of snippet languages.
Thus, future work could create a fine-tuned LLM to replace Guesslang for the language detection task.

\paragraph{Security classification.}
Several studies rely on a security classification of code snippets by either building/using dedicated tools for specific languages (\textbf{D2} in \Cref{tab:comparision_table}) or, in the absence of a generic code security classifier, leveraging the context of snippets (i.e., comments and commit messages; \textbf{D4} in \Cref{tab:comparision_table}).
We used the tool by Jallow et al.~\cite{jallow_sp24} from the latter category. 
However, this tool has a potentially high false positive rate due to its keyword-based detection.
Future work could investigate a more reliable context-based tool, e.g., taking inspiration from other NLP-based solutions~\cite{hark_oakland,hark_usenix24}.

\paragraph{User studies.}
Several studies involved user studies~\cite{acar2016YouGetWhereYouLook,toxicCodeSnippets,bai_insecure_code_propagration,Ren_Demystify_API_Usage_Directives,Fischer_Effect_of_Google_Search_on_Software_Security,Mahajan_Recommending_stack_overflow_posts,Chen_crowd_debugging,Mahajan_repairing_runtime_exceptions,rahman_reusability_insight} (\textbf{R4}, \Cref{tab:comparision_table}) and it is a limitation of this work that we did not replicate any human-centered research.
Some of these prior works conducted developer surveys~\cite{toxicCodeSnippets,bai_insecure_code_propagration,Fischer_Effect_of_Google_Search_on_Software_Security,Chen_crowd_debugging} and are more amenable to replication.
In contrast, other works involved developer studies to solve programming challenges with the help of StackOverflow~\cite{acar2016YouGetWhereYouLook} or to evaluate their tools~\cite{Ren_Demystify_API_Usage_Directives,Mahajan_Recommending_stack_overflow_posts,Mindermann_fluid_intelligence}.
These studies require a carefully crafted replication to isolate differences between participant groups from the impact of Stack Overflow evolution.

\section{Related Works}\label{sec:related_work}
We briefly present related works that explore meta-research studies and investigate the evolution of data.

\subsection{Meta-research}\label{sec:related_works:meta-research}
Meta-research is an important tool for evaluating and improving research practices.
It helps to identify which research methods provide reliable and reproducible answers.
Ioannidis~\cite{Ioannidis_why_most_findings_are_false} investigated prior research findings and found that most published research turned out to be false.
One factor contributing to this crisis is the lack of code and data to verify research claims.
Ioannidis et al.~\cite{Ioannidis_meta-research} introduced a framework that serves as a benchmark to holistically evaluate and improve research practices to make results more reliable.
Demir et al.~\cite{demir2022reproducibility} examined the reproducibility and replicability of web measurement studies.
They discovered that many studies lacked proper documentation of their experimental setups, which is essential for accurately reproducing and replicating results.
In particular, they found that even slight variations in experimental setup could lead to significant differences in results.
The authors recommended proper documentation and adopting standardized practices to make web measurement findings more reliable.
Weber et al.~\cite{weber2019essential} conducted a study of benchmarking studies to develop guidelines to help computational scientists conduct better benchmarking studies.
While their guidelines are for a different target group than ours, some of them translate to our work.
For instance, the guideline on selecting datasets and reproducible research best practices can be transferred, as it shows that security studies focused on Stack Exchange datasets should be measured using different dataset versions and that the code and data for studies should be made publicly available to facilitate replication studies.

\subsection{Dataset Evolution}\label{sec:related_works_dataset:evolution}

Ceroni et al.~\cite{ceroni2014information} investigated Wikipedia's dynamic nature to examine how its content changes over time, specifically looking at page edits.
Their research highlights the significance of understanding content evolution to evaluate the quality and accuracy of information on Wikipedia.
In a follow-up, Tran et al.~\cite{tran2014wikipevent} leverage the evolution of data on Wikipedia to analyze the history of user edits to extract and represent complex events.
Fetterly et al.~\cite{fetterly2003large} provides a detailed investigation of how web pages evolve over time to understand how that evolution might impact crawling and indexing processes for the web. 
Their study shed light on the importance of understanding the evolution of web pages for improving web crawling and indexing processes. 
Jallow et al. ~\cite{jallow_sp24} is closely related to us.
They studied the impact of evolving Stack Overflow code snippets on software developers.
In contrast, we focus on how code evolution might impact prior research findings.

Conducting time-series analysis to better understand data that can change over time is explained in the textbooks of several other research disciplines, such as economics~\cite{hamilton1994series}, environmental science~\cite{wilks-statistical_2011}, medicine~\cite{ts_shumway}, finance~\cite{ts_tsay}, social sciences~\cite{Box-Steffensmeier_Freeman_Hitt_Pevehouse_2014}, or engineering~\cite{ts_ljung}.

\section{Conclusions for Future Studies}\label{sec:conclusion}

\paragraph{Time-Series Analysis.}
Our systematization showed that security-focused Stack Overflow studies relied on specific dataset versions and specific, non-stationary aspects.
Among the 42 relevant papers we identified, only Ragkhitwetsagul et al.~\cite{toxicCodeSnippets} discussed the potential impact of evolution on their results.
Four works~\cite{dicos,selvaraj_collaborative_editing,zhang_code_weaknesses,ManesBaysal_MSR21} use code evolution in their methodology but do not discuss its consequences for their findings.
With our six replication studies, we revealed that the findings of four studies differ for newer dataset versions.
This does \textit{not} mean that the results of prior research are wrong---but missing context.
Researchers are advised to provide additional context around their results by treating the Stack Overflow dataset as the time-series data it is.
Considering the temporal dimension of posts and conducting trend analysis can help better understand such results, e.g., whether issues are recurring or stable, and can address temporal dynamics, such as differentiating short-lived trends from long-term changes.
The StackOverflow dataset is particularly suited to this practice since the versioning of posts allows analysis of the platform's prior states.
While it is unrealistic to predict future changes with certainty, considering multiple dataset versions and their aspects' stationarity provides insights into how changes on Stack Overflow might affect the findings.
For example, we found that CWE types of C/C++ code snippets shifted over time while crypto API misuse in Java snippets seemed stable or even increasing.
Trend analysis is common in other research disciplines, where drawing conclusions from a single observation would be insufficient, if not misleading.
This is particularly true in fields such as economics~\cite{hamilton1994series} (studying variables like GDP, inflation, or stock prices), environmental science~\cite{wilks-statistical_2011} (analyzing climate data or temperature changes), or social sciences~\cite{Box-Steffensmeier_Freeman_Hitt_Pevehouse_2014} (examining voting patterns, crime rates, or demographic shifts).
We suggest that our discipline adopt such a longitudinal analysis methodology when studying data that changes over time.

\paragraph{Open Science Best Practices.}
Although not a goal at the outset of our work, we want to re-iterate some best practices for open science and reproducibility that we found disregarded during our replication studies.
We found that many artifacts were not or only partially available (on request) or non-functional.
Making code and data available on request is often insufficient~\cite{weber2019essential}, and experience shows that paper authors can often ignore such requests.
Thus, our work underlines the need to establish better efforts for open science, and we welcome the recent steps to introduce artifact evaluation committees and require artifacts for accepted papers.
Moreover, we also found that, in most cases, the information provided in the papers was insufficient to allow others to re-implement the methods when artifacts were unavailable.
Both the ACM SIGPLAN Empirical Evaluation Guidelines~\cite{sigplan_metrics} and van der Kouwe et al.'s benchmarking flaws~\cite{benchmarking_flaws} remark that lack of such information leads to a lack of reproducibility and can hinder scientific progress.
Thus, we urge researchers to report the software versions used in their methods.
Containerization tools, like Docker, to encapsulate software environments and preserve package versions and dependencies could ease replication efforts and help with artifact releases.

\section{Ethics Considerations and Open Science}
\subsection*{Open Science and Availability}
The artifact for this paper is publicly accessible and permanently hosted on Zenodo~\cite{jallow_usenix2025_artifact}.

\subsection*{Ethics Considerations}

\subsubsection*{Reproducibility and Reliability}
We tried to consider the implications of our findings for the broader field.
Where our replication studies produced different results from the original study, we tried to discuss the possible reasons for these differences and their impact on the reliability of the original findings.
Sections~\ref{sec:code_based_studies} and \ref{sec:knowledge_evolution} aim to identify the potential causes for diverging results before our replication study in a systematic way.

\subsubsection*{Originality and Plagiarism}
We clearly acknowledge the original studies we replicated. We avoid plagiarism by properly crediting the original authors and citing their work appropriately.

\subsubsection*{Respect for Original Research}
We tried to be respectful of the original researchers’ work. Critiques or discussions regarding the original study are phrased constructively and based on the data and findings.

\subsubsection*{Consent for the Use of Original Data}
All data and code artifacts of the prior studies were either publicly available (on request) or re-implemented based on the published results.
We credited the original researchers appropriately if their artifacts were reused, and our systematization in Section~\ref{sec:code_based_studies} highlights which artifacts were available.

\subsubsection*{Communication with Original Researchers}
We engaged with the original researchers, especially when we could not reproduce the original results and when artifacts were unavailable or seemingly flawed.
This was intended to help ensure that any discrepancies are thoroughly understood and addressed in a collaborative manner.

\subsubsection*{Responsible Disclosure and Report Findings}

Some of the replicated studies~\cite{zhang_code_weaknesses,fischer2017SOConsideredHarmful} aim at discovering vulnerable code snippets.
Replicating their methodology on a newer version of the Stack Overflow data set can result in discovering additional snippets with weaknesses.
While discovered vulnerabilities in other code bases (e.g., production or open-source software) should be responsibly disclosed to vendors/developers, there are no guidelines about disclosure for publicly posted code snippets.
While the technical option exists to add a corresponding comment\footnote{\url{https://api.stackexchange.com/docs/create-comment}} on each post with a vulnerable snippet, we found this to be an ethical dilemma.
First, the vulnerability does not affect Stack Overflow but only developers that copy such snippets into their code bases.
Thus, the disclosure might not reach the affected parties.
Second, the employed methodologies only discover weaknesses, but without a concrete explanation of the vulnerability and the corresponding fix, a comment would not benefit the community.
Third, and more crucial, without careful consideration of the context of the snippet, commenting on weaknesses can easily create noise and even be considered offensive/useless (e.g., if the original question asked about an explanation of a vulnerability, i.e., if the vulnerability was posted on purpose).
Without being able to address the second and third concerns at scale, such auto-commenting of weaknesses would also violate the Stack Overflow code of conduct.\footnote{\url{https://stackoverflow.com/help/privileges/comment}}
We further consulted the Stack Exchange Meta discourse\footnote{\url{https://meta.stackexchange.com/}} about how vulnerabilities in code snippets should be reported.
The meta-discussions mirror, to some extent, our concerns.
For example, that mass commenting will create noise\footnote{\url{https://meta.stackexchange.com/questions/258328/right-approach-to-crawl-and-identify-bad-code}} and that such discoveries should be accompanied by code fixes\footnote{\label{footnote:so1}\url{https://meta.stackexchange.com/questions/9460/how-to-deal-with-questions-answers-with-a-security-vulnerability}}.
The community also discussed other measures, such as introducing a ``security'' or ``caution'' flag to Stack Overflow posts with vulnerabilities or outdated security measures.\footnote{\url{https://meta.stackexchange.com/questions/301592/keeping-answers-related-to-security-up-to-date}, \url{https://meta.stackexchange.com/questions/89469/what-to-do-with-questions-with-harmful-content/89474}}
However, the community also agrees on the sentiment that while code snippets could be copied\&pasted, Stack Overflow is a teaching platform for developers to learn how to solve their specific problems and that authored code should be owned, hence, not just pasted and instead, the solutions on the platform should be scrutinized and lead to further education and ultimately custom solutions.\footnote{\url{https://meta.stackexchange.com/questions/334811/stack-overflow-made-the-bbc-news-copycat-coders-create-vulnerable-apps}}
Since it is currently impossible to make comments/edits of the required quality, we did not create fixes for newly discovered posts with weaknesses for the above-stated reasons.
We note that neither Zhang et al.~\cite{zhang_code_weaknesses} nor Fischer et al.~\cite{fischer2017SOConsideredHarmful} mentioned any attempted disclosure on Stack Overflow---Fischer et al., however, built-in follow-up work~\cite{FischerUsenix2019} a browser plugin to warn Stack Overflow users of insecure snippets.

{\small
\bibliographystyle{IEEEtranS}
\bibliography{main}
}
\appendix
\clearpage
\section{Systematization and Evolution}

\subsection{OpenAI GPT4o Evaluation}\label[secinapp]{sec:appendix:llm}

To assess GPT4o’s performance in screening papers based on their relevance to the security and bugs of code snippets, we tested it with two sets of manually labeled studies: 10 true positive cases fitting these criteria and 10 false positive cases. 
We carefully selected the negative samples to include studies that almost matched the inclusion criteria, helping us evaluate how well the model distinguishes between relevant and irrelevant studies.
We iteratively refined our LLM prompt until we were confident it would correctly identify studies that meet our criteria and exclude those that do not. The results of this evaluation are provided in \Cref{tab:llmtestpapers} and show that GPT4o performed well in distinguishing the papers.
The optimized LLM prompt is listed in \Cref{lst:llmprompt}.

\begin{table}[h]
    \centering
    \caption{Papers used to test the GPT4o performance}
    \label{tab:llmtestpapers}
\fontsize{6}{7}\selectfont
\rowcolors{2}{gray!25}{white}
    \begin{tabularx}{\linewidth}{|X|c|c|}
\hline
\textbf{Paper Title} &
\textbf{Truth Label} &
\textbf{GPT4o Label} \\ \hline
Dicos: Discovering Insecure Code Snippets from Stack Overflow Posts by Leveraging User Discussions  & True & True \\ \hline
Toxic Code Snippets on Stack Overflow  & True & True \\ \hline
Does Collaborative Editing Help Mitigate Security Vulnerabilities in Crowd-Shared IoT Code Examples?  & True & True \\ \hline
Mining Rule Violations in JavaScript Code Snippets  & True & True \\ \hline
Snakes in Paradise?: Insecure Python-Related Coding Practices in Stack Overflow  & True & True \\ \hline
How Reliable is the Crowdsourced Knowledge of Security Implementation?  & True & True \\ \hline
Secure Coding Practices in Java: Challenges and Vulnerabilities  & True & True \\ \hline
You Get Where You're Looking for: The Impact of Information Sources on Code Security  & True & True \\ \hline
Stack Overflow Considered Harmful? The Impact of Copy\&Paste on Android Application Security  & True & True \\ \hline
Stack Overflow Considered Helpful! Deep Learning Security Nudges Towards Stronger Cryptography  & True & True \\ \hline
Understanding Privacy-Related Questions on Stack Overflow  & False & False \\ \hline
Exploring Technical Debt in Security Questions on Stack Overflow  & False & False \\ \hline
Crowd-GPS-Sec: Leveraging Crowdsourcing to Detect and Localize GPS Spoofing Attacks  & False & False \\ \hline
An Investigation of Security Conversations in Stack Overflow: Perceptions of Security and Community Involvement  & False & False \\ \hline
An Anatomy of Security Conversations in Stack Overflow  & False & False \\ \hline
Is Reputation on Stack Overflow Always a Good Indicator of Users' Expertise? No  & False & False \\ \hline
Mobile App Security Trends and Topics: An Examination of Questions from Stack Overflow  & False & False \\ \hline
Challenges in Docker Development: A Large-Scale Study Using Stack Overflow  & False & False \\ \hline
What Security Questions Do Developers Ask? A Large-Scale Study of Stack Overflow Posts  & False & False \\ \hline
Hurdles for Developers in Cryptography  & False & False \\ \hline
\end{tabularx}
\end{table}

\begin{lstlisting}[caption="Final GPT4o prompt",label=lst:llmprompt,basicstyle=\scriptsize,language=json]
{
   "messages": [
      {
         "role": "system",
         "content": "You are an expert in analyzing academic research focused on the security or correctness (e.g., buggy or faulty) of code snippets on Stack Overflow. Do not consider studies that focus on general security discussions or questions about security, unless they specifically analyze the security of the code snippets themselves, or identify and address issues such as bugs or faults in the code snippets."
      },
      {
         "role": "user",
         "content": "Does this paper study the security of code snippets on Stack Overflow or identify and address issues such as bugs or faults in the code snippets? Provide a 'Yes' or 'No' and a brief justification in the following JSON format: { 'SecurityRelevant': 'True' or 'False', 'Justification': 'brief justification' }: \n\n{paper_abstract}"
      }
   ]
}
\end{lstlisting}

\subsection{Search Terms}\label[secinapp]{sec:appendix:systematization:keywords}

We used the following search terms to identify research studies focused on Stack Overflow. To identify studies containing the keywords in their titles or abstracts, we normalized all keywords, titles, and abstracts to lowercase before performing the search.
For example, "Stack Overflow" or "StackOverflow" was normalized to "stack overflow" and "stackoverflow"  to ensure consistent matching.

\begin{lstlisting}[language=bash,basicstyle=\scriptsize]
"stackoverflow" OR "stack overflow" OR "crowd knowledge" OR
"crowdsource knowledge" OR "crowd-source knowledge" OR "Q&A websites" OR
"Q&A sites" OR "social Q&A websites" OR  "online Q&A communities" OR
"community question answering" OR "knowledge sharing" OR
"knowledge-sharing" OR "crowdsourcing" OR "Q&A Forums" OR
"Online Code Snippets"
\end{lstlisting}

\subsection{Venues}\label[secinapp]{sec:appendix:systematization:venues}

\Cref{tab:venues} lists the venues in whose proceedings and volumes we search for candidate papers that study Stack Overflow using the search terms presented in \Cref{sec:appendix:systematization:keywords}.

\begin{table}[!htbp]
\caption{List of venues for literature search.}
\label{tab:venues}
\centering
\fontsize{6}{7}\selectfont
\rowcolors{2}{gray!25}{white}
\begin{tabularx}{\linewidth}{|X|l|c|}
\hline
\textbf{Venue} & \textbf{Cat.} & \textbf{Type} \\
\hline
ACM Conference on Computer and Communications Security (CCS) & SP & C \\
\hline
IEEE Symposium on Security and Privacy (SP) & SP & C \\
\hline
USENIX Security Symposium & SP & C \\
\hline
Network and Distributed System Security Symposium (NDSS) & SP & C \\
\hline
Annual Computer Security Applications Conference (ACSAC) & SP & C \\
\hline
ACM ASIA Conference on Computer and Communications Security & SP & C \\
\hline
IEEE European Symposium on Security and Privacy & SP & C \\
\hline
\hline
Detection of Intrusions and Malware \& Vulnerability Assessment (DIMVA) & HCl & C \\
\hline
ACM Conference on Human Factors in Computing Systems (CHI) & HCl & C \\
\hline
ACM Conference on Computer-Supported Cooperative Work and Social Computing (CSCW) & HCl & C \\
\hline
ACM Symposium on User Interface Software and Technology (UIST) & SE & C \\
\hline
ACM SIGSOFT Conference on the Foundations of Software Engineering (FSE) & SE & C \\
\hline
International Conference on Software Engineering (ICSE) & SE & C \\
\hline
International Conference on Automated Software Engineering (ASE) & SE & C \\
\hline
IEEE International Conference on Software Maintenance (ICSM) & SE & C \\
\hline
International Symposium on Software Reliability Engineering (ISSRE) & SE & C \\
\hline
International Conference on Software Testing, Verification and Validation (ICST) & SE & C \\
\hline
International Conference on Program Comprehension (ICPC) & SE & C \\
\hline
International Conference on Mining Software Repositories (MSR) & SE & C \\
\hline
Empirical Software Engineering and Measurement (ESEM) & SE & C \\
\hline
International Conference on Software Analysis, Evolution, and Reengineering (SANER) & SE & C \\
\hline
European Software Engineering Conference (ESEC) & SE & C \\
\hline
IEEE International Conference on Software Maintenance and Evolution (ICSME) & SE & C \\
\hline
IEEE Conference on Reverse Engineering (WCRE) & SE & C \\
\hline
International Symposium on Software Testing and Analysis (ISSTA) & SE & C \\
\hline
Fundamental Approaches to Software Engineering (FASE) & SE & C \\
\hline
IEEE International Conference on Program Comprehension (ICPC) & SE & C \\
\hline
The Web Conference (WWW) & SE & J \\
\hline
IEEE Transactions on Software Engineering (TSE) & SE & J \\
\hline
Empirical Software Engineering (EMSE) & SE & J \\
\hline
\multicolumn{3}{c}{C = Conference; J = Journal}
\end{tabularx}
\end{table}

\subsection{Relevance of Stack Overflow Evolution}\label[secinapp]{sec:appendix:systematization}

We provide further information about how the evolution of Stack Overflow can affect the non-replicated papers from \Cref{tab:comparision_table}.

Verdi et al.~\cite{verdi19} examined the insecure reuse of C++ code snippets in open-source GitHub projects (\textbf{D1}: C/C++). They manually assessed the security of the snippets (\textbf{D2}: \resolve{no}) and focused only on those explicitly attributed within projects (\textbf{D5}: \resolve{yes}).
The study did not consider the evolution of these snippets (\textbf{D3}: \resolve{no}) or their surrounding contexts (\textbf{D4}: \resolve{no}).

Selvaraj et al.~\cite{selvaraj_collaborative_editing} extended Zhang et al.'s~\cite{zhang_code_weaknesses} examine if revisions to C/C++ IoT code snippets (\textbf{D1}: C) could reduce vulnerabilities on Stack Overflow, Arduino, and Raspberry Pi Stack Exchange sites. Their methodology closely mirrors Zhang et al.'s (\textbf{D2}: \resolve{yes}, \textbf{D3}: \resolve{yes}, \textbf{D4}: \resolve{no}) but focuses solely on IoT code snippets (\textbf{D5}: \resolve{yes}).

Acar et al.~\cite{acar2016YouGetWhereYouLook} studied how Android app developers (\textbf{D1}: Java) use code snippets from Stack Overflow to solve programming challenges. There were no restrictions on the reusable snippets (\textbf{D5}: \resolve{no}), and their security was manually determined (\textbf{D2}: \resolve{no}). Neither code evolution nor the surrounding context was considered in the study (\textbf{D3}: \resolve{no}); (\textbf{D4}: \resolve{no}).

Chen et al.~\cite{Chen_reliable_crowd_source_knowledge} studied the security of Java code snippets (\textbf{D1}: Java), focusing on crypto APIs (\textbf{D5}: \resolve{yes}). They manually labeled snippets as secure/insecure (\textbf{D2}: \resolve{no}) without considering code evolution or context (\textbf{D3}, \textbf{D4}: \resolve{no}).

Meng et al. studied Java code snippets (\textbf{D1}: Java) to explore the challenges in implementing secure solutions. They manually assessed security properties (\textbf{D2}: \resolve{no}) and focused on posts with security tags (\textbf{D5}: \resolve{yes}), considering the surrounding context (\textbf{D4}: \resolve{yes}) but not code evolution (\textbf{D3}: \resolve{no}).

Ragkhitwetsagul et al.~\cite{toxicCodeSnippets} examined Java snippets (\textbf{D1}: Java) copied from GitHub to see if they remain outdated on Stack Overflow. They focused on accepted answers (\textbf{D5}: \resolve{yes}) and code evolution to assess outdatedness (\textbf{D3}: \resolve{yes}), without considering surrounding context (\textbf{D4}: \resolve{no}) or using a code scanner (\textbf{D2}: \resolve{no}).

Bai et al.~\cite{bai_insecure_code_propagration} examined Java code snippets (\textbf{D1}: Java) to investigate the propagation of insecure code from Stack Overflow. They relied on manual analysis (\textbf{D2}: \resolve{no}) to classify code snippets from answers only (\textbf{D5}: \resolve{yes}). The authors did not consider the evolution of code snippets (\textbf{D3}: \resolve{no}), nor the context surrounding them (\textbf{D4}: \resolve{no}).

Bagherzadeh et al.~\cite{Bagherzadeh_akka_actor_bugs} analyzed 186 real-world Akka actor bugs in Java and Scala code snippets (\textbf{D1}: Java/Scala) to understand and classify the symptoms, root causes, and their associated API usages. The Akka-related bugs and vulnerabilities were detected manually (\textbf{D2} = \resolve{no}) and code evolution was not considered in their study (\textbf{D3}: \resolve{no}). However, the authors analyzed the context surrounding code snippets, as it includes analysis of developer discussions in Stack Overflow and GitHub (\textbf{D4}: \resolve{yes}). The study focuses only on Akka-related code snippets (\textbf{D5}: \resolve{yes}).

Chen et al.~\cite{chen_intelligent_detection_system} analyzed Java code snippets (\textbf{D1}: Java) to detect insecure snippets from answer posts using a novel hierarchical attention-based sequence learning model (\textbf{D2}: \resolve{orange}, M; \textbf{D5}: \resolve{yes}). The authors did not consider the evolution of code snippets (\textbf{D3}: \resolve{no}), nor did they analyze the context surrounding them (\textbf{D4}: \resolve{no}).

Zhang et al.~\cite{ZhangSoApiMisuse} analyzed Java code snippets (\textbf{D1}: Java) with the ExampleCheck custom tool (\textbf{D2}: \resolve{orange}, P) to assess the reliability of code examples in terms of API misuse. The study did not consider the evolution of code snippets (\textbf{D3}: \resolve{no}) nor their surrounding context (\textbf{D4}: \resolve{no}) and excluded code snippets from question posts (\textbf{D5}: \resolve{yes}).

Rahman et al.~\cite{rahman_reusability_insight} investigated the impact of reusing Java code snippets (\textbf{D1}: Java) by manually labeling (\textbf{D2}: \resolve{no}) the reusability, stability, and bug-proneness of code snippets when reused in mobile applications.
They did not consider code evolution (\textbf{D3}: \resolve{no}) nor the context surrounding the code snippets considered in their study (\textbf{D4}: \resolve{no}), and they considered only code snippets in answer posts (\textbf{D5}: \resolve{yes}).

Reinhardt et al.~\cite{reinhardt_augmenting_stack_overflow} analyzed Java code snippets (\textbf{D1}: Java) to detect API misuse using ExampleCheck (\textbf{D2}: \resolve{yes}). The study did not consider code evolution (\textbf{D3}: \resolve{no}) nor the context around code snippets (\textbf{D4}: \resolve{no}). The study only considered code snippets that used specific Java APIs (\textbf{D5}: \resolve{yes}).

Licorish et al.~\cite{Licorish_contextual_profiling} analyzed Java code snippets (\textbf{D1}: Java) for security vulnerabilities using FindBugs (\textbf{D2}: \resolve{yes}). The study did not consider code evolution (\textbf{D3}: \resolve{no}) but did analyze the context surrounding the code snippets, including discussions and comments, to gauge awareness of security issues (\textbf{D4}: \resolve{yes}). The study did not filter out specific code snippets, analyzing all relevant code from Stack Overflow posts~(\textbf{D5}:~\resolve{no}).

Schmidt et al.~\cite{Schmidt_CopypastaVulGuard} analyzed JavaScript and PHP Code snippets (\textbf{D1}: JavaScript/PHP) using CopypastaVulGuard, a tool that primarily relies on SQL queries and regular expressions to identify specific vulnerabilities, such as SQL injection, remote code execution, and deprecated functions in source snippets (\textbf{D2}: \resolve{no}). The study did not focus on code evolution (\textbf{D3}: \resolve{no}) but considered the surrounding context of the posts (\textbf{D4}: \resolve{yes}). The study considered all relevant code snippets from Stack Overflow without filtering (\textbf{D5}: \resolve{no}).

 Liu et al.~\cite{Liu_deepanna} analyzed Java code snippets (\textbf{D1}: Java) to recommend and detect annotation misuses using a custom tool based on deep learning and multi-label classification (\textbf{D2}: \resolve{orange},M). The study did not consider code evolution (\textbf{D3}: \resolve{no}) but incorporated both structural and textual context (e.g., code comments) for annotation recommendation and misuse detection (\textbf{D4}: \resolve{yes}). The study analyzed all Java code snippets relevant to its focus without filtering any specific snippets~(\textbf{D5}:~\resolve{no}).

Ren et al.~\cite{Ren_Demystify_API_Usage_Directives} analyzed Java code snippets (\textbf{D1}: Java) to identify API misuse scenarios and generate demystification reports using a custom tool(\textbf{D2}: \resolve{orange}, M) based on text mining. Although code evolution was not part of their methodology (\textbf{D3}: \resolve{no}), they did analyze the surrounding context of the code snippets, including discussions and API usage directives, to create comprehensive reports (\textbf{D4}: \resolve{yes}). The study filtered out specific discussion threads by focusing on those that contained relevant API mentions, excluding other threads (\textbf{D5}: \resolve{yes}).

Licorish et al.~\cite{Licorish_dissecting_copy_delete_replace_swap_mutations} analyzed Java code snippets (\textbf{D1}: Java) to investigate the effects of mutating Java code to fix performance-related issues detected by PMD (\textbf{D2}: \resolve{yes}).
The study did not consider the evolution of the code (\textbf{D3}: \resolve{no}) nor its surrounding context (\textbf{D4}: \resolve{no}). However, the authors analyzed all relevant code snippets without additional filtering~(\textbf{D5}:~\resolve{no}).

Rangeet Pan~\cite{Pan_deep_learning_bug_fixing} analyzed Python code snippets (\textbf{D1}: Python) to examine the impact of bug-fixing on the robustness of deep learning models. The authors relied on a manual classification scheme to categorize bug-fix posts (\textbf{D2}: \resolve{no}) to classify code snippets and did not consider the evolution of code snippets (\textbf{D3}: \resolve{no}). However, they leveraged user discussions (\textbf{D4}: \resolve{yes}) in the classification and filtered out bug-fix posts related to deep learning (\textbf{D5}: \resolve{yes}).

Ye et al.~\cite{Ye_Insecure_Code_Snippet_Detection} analyzed Java code snippets (\textbf{D1}: Java) to detect insecure code using a heterogeneous information network (HIN) and a novel network embedding model (snippet2vec) (\textbf{D2}: \resolve{orange}, M). The study did not focus on code evolution (\textbf{D3}: \resolve{no}) nor considered the context surrounding around code snippets (\textbf{D4}: \resolve{no}) and focused on Android (\textbf{D5}: \resolve{yes}).

Chen et al.~\cite{Chen_crowd_debugging} analyzed Java code snippets (\textbf{D1}: Java) to develop a \textit{crowd debugging} technique based on machine learning to identify defective code fragments (\textbf{D2}: \resolve{orange}, M). The study did not consider code evolution (\textbf{D3}: \resolve{no}), but the technique relies on explanations and suggestions provided by users in answer posts (\textbf{D4}: \resolve{yes}). Only code snippets with meaningful code explanations were selected (\textbf{D5}: \resolve{yes}).

Zhang et al.~\cite{Zhang_TensorFlow_program_bugs} manually collected 87 TensorFlow-related bug reports from Stack Overflow TensorFlow-related code snippets (\textbf{D1}: Python, \textbf{D2}: \resolve{no}) to identify bugs and their root causes. 
The study did not consider the evolution of snippets (\textbf{D3}: \resolve{no}), nor their surrounding context (\textbf{D4}: \resolve{no}). However, they did filter out irrelevant discussions, focusing on bugs related to TensorFlow (\textbf{D5}: \resolve{yes}).

Alhanahnah et al.~\cite{Alhanahnah_best_secure_coding_practice} analyzed Java code snippets (\textbf{D1}: Java) to detect insecure SSL/TLS implementation patterns, specifically targeting certificate and hostname validation vulnerabilities. They used the PMD tool with a custom ruleset designed to detect SSL/TLS security flaws (\textbf{D2}: \resolve{yes}). 
The study did not consider the evolution of code snippets (\textbf{D3}: \resolve{no}) but considered the context surrounding them to identify insecure implementation patterns (\textbf{D4}: \resolve{yes}) and focused specifically on snippets related to SSL/TLS (\textbf{D5}: \resolve{yes}).

Imai et al.~\cite{Imai_Time_Series_Analysis_Copy_and_Paste_Impact} analyzed the impact of reusing potentially vulnerable Java snippets (\textbf{D1}: Java) into Android applications. The study uses a custom tool based on pattern recognition (\textbf{D2}: \resolve{orange},P) to detect vulnerable snippets. The study did not examine the evolution of code snippets (\textbf{D3}: \resolve{no}) nor their surrounding context (\textbf{D4}: \resolve{no}) and focused exclusively on Android snippets~(\textbf{D5}:~\resolve{yes}).

Fischer et al.~\cite{Fischer_Effect_of_Google_Search_on_Software_Security} analyzed Java code snippets (\textbf{D1}: Java) to assess how \textit{Google Search rankings} affect the security of cryptographic code examples that developers reuse. The study introduced a security-based re-ranking system, based on machine learning, to improve the visibility of secure code by altering the ranking of search results (\textbf{D2}: \resolve{orange},M). The authors did not consider the evolution of the cryptographic code snippets (\textbf{D3}: \resolve{no}) nor their context (\textbf{D4}: \resolve{no}). Similar to Fischer et al.~\cite{fischer2017SOConsideredHarmful, FischerUsenix2019}, the study focused on cryptographic code examples (\textbf{D5}: \resolve{yes}).

Almeida et al.~\cite{Almeida_Rexstepper} developed REXSTEPPER, a debugger for JavaScript regular expressions (\textbf{D1}: JavaScript) designed to troubleshoot common issues. The study utilized \textit{REXREF}, an off-the-shelve tool for detecting issues in the regular expressions (\textbf{D2}: \resolve{yes}). The study focused on debugging 18 faulty regular expressions sourced from Stack Overflow (\textbf{D5}: \resolve{yes}) but did not consider code evolution (\textbf{D3}: \resolve{no}) or their surrounding context (\textbf{D4}: \resolve{no}).

Islam et al.~\cite{Islam_repairing_deep_neural_networks} analyzed code snippets related to deep neural networks from both Stack Overflow and GitHub (\textbf{D1}: Python) to investigate bug fix patterns to understand the common challenges and solutions used by developers. The study relied on manual classification of bug fixes (\textbf{D2}: \resolve{no}) and did not consider code evolution (\textbf{D3}: \resolve{no}) nor their surrounding context (\textbf{D4}: \resolve{no}). 
The authors only considered Python code snippets that use deep learning libraries (\textbf{D5}: \resolve{yes}).

Mahajan et al.~\cite{Mahajan_Recommending_stack_overflow_posts} analyzed Java code snippets (\textbf{D1}: Java) to recommend relevant Stack Overflow posts for fixing runtime exceptions by using a custom-built tool based on Abstract Program Graphs (APGs) (\textbf{D2}: \resolve{orange}, A). The study did not consider the evolution of code snippets (\textbf{D3}: \resolve{no}) nor their surrounding context (\textbf{D4}: \resolve{no}) but filtered relevant posts based on certain criteria, e.g., question post must have an accepted answer (\textbf{D5}: \resolve{yes}).

Mahajan et al.~\cite{Mahajan_repairing_runtime_exceptions} analyzed Java code snippets (\textbf{D1}: Java) to automatically recommend real-time fixes for runtime exceptions (\textbf{D2}: \resolve{yes}). The study did not consider the evolution of code snippets (\textbf{D3}: \resolve{no}) nor their surrounding context (\textbf{D4}: \resolve{no}) but considered question posts with an accepted answer~(\textbf{D5}:~\resolve{yes}).

Yadavally et al.~\cite{Yadavally_program_dependence_learning} analyzed pre-labelled Java and C/C++ code snippets (\textbf{D1}: Java, C/C++) from Stack Overflow (\textbf{D2}: \resolve{no}) to detect vulnerable code snippets.
The study did not consider code snippet evolution(\textbf{D3}: \resolve{no}) nor the context surrounding them (\textbf{D4}: \resolve{no}). Additionally, they filtered out incomplete code snippets and focused on analyzing method-level code with a maximum of 8 LoC (\textbf{D5}: \resolve{yes}).

Firouzi et al.~\cite{Firouzi_unsafe_code_context} analyzed C\# code snippets that use the \textit{unsafe} keyword on Stack Overflow (\textbf{D1}: C\#) using regular expressions and manual checks to extract and analyze the snippets for potential security vulnerabilities (\textbf{D2}: \resolve{no}). The authors did not consider code evolution (\textbf{D3}: \resolve{no}), nor did it their surrounding context (\textbf{D4}: \resolve{no}), and focused on extracting only code snippets that used the unsafe keyword (\textbf{D5}: \resolve{yes}).

Ghanbari et al.~\cite{Ghanbari_fault_localization} analyzed deep neural networks (\textbf{D1}: Python)  to locate faults using mutation-based fault localization (\textbf{D2}: \resolve{orange}, F) to localize faults in pre-trained DNN models. The study did not consider the evolution of code snippets (\textbf{D3}: \resolve{no}), nor their surrounding context (\textbf{D4}: \resolve{no}) and focused solely on deep-learning-related code snippets (\textbf{D5}: \resolve{yes}).

Gao et al.~\cite{Gao_fixing_recurring_crash_bugs} proposed an automatic approach to fixing recurring crash bugs in Java code snippets (\textbf{D1}: Java).
They used off-the-shelve tools for partial parsing, tree-based code differencing, and edit script generation to identify and apply fixes for crash bugs in code snippets (\textbf{D2}: \resolve{yes}). 
They did not consider the evolution of code snippets (\textbf{D3}: \resolve{no}), nor their surrounding context (\textbf{D4}: \resolve{no}).
Only code snippets containing fixes to crash bugs were sampled (\textbf{D5}: \resolve{yes}).

M. Chakraborty\cite{Chakraborty_bug_propagation} analyzed the reuse of the pre-trained BERT model in Python code snippets (\textbf{D1}: Python) to identify bugs related to the reuse of BERT. The authors used a manual analysis approach to identify those bugs (\textbf{D2}: \resolve{no}) and did not consider the evolution of these BERT-related buggy snippets (\textbf{D3}: \resolve{no}), nor their surrounding context (\textbf{D4}: \resolve{no}). Posts tagged with BERT or related keywords with a score $ge$ 2 were considered (\textbf{D5}: \resolve{yes}).

Moradi Moghadam et al.~\cite{Moghadam_mutation_testing} proposed $\mu$Akka, a mutation testing framework for actor concurrency in the Akka system, using 130 real-world Akka bugs sourced from Java code snippets  (\textbf{D1}: Java). The authors manually reviewed (\textbf{D2}: \resolve{no}) the 130 code snippets containing those bugs.
The evolution of the snippets was not considered (\textbf{D3}: \resolve{no}), nor took into account their surrounding context (\textbf{D4}: \resolve{no}) and focused specifically on Akka actor bugs on Stack Overflow (\textbf{D5}: \resolve{yes}).

Magnus Madsen et al.~\cite{Madsen_javaScript_promises} analyzed JavaScript promises (\textbf{D1}: JavaScript) to detect promise-related errors, using a promise graph (\textbf{D2}: \resolve{orange}, S). They did not examine the evolution of code snippets (\textbf{D3}: \resolve{no}), but they extensively studied the surrounding context by manually inspecting StackOverflow discussions to understand errors related to promise usage (\textbf{D4}: \resolve{yes}). They also applied filters by focusing on StackOverflow posts that were directly related to promise errors rather than all JavaScript code snippets (\textbf{D5}: \resolve{yes}).

Jhoo et al.~\cite{Jhoo_static_analyzer} analyzed Python code snippets (\textbf{D1}: Python) to detect tensor shape errors in PyTorch snippets using a custom static analyzer tool called PyTea (\textbf{D2}: \resolve{orange}, S). The authors did not consider code evolution (\textbf{D3}: \resolve{no}) nor the context surrounding them (\textbf{D4}: \resolve{no}). Only PyTorch-specific code snippets were considered (\textbf{D5}: \resolve{yes}).

\subsection{Evolution of Stack Overflow}\label[secinapp]{sec:appendix:systematization:evolution}

\begin{figure}[h]
    \centering
    \includegraphics[width=.8\linewidth]{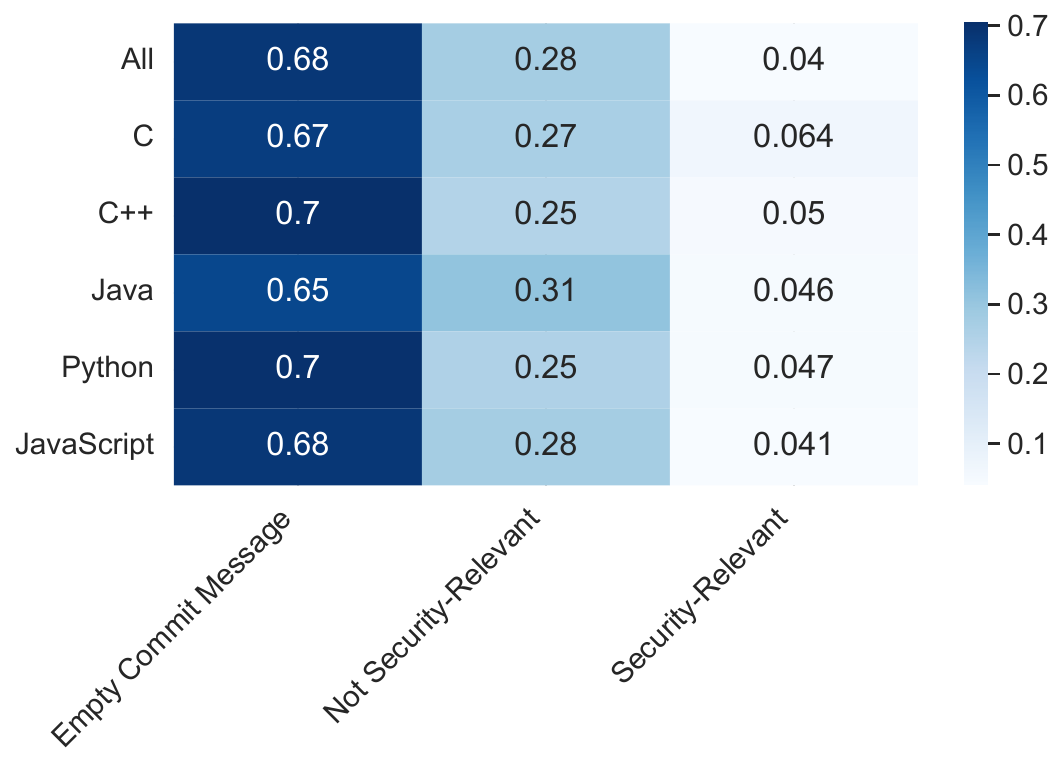}
    \caption{Normalized confusion matrix for \Cref{tab:appendix:commitmsg}}
    \label{fig:appendix:cm_all}
\end{figure}

\begin{figure}[h]
    \centering
    \includegraphics[width=.8\linewidth]{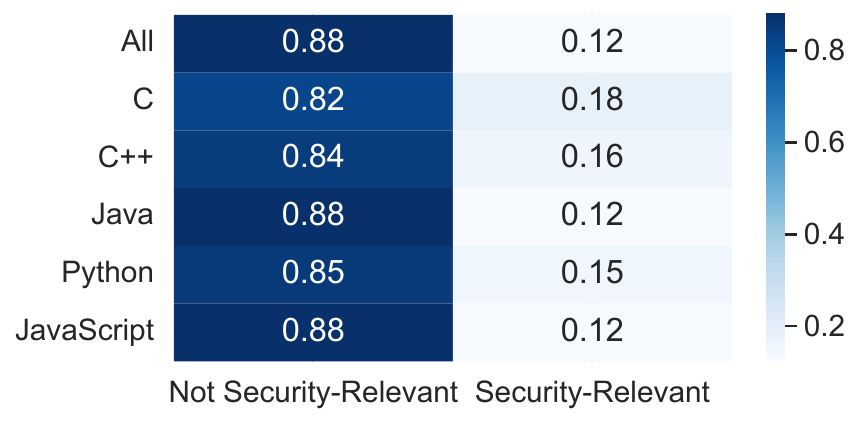}
    \caption{Normalized confusion matrix for \Cref{tab:appendix:commitmsg} excluding empty commit messages.}
    \label{fig:appendix:cm_nonempty}
\end{figure}

\begin{figure}[h]
\centering
\begin{subfigure}[b]{\linewidth}
    \includegraphics[width=1\linewidth]{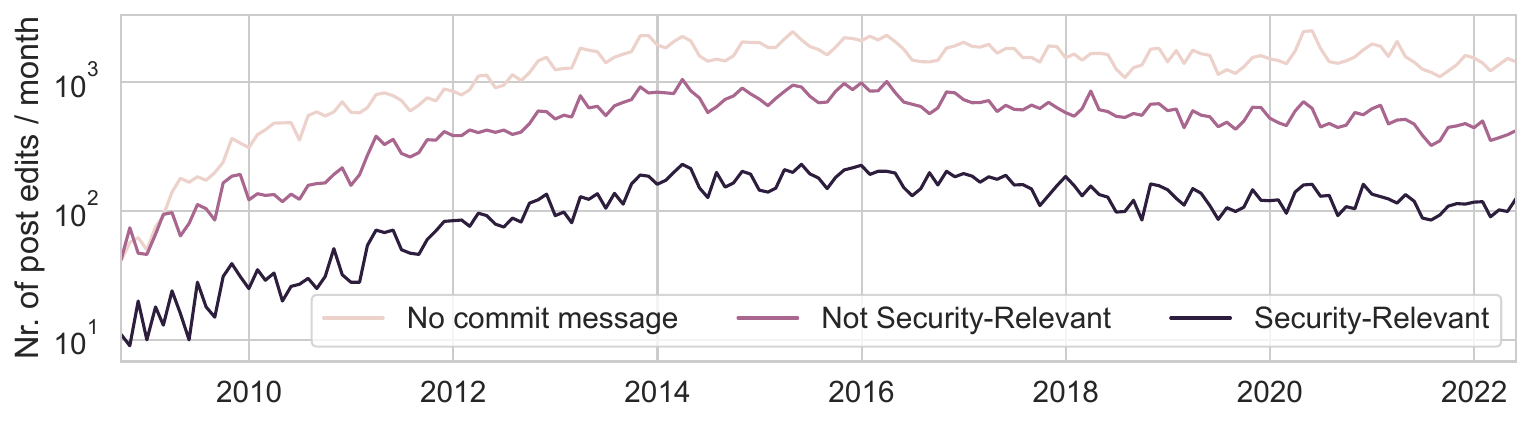}
    \caption{Nr.~of monthly post edits categorized by their security relevance.}
\end{subfigure}

\begin{subfigure}[b]{\linewidth}
   \includegraphics[width=1\linewidth]{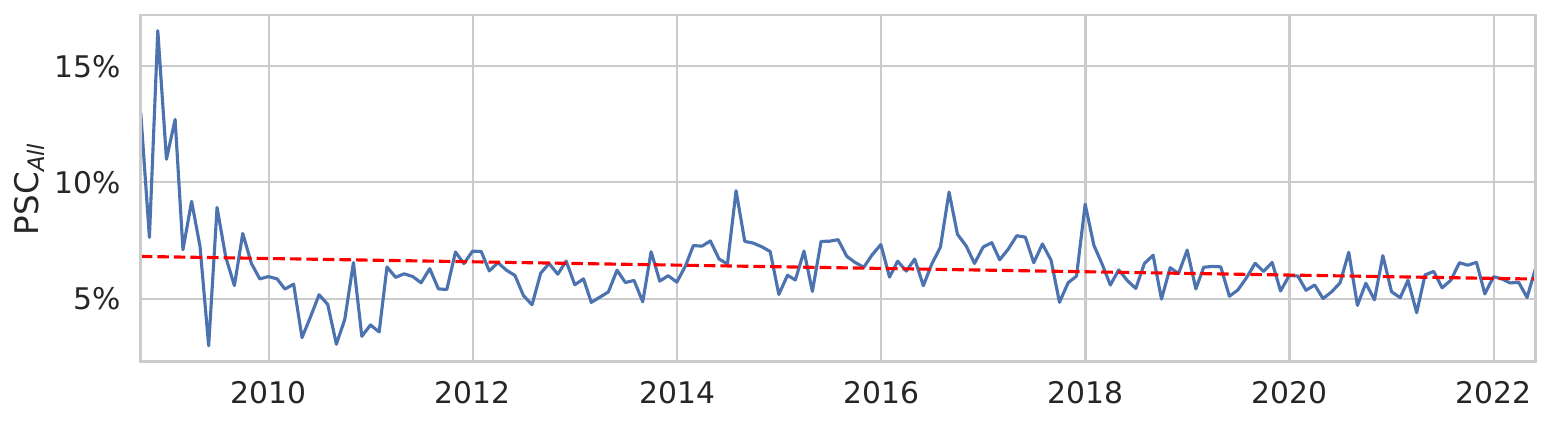}
   \caption{Including empty commit messages.}
\end{subfigure}

\begin{subfigure}[b]{\linewidth}
   \includegraphics[width=1\linewidth]{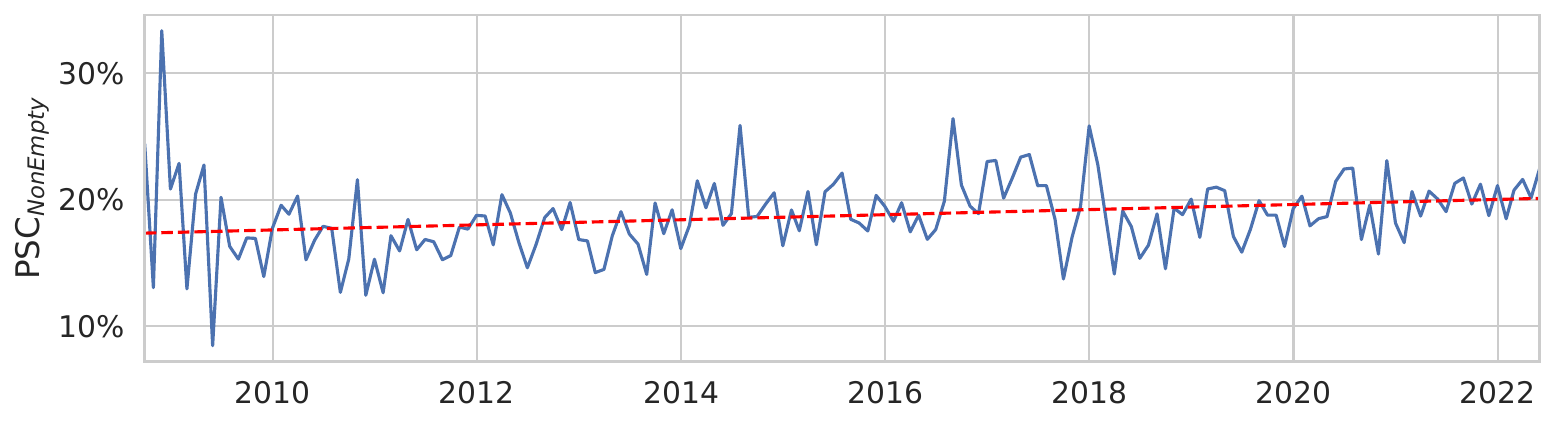}
   \caption{Excluding empty commit messages.}
\end{subfigure}

\caption{\textbf{C language:} PSC in monthly intervals. Dashed line is the fitted linear regression.}
\label{fig:appendix:pcs_c}
\end{figure}

\begin{figure}[h]
\centering
\begin{subfigure}[b]{\linewidth}
    \includegraphics[width=1\linewidth]{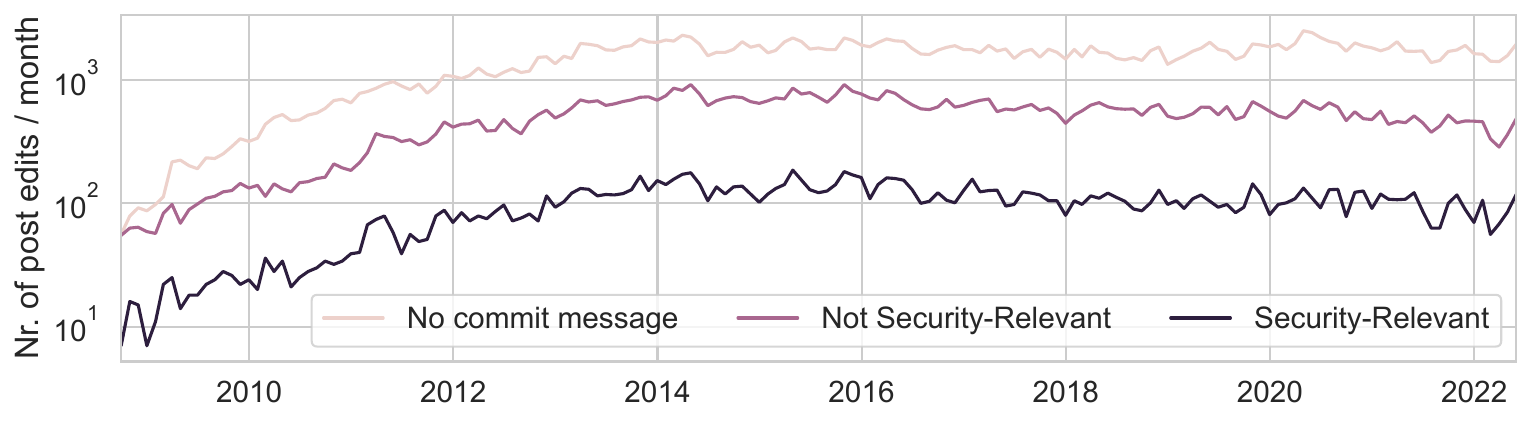}
    \caption{Nr.~of monthly post edits categorized by their security relevance.}
\end{subfigure}

\begin{subfigure}[b]{\linewidth}
   \includegraphics[width=1\linewidth]{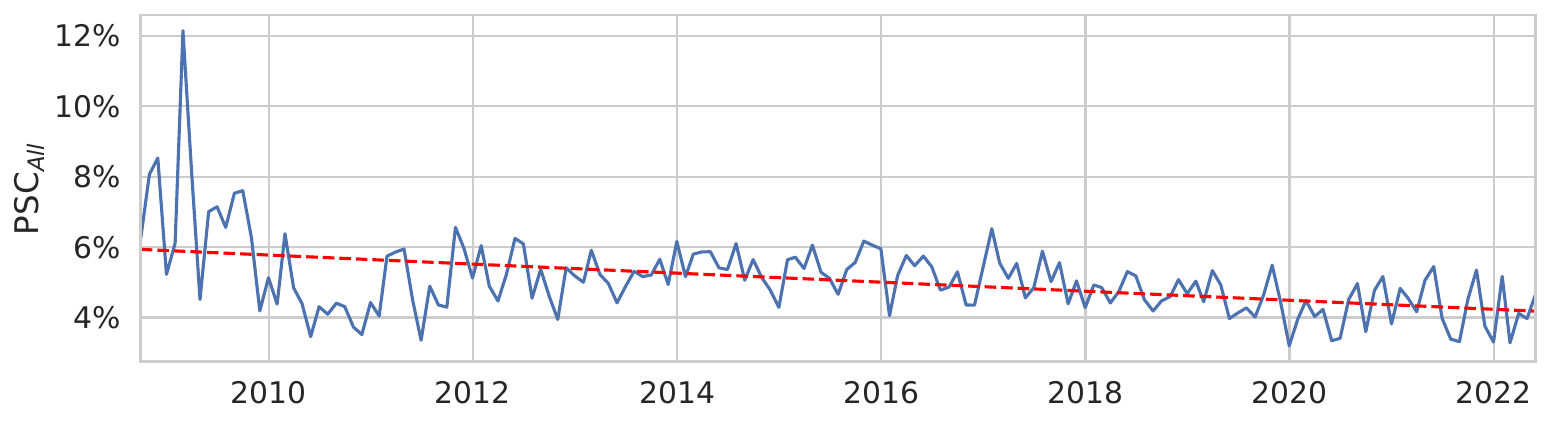}
   \caption{Including empty commit messages.}
\end{subfigure}

\begin{subfigure}[b]{\linewidth}
   \includegraphics[width=1\linewidth]{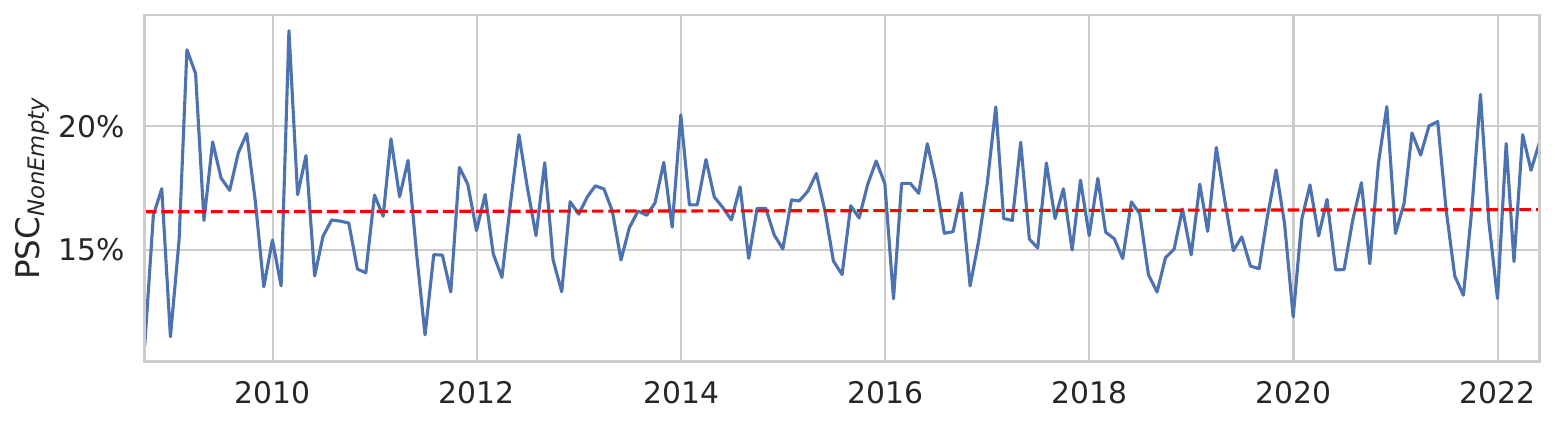}
   \caption{Excluding empty commit messages.}
\end{subfigure}

\caption{\textbf{C++ language:} PSC in monthly intervals. Dashed line is the fitted linear regression.}
\label{fig:appendix:pcs_cpp}
\end{figure}

\begin{figure}[h]
\centering
\begin{subfigure}[b]{\linewidth}
    \includegraphics[width=1\linewidth]{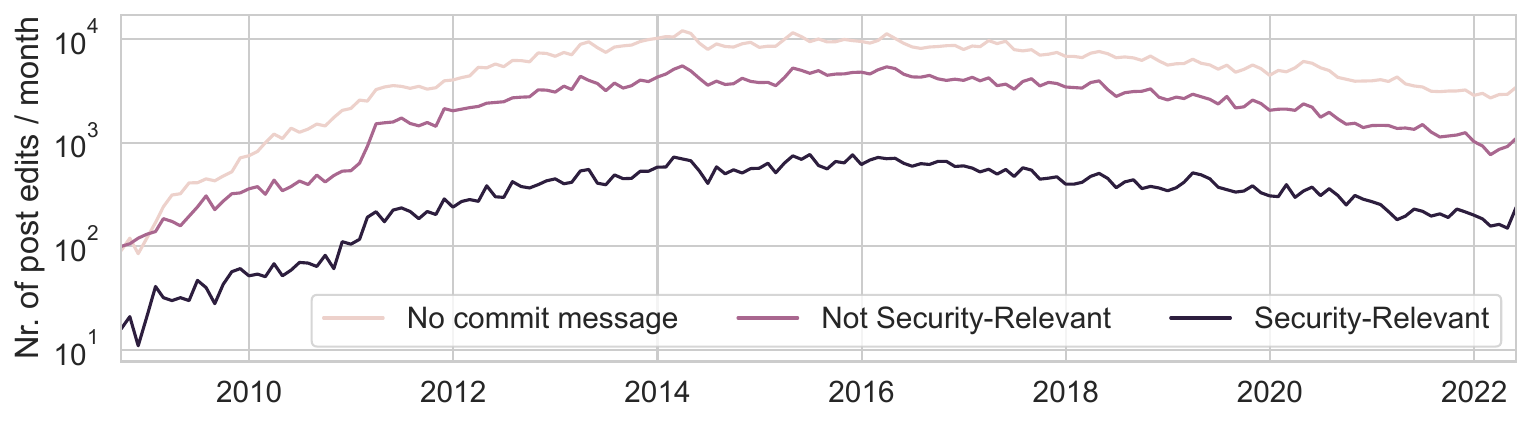}
    \caption{Nr.~of monthly post edits categorized by their security relevance.}
\end{subfigure}

\begin{subfigure}[b]{\linewidth}
   \includegraphics[width=1\linewidth]{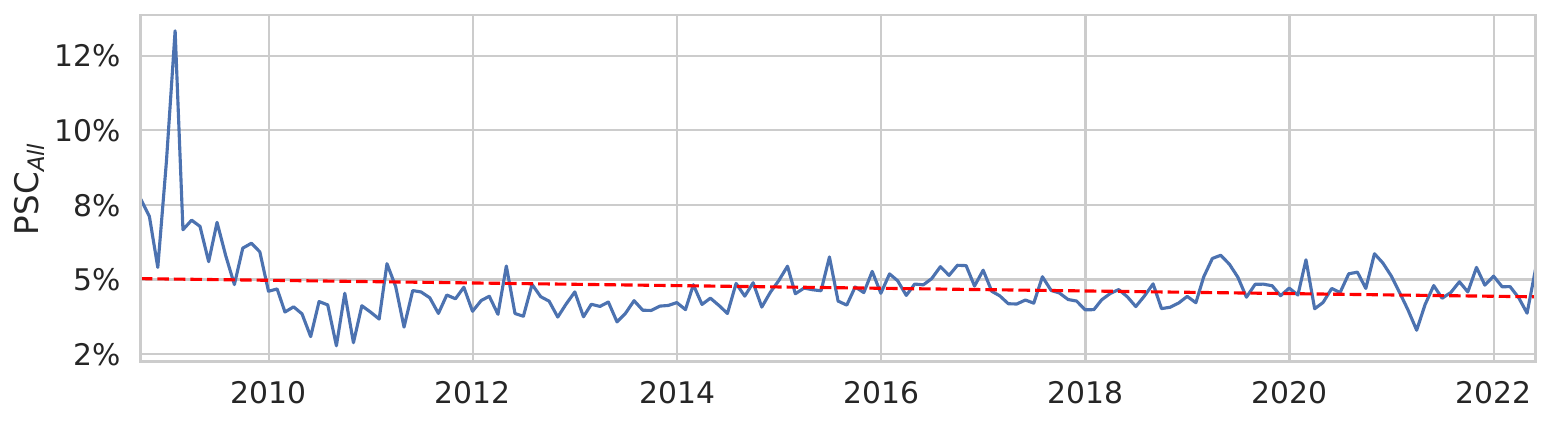}
   \caption{Including empty commit messages.}
\end{subfigure}

\begin{subfigure}[b]{\linewidth}
   \includegraphics[width=1\linewidth]{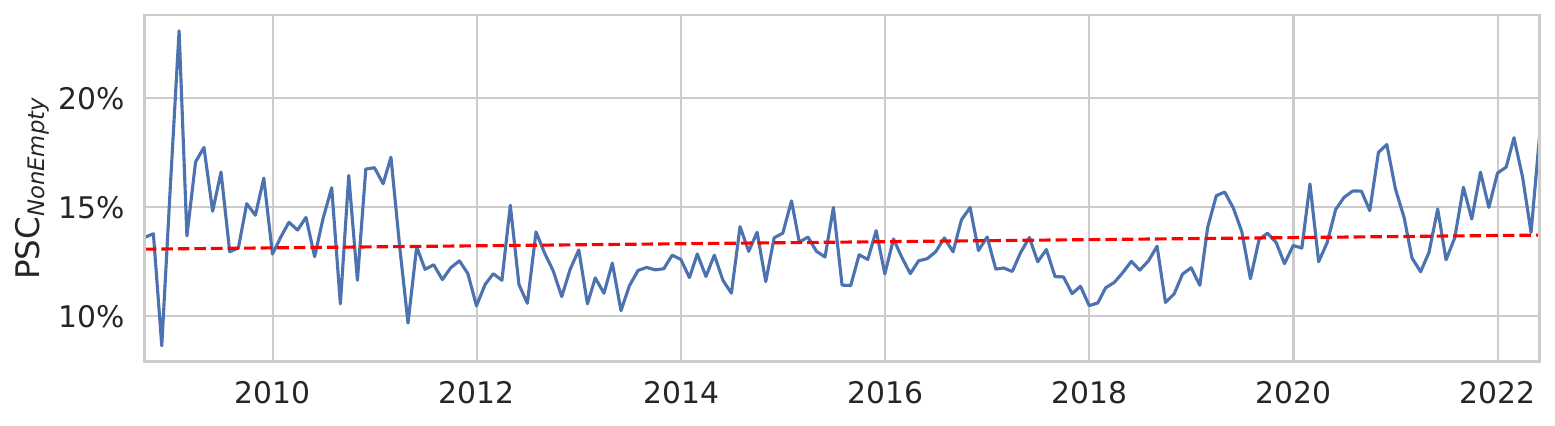}
   \caption{Excluding empty commit messages.}
\end{subfigure}

\caption{\textbf{Java language:} PSC in monthly intervals. Dashed line is the fitted linear regression.}
\label{fig:appendix:pcs_java}
\end{figure}

\begin{figure}[h]
\centering
\begin{subfigure}[b]{\linewidth}
    \includegraphics[width=1\linewidth]{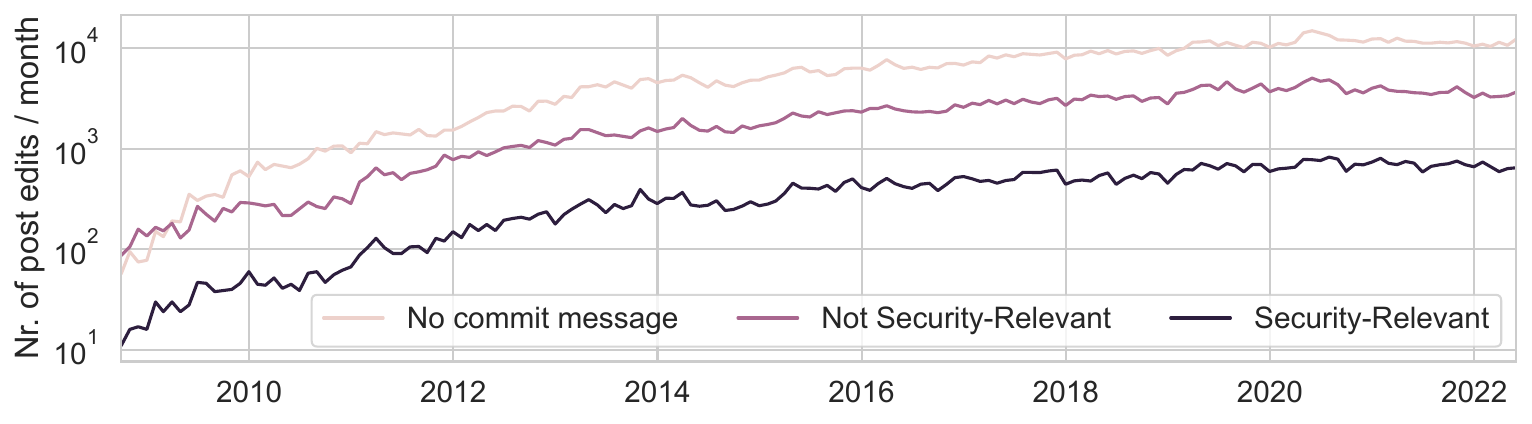}
    \caption{Nr.~of monthly post edits categorized by their security relevance.}
\end{subfigure}

\begin{subfigure}[b]{\linewidth}
   \includegraphics[width=1\linewidth]{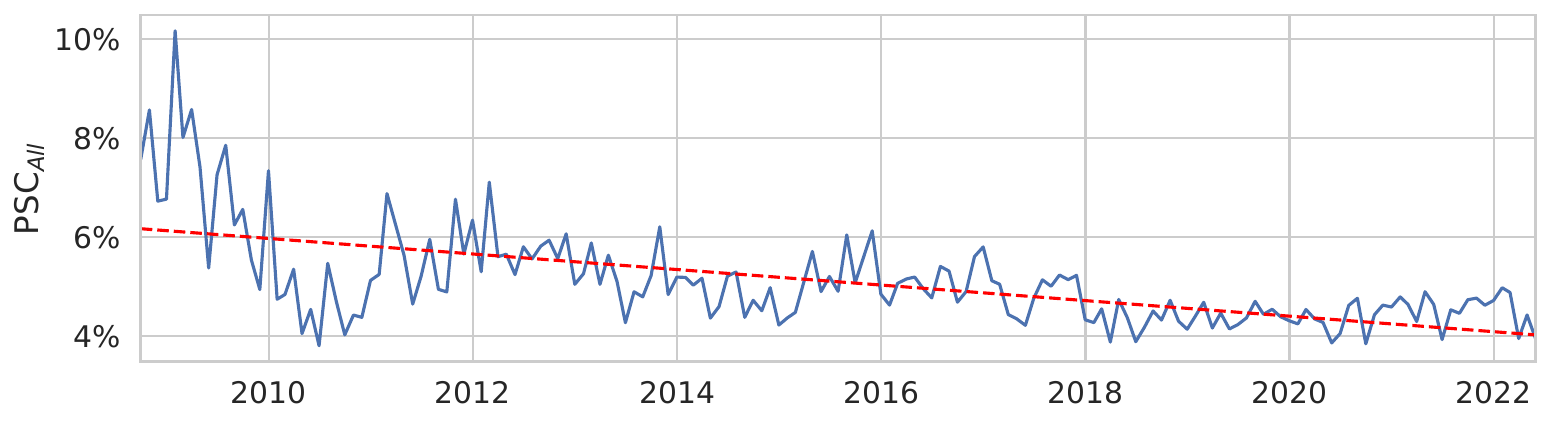}
   \caption{Including empty commit messages.}
\end{subfigure}

\begin{subfigure}[b]{\linewidth}
   \includegraphics[width=1\linewidth]{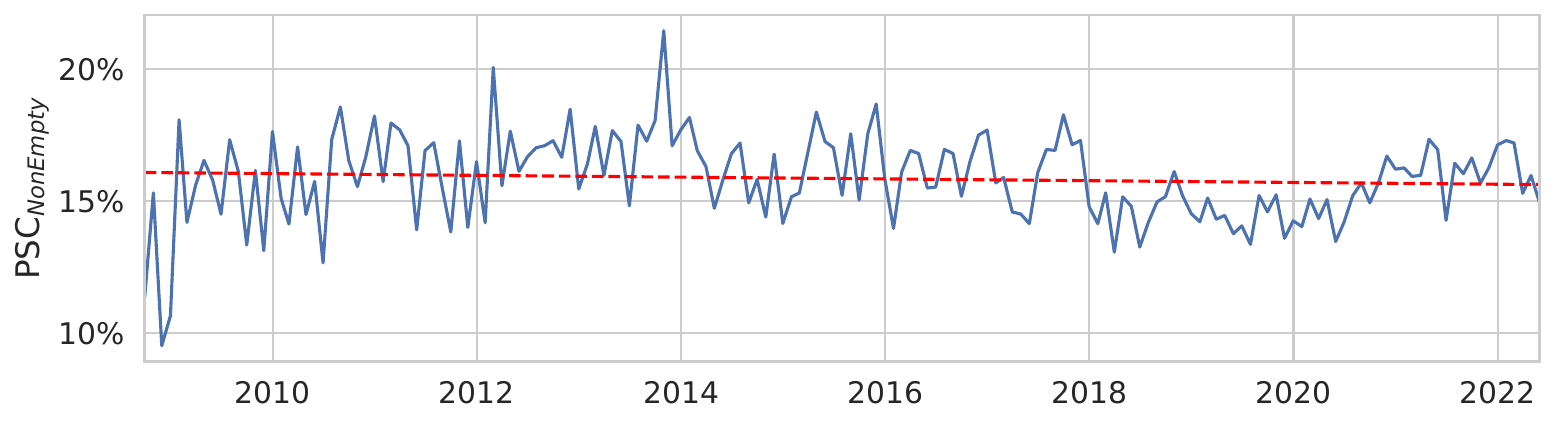}
   \caption{Excluding empty commit messages.}
\end{subfigure}

\caption{\textbf{Python language:} PSC in monthly intervals. Dashed line is the fitted linear regression.}
\label{fig:appendix:pcs_python}
\end{figure}

\begin{figure}[h]
\centering
\begin{subfigure}[b]{\linewidth}
    \includegraphics[width=1\linewidth]{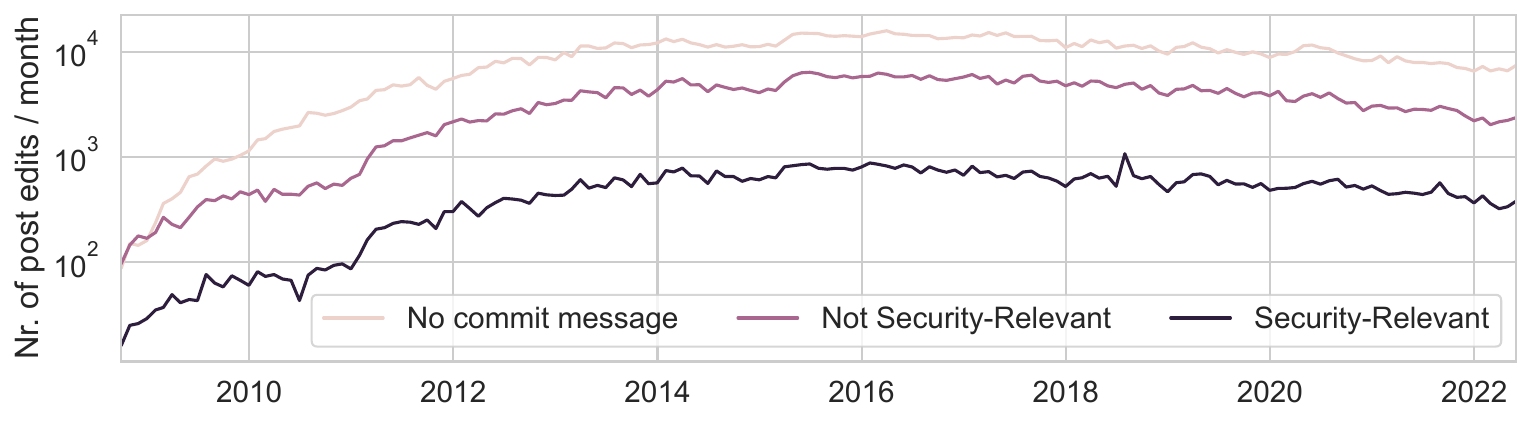}
    \caption{Nr.~of monthly post edits categorized by their security relevance.}
\end{subfigure}

\begin{subfigure}[b]{\linewidth}
   \includegraphics[width=1\linewidth]{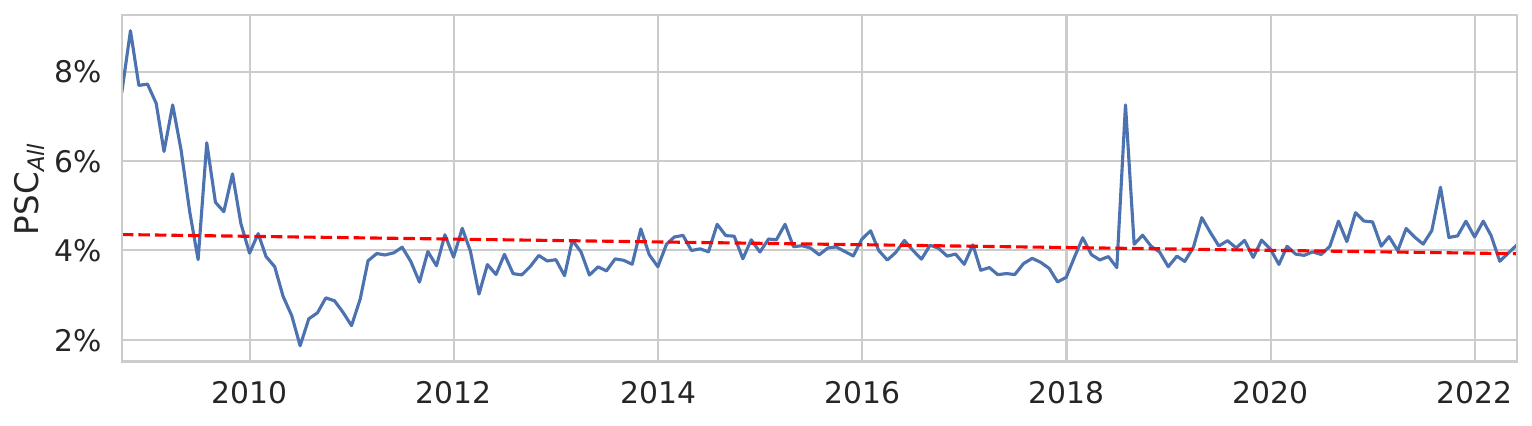}
   \caption{Including empty commit messages.}
\end{subfigure}

\begin{subfigure}[b]{\linewidth}
   \includegraphics[width=1\linewidth]{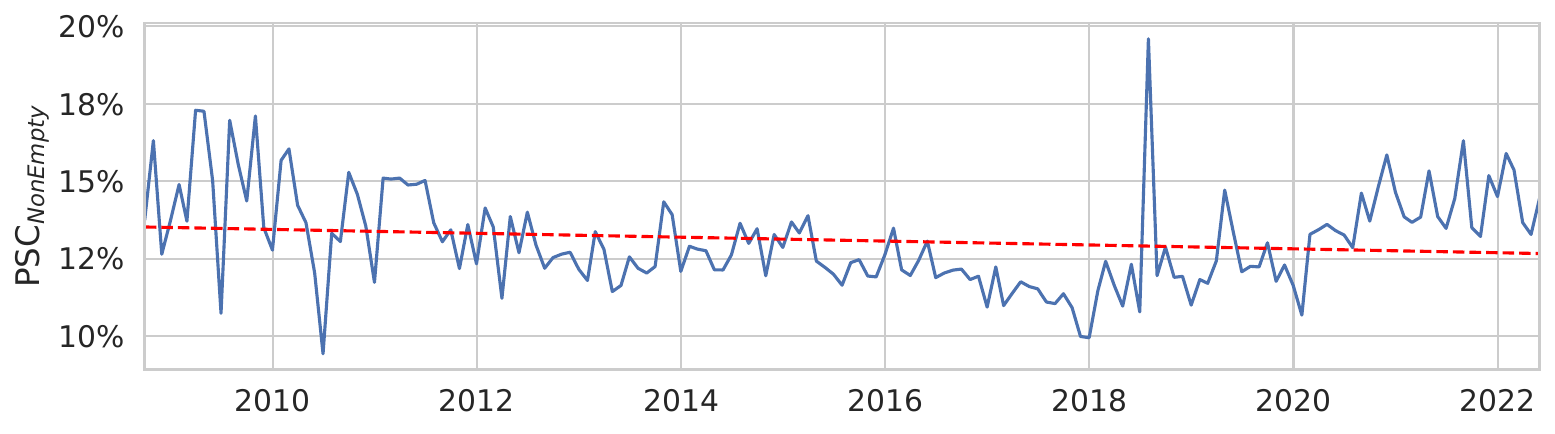}
   \caption{Excluding empty commit messages.}
\end{subfigure}

\caption{\textbf{JavaScript language:} PSC in monthly intervals. Dashed line is the fitted linear regression.}
\label{fig:appendix:pcs_js}
\end{figure}

\begin{figure}[h]
\centering
\begin{subfigure}[b]{\linewidth}
    \includegraphics[width=1\linewidth]{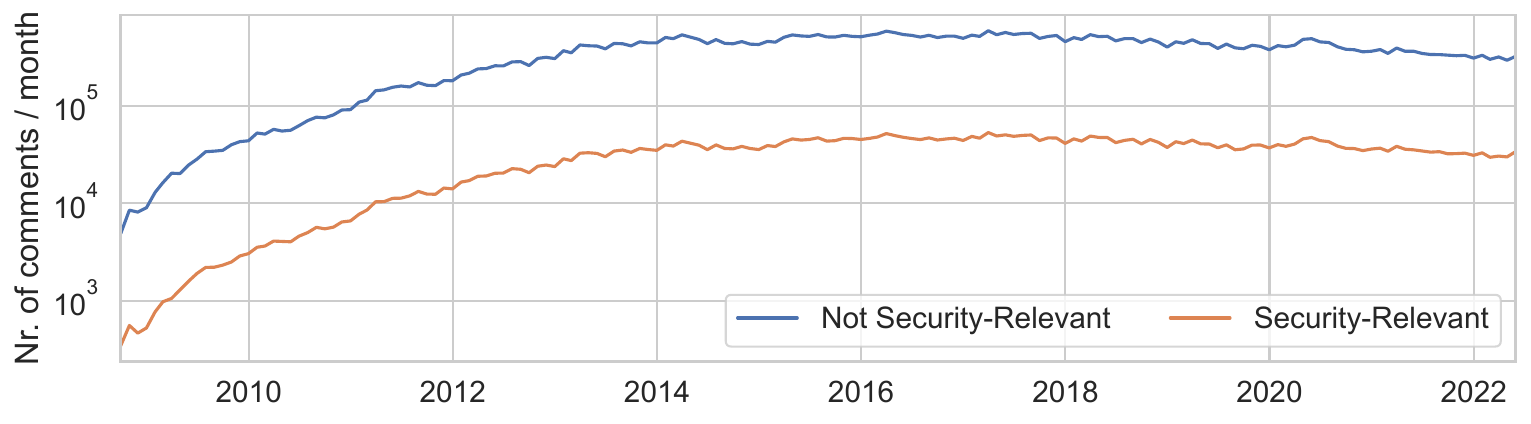}
    \caption{Nr.~of monthly comments categorized by their security relevance.}
\end{subfigure}

\begin{subfigure}[b]{\linewidth}
   \includegraphics[width=1\linewidth]{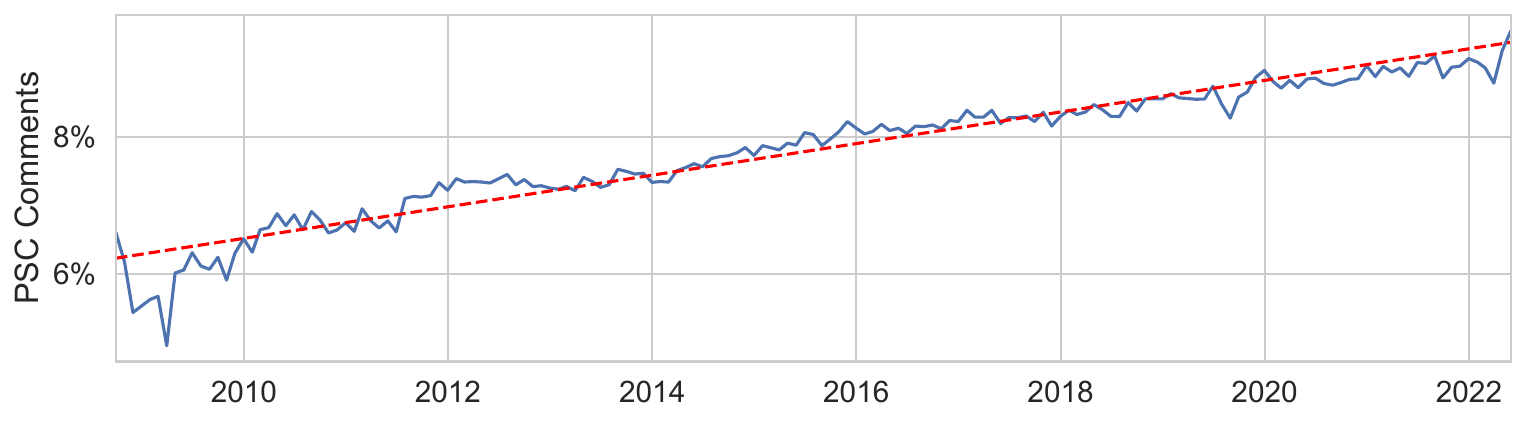}
   \caption{Percentage of new security-relevant comments per monthly interval. Dashed line is the fitted linear regression.}
\end{subfigure}

\caption{Security-relevant comments on Stack Overflow}
\label{fig:appendix:pcs_comments_all}
\end{figure}

\begin{figure}[h]
\centering
\begin{subfigure}[b]{\linewidth}
    \includegraphics[width=1\linewidth]{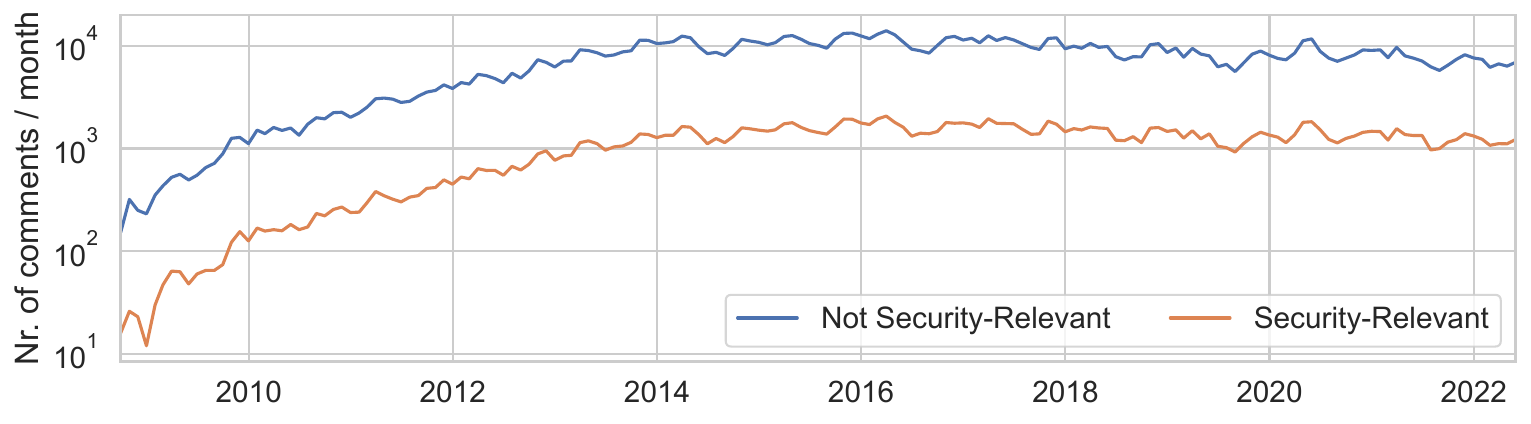}
    \caption{Nr.~of monthly comments categorized by their security relevance.}
\end{subfigure}

\begin{subfigure}[b]{\linewidth}
   \includegraphics[width=1\linewidth]{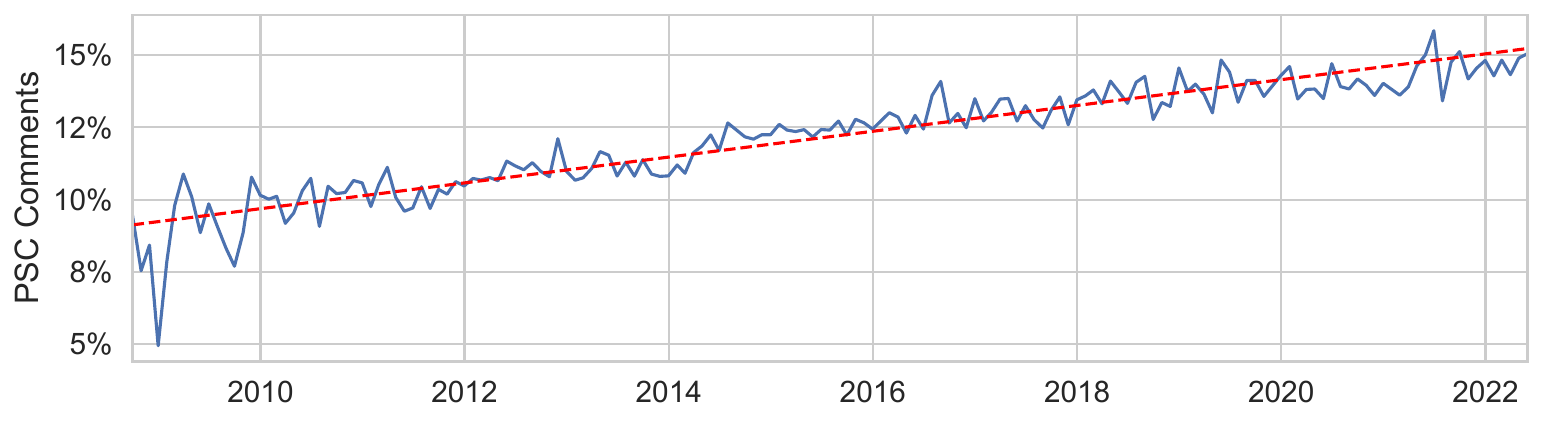}
   \caption{Percentage of new security-relevant comments per monthly interval. Dashed line is the fitted linear regression.}
\end{subfigure}

\caption{Security-relevant comments on Stack Overflow for posts with snippets in the \textbf{C language}.}
\label{fig:appendix:pcs_comments_c}
\end{figure}

\begin{figure}[h]
\centering
\begin{subfigure}[b]{\linewidth}
    \includegraphics[width=1\linewidth]{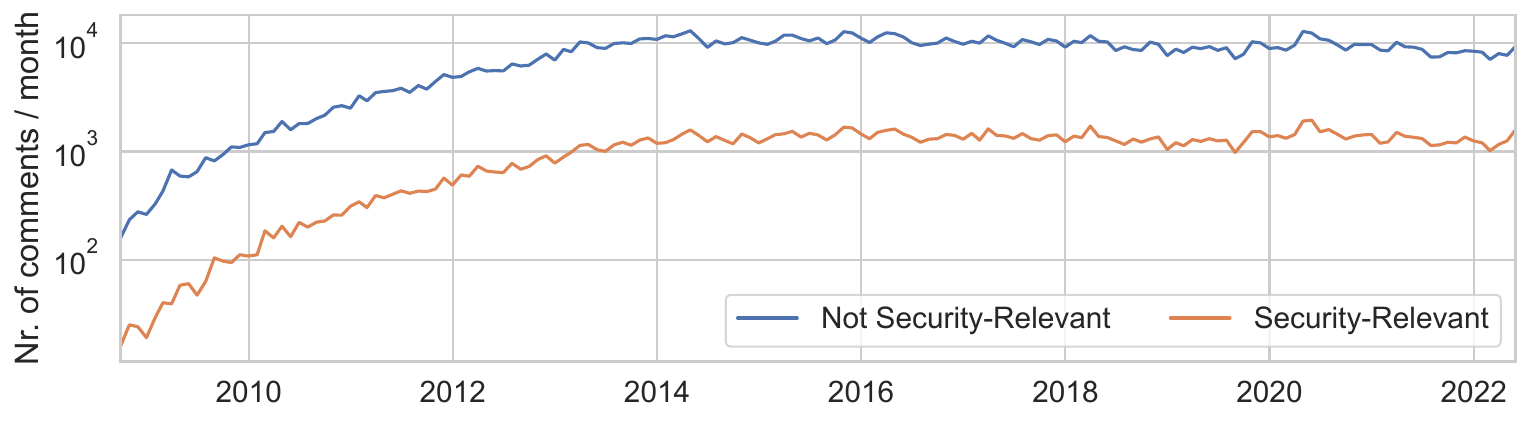}
    \caption{Nr.~of monthly comments categorized by their security relevance.}
\end{subfigure}

\begin{subfigure}[b]{\linewidth}
   \includegraphics[width=1\linewidth]{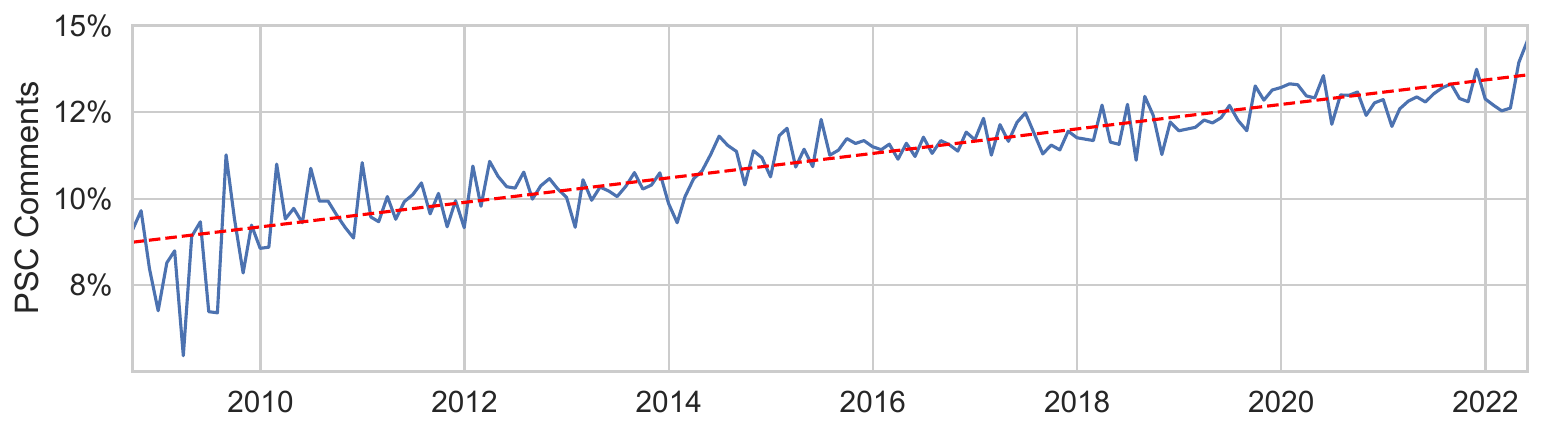}
   \caption{Percentage of new security-relevant comments per monthly interval. Dashed line is the fitted linear regression.}
\end{subfigure}

\caption{Security-relevant comments on Stack Overflow for posts with snippets in the \textbf{C++ language}.}
\label{fig:appendix:pcs_comments_cpp}
\end{figure}

\begin{figure}[h]
\centering
\begin{subfigure}[b]{\linewidth}
    \includegraphics[width=1\linewidth]{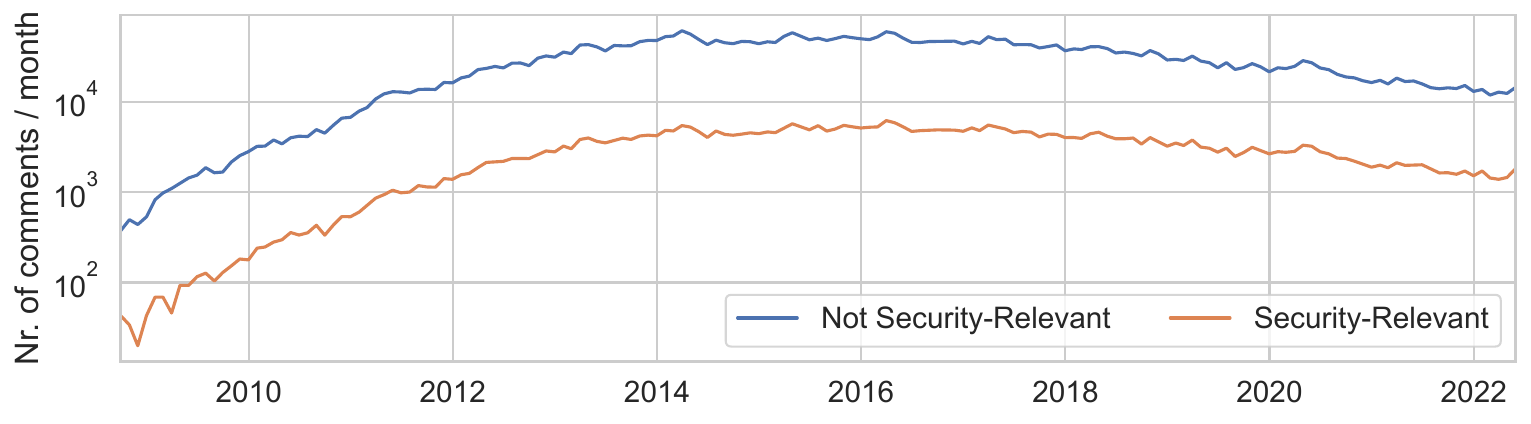}
    \caption{Nr.~of monthly comments categorized by their security relevance.}
\end{subfigure}

\begin{subfigure}[b]{\linewidth}
   \includegraphics[width=1\linewidth]{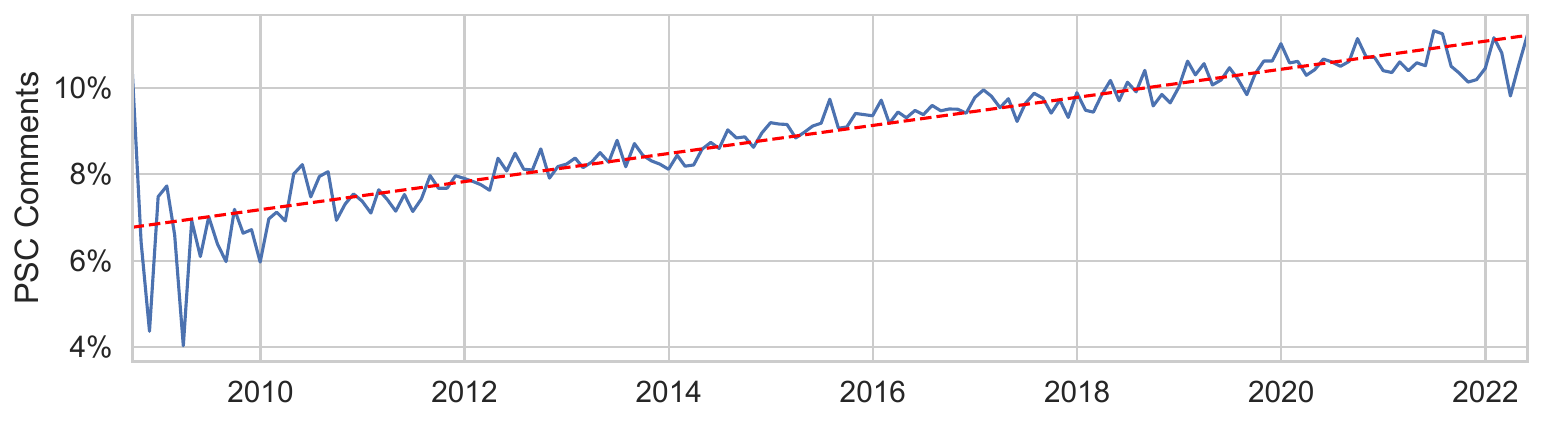}
   \caption{Percentage of new security-relevant comments per monthly interval. Dashed line is the fitted linear regression.}
\end{subfigure}

\caption{Security-relevant comments on Stack Overflow for posts with snippets in the \textbf{Java language}.}
\label{fig:appendix:pcs_comments_java}
\end{figure}

\begin{figure}[h]
\centering
\begin{subfigure}[b]{\linewidth}
    \includegraphics[width=1\linewidth]{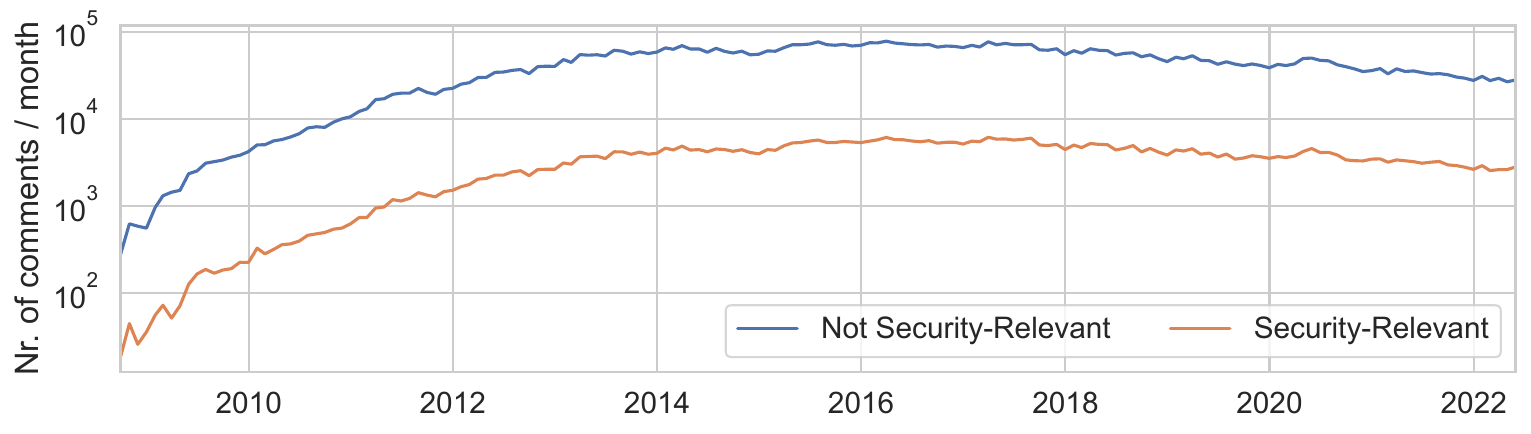}
    \caption{Nr.~of monthly comments categorized by their security relevance.}
\end{subfigure}

\begin{subfigure}[b]{\linewidth}
   \includegraphics[width=1\linewidth]{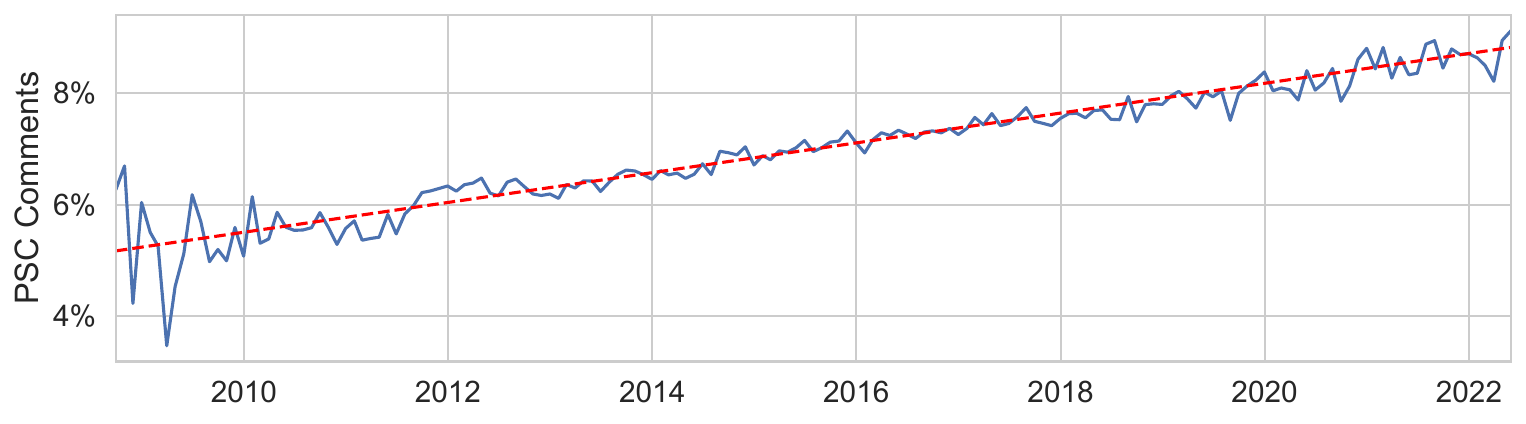}
   \caption{Percentage of new security-relevant comments per monthly interval. Dashed line is the fitted linear regression.}
\end{subfigure}

\caption{Security-relevant comments on Stack Overflow for posts with snippets in the \textbf{JavaScript language}.}
\label{fig:appendix:pcs_comments_javascript}
\end{figure}

\begin{figure}[h]
\centering
\begin{subfigure}[b]{\linewidth}
    \includegraphics[width=1\linewidth]{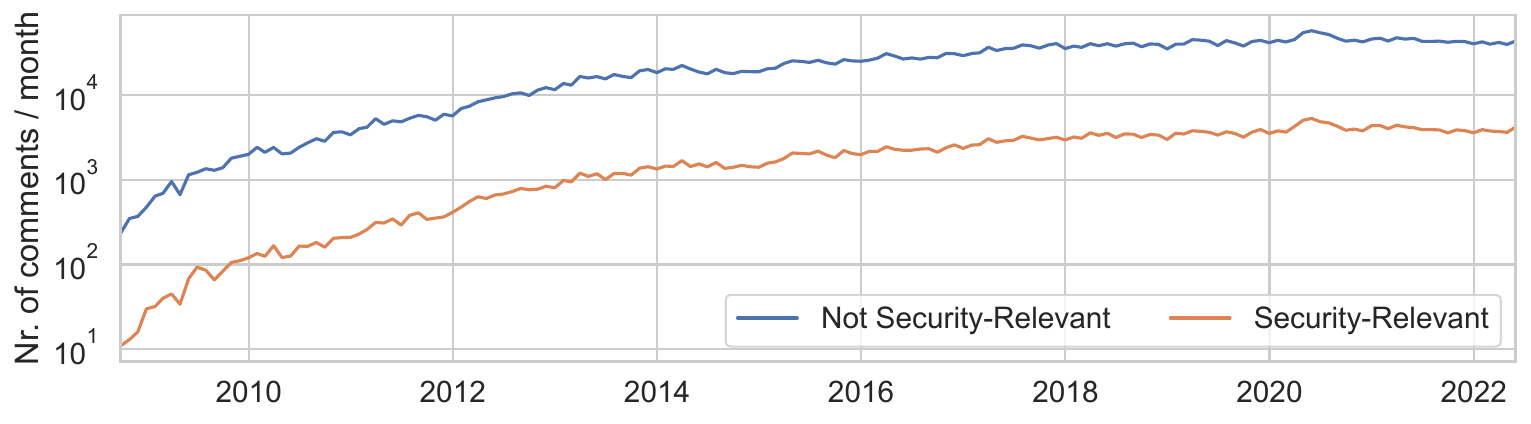}
    \caption{Nr.~of monthly comments categorized by their security relevance.}
\end{subfigure}

\begin{subfigure}[b]{\linewidth}
   \includegraphics[width=1\linewidth]{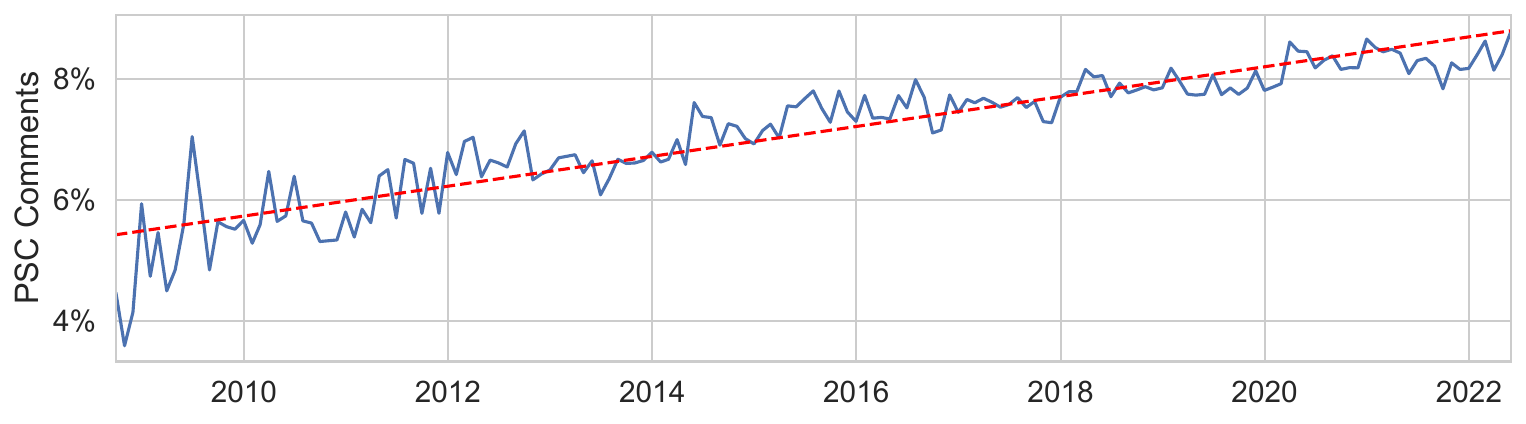}
   \caption{Percentage of new security-relevant comments per monthly interval. Dashed line is the fitted linear regression.}
\end{subfigure}

\caption{Security-relevant comments on Stack Overflow for posts with snippets in the \textbf{Python language}.}
\label{fig:appendix:pcs_comments_python}
\end{figure}

\Cref{tab:appendix:editslang} provides an overview of the edits per code snippet on Stack Overflow, broken down by the snippets' programming languages.
\Cref{tab:appendix:commitmsg} lists the number of commit messages by programming language and by their type (i.e., no commit message, not security-relevant, security-relevant).
\Cref{fig:appendix:cm_all} and \ref{fig:appendix:cm_nonempty} depict the normalized confusion matrices for Table~\ref{tab:appendix:commitmsg}.
A $\chi^2_{\textrm{All}}(10, N=19,311,100)=16288$, $p_{\textrm{All}} < 0.001$, Cramér's $V_{\textrm{All}}=0.021$ for all commit messages and $\chi^2_{\textrm{NonEmpty}}(5, N=6,134,834)=6624$, $p_{\textrm{NonEmpty}} < 0.001$, Cramér's $V_{\textrm{NonEmpty}}=0.023$ shows a significant relationship between programming languages and type of commit message.
\Cref{tab:appendix:languagestats} shows the mean monthly PSC and a fitted linear regression per programming language.
Figures~\ref{fig:appendix:pcs_c} to \ref{fig:appendix:pcs_js} illustrate the corresponding distributions of commit messages and PSC per language.
\Cref{tab:appendix:kpssadf} shows the Kwiatkowski–Phillips–Schmidt–Shin (KPSS) and Augmented Dickey-Fuller (ADF) test statistics for the PSC with and without empty commit messages.
\Cref{tab:appendix:psc_comments} shows the mean monthly percentage of newly added security-relevant comments and the fitted linear regression on this percentage, broken down by the programming language of the code snippets in the commented post.
\Cref{tab:appendix:kpssadf_comments} presents the KPSS and ADF test statistics for the time series of this monthly percentage.
\Cref{fig:appendix:pcs_comments_all} to \ref{fig:appendix:pcs_comments_python} depicts the time series data for comments.

\begin{table}[h]
    \caption{Edits per code snippet in different languages.}
    \label{tab:appendix:editslang}
    \centering
    \footnotesize
    \rowcolors{2}{gray!15}{white}
    \begin{tabular}{lrS[table-format=1.3+-1.3]rrrr}
    \toprule
    Language & Count & \multicolumn{1}{c}{Mean} & Min & 75\% & 99\% & Max \\
    \midrule
    C & 82,456 & 1.474(0.007) & 1 & 2 & 5 & 36 \\
    C++ & 90,264 & 1.480(0.007) & 1 & 2 & 5 & 34 \\
    Java & 359,257 & 1.307(0.002) & 1 & 1 & 4 & 27 \\
    JavaScript & 700,592 & 1.311(0.002) & 1 & 1 & 4 & 53 \\
    Python & 455,606 & 1.410(0.003) & 1 & 2 & 5 & 58 \\
    \bottomrule
    \end{tabular}

\end{table}

\begin{table}[h]
    \caption{Nr.~of commit messages by language and type.}
    \label{tab:appendix:commitmsg}
    \centering
    \scriptsize
    \rowcolors{2}{gray!15}{white}
    \begin{tabular}{lrrr}
        \toprule
        Language & Empty Message & Not Security-Relevant & Security-Relevant \\
        \midrule
        All & 12,492,209 & 4,875,276 & 688,157 \\
        C & 222,398 & 85,173 & 19,133 \\
        C++ & 239,386 & 81,796 & 15,663 \\
        Java & 931,999 & 419,725 & 59,261 \\
        Python & 1,004,514 & 352,041 & 62,442 \\
        JavaScript & 1,463,354 & 558,779 & 78,482 \\
        \bottomrule
    \end{tabular}

\end{table}

\begin{table}[h]
    \caption{PSC broken down by language. Monthly average PSC (with $CI=95\%$) plus $R^2$ and significance for a fitted linear regression.}
    \label{tab:appendix:languagestats}
    \centering
    \scriptsize
    \rowcolors{2}{gray!15}{white}
    \begin{tabular}{l|S[table-format=1.3+-1.3]S[table-format=1.3, table-space-text-post ={****}]|S[table-format=2.3+-1.3]S[table-format=1.3, table-space-text-post ={****}]}
\toprule
 & \multicolumn{2}{c|}{\textbf{All Commit Messages}} & \multicolumn{2}{c}{\textbf{Non-Empty Commit Messages}}\\\midrule
 Language & \multicolumn{1}{c}{{Mean PSC}} & {R$^2$} & \multicolumn{1}{c}{{Mean PSC}} & {R$^2$}\\
\midrule
\textbf{All} & 4.136(0.107) & 0.128*** & 12.863(0.167) & 0.002 \\
\textbf{C} & 6.331(0.242) & 0.031* & 18.699(0.457) & 0.069*** \\
\textbf{C++} & 5.069(0.169) & 0.207*** & 16.57(0.325) & 0.000 \\
\textbf{Java} & 4.729(0.163) & 0.027* & 13.372(0.309) & 0.008 \\
\textbf{Python} & 5.118(0.152) & 0.382*** & 15.852(0.249) & 0.006 \\
\textbf{JavaScript} & 4.144(0.147) & 0.016 & 13.1(0.24) & 0.024* \\
\bottomrule
\rowcolor{white}\multicolumn{5}{c}{* $p<0.05$, ** $p<0.01$, ***$p<0.001$}
\end{tabular}
\end{table}

\begin{table*}[h]
    \caption{Kwiatkowski–Phillips–Schmidt–Shin (KPSS) and Augmented Dickey-Fuller (ADF) test statistics for PSC.}
    \label{tab:appendix:kpssadf}
    \centering
    \small
    \rowcolors{2}{gray!15}{white}
    \begin{tabular}{l|S[table-format=1.3, table-space-text-post ={***}]S[table-format=-1.3, table-space-text-post ={***}]l|S[table-format=-1.3, table-space-text-post ={***}]S[table-format=-2.3, table-space-text-post ={***}]l}
    \toprule
    & \multicolumn{3}{c|}{\textbf{All Commit Messages}} & \multicolumn{3}{c}{\textbf{Non-Empty Commit Messages}}\\\midrule
     & {KPSS} & {ADF} & Stationary & {KPSS} & {ADF} & Stationary \\
    \midrule
    \textbf{All} & 0.417 & -4.848*** & Stationary & 0.386 & -2.004 & \cellcolor{red!15}Trend Stationary \\
    \textbf{C} & 0.192 & -5.918*** & Stationary & 0.780* & -4.508*** & \cellcolor{red!15}Difference Stationary \\
    \textbf{C++} & 0.876* & -4.409*** & \cellcolor{red!15}Difference Stationary & 0.074 & -11.060*** & Stationary \\
    \textbf{Java} & 0.211 & -6.152*** & Stationary & 0.356 & -3.525** & Stationary \\
    \textbf{Python} & 1.139* & -4.130*** & \cellcolor{red!15}Difference Stationary & 0.306 & -5.173*** & Stationary \\
    \textbf{JavaScript} & 0.171 & -5.367*** & Stationary & 0.488* & -2.122 & \cellcolor{red!15}Non-Stationary \\
    \bottomrule
    \rowcolor{white}\multicolumn{7}{c}{* $p<0.05$, ** $p<0.01$, ***$p<0.001$}
    \end{tabular}
\end{table*}

\begin{table}[h]
    \caption{Monthly average percentage of created security-relevant comments (with $CI=95\%$) as well as $R^2$ and significance for a fitted linear time series regression .}
    \label{tab:appendix:psc_comments}
    \centering
    \small
    \rowcolors{2}{gray!15}{white}
    \begin{tabular}{lS[table-format=1.3+-1.3]S[table-format=1.3, table-space-text-post ={***}]}
    \toprule
     & {Mean PSC} & {R$^2$} \\
    \midrule
    \textbf{All} & 7.781(0.143) & 0.935*** \\
    \textbf{C} & 12.134(0.286) & 0.874*** \\
    \textbf{C++} & 11.125(0.237) & 0.799*** \\
    \textbf{Java} & 8.959(0.212) & 0.844*** \\
    \textbf{Python} & 7.091(0.161) & 0.846*** \\
    \textbf{JavaScript} & 6.971(0.167) & 0.917*** \\
    \bottomrule
    \rowcolor{white}\multicolumn{3}{c}{* $p<0.05$, ** $p<0.01$, ***$p<0.001$}
    \end{tabular}

\end{table}

\begin{table}[h]
    \caption{Kwiatkowski–Phillips–Schmidt–Shin (KPSS) and Augmented Dickey–Fuller (ADF) test statistics for the percentage of security-relevant comments.}
    \label{tab:appendix:kpssadf_comments}
    \centering
    \small
    \rowcolors{2}{gray!15}{white}
    \begin{tabular}{lS[table-format=1.3, table-space-text-post ={***}]S[table-format=1.3, table-space-text-post ={***}]l}
    \toprule
     & {KPSS} & {ADF} & {Stationary} \\
    \midrule
    \textbf{All} & 1.851* & -3.778** & Difference Stationary \\
    \textbf{C} & 1.904* & -1.341 & Non-Stationary \\
    \textbf{C++} & 1.831* & -1.467 & Non-Stationary \\
    \textbf{Java} & 1.845* & -2.177 & Non-Stationary \\
    \textbf{Python} & 1.858* & -2.781 & Non-Stationary \\
    \textbf{JavaScript} & 1.882* & -0.323 & Non-Stationary \\
    \bottomrule
    \rowcolor{white}\multicolumn{4}{c}{* $p<0.05$, ** $p<0.01$, ***$p<0.001$}
    \end{tabular}
\end{table}

\clearpage
\section{Case Study 1: C/C++ Code Weaknesses}\label[secinapp]{sec:appendix:casestudy_1}
\begingroup
    \sisetup{detect-weight,
            output-decimal-marker={,},
            group-minimum-digits=2,
            group-separator={.}
            }
\begin{table*}[t]
    \caption{Comparison of \hlblue{results by Zhang et al.}
    (cf.~Table~1 in \cite{zhang_code_weaknesses}) and our \hlgreen{evaluation} using Cppcheck v1.86 for both language and security weakness detection.}
    \label{tab:appendix_zhang_evaluation}
    \centering
    \small
    \begin{tabular}{lccc|lccc}\toprule
    \multicolumn{4}{c|}{\textbf{SOTorrent18 (\hlblue{Original})}} & \multicolumn{4}{c}{\textbf{SOTorrent18 (\hlgreen{Evaluation})}}\\
    \midrule
     & \textit{Answer \#} & \textit{Code Snippet \#} & \textit{Code Version \#} &  & \textit{Answer \#} & \textit{Code Snippet \#} & \textit{Code Version \#}\\\midrule
    SOTorrent & 867,734 & 1,561,550 & 1,833,449 & SOTorrent &  867,734 & 1,561,550 & 1,833,449 \\
    \rowcolor{gray!15}LOC $>=$ 5 & 527,932 & 724,784 & 919,947 & LOC $>=$ 5 &  527,932 & 724,784 & 919,947 \\
    Guesslang & 490,778 & 646,716 & 826,520 & Cppcheck $v1.86$ & 141,215 & 170,974 & 206,582 \\
    \rowcolor{gray!15}$Code_w$ & \cellcolor{blue!25} \textbf{11,235} & \cellcolor{blue!25} \textbf{11,748} & \cellcolor{blue!25}
    \textbf{14,934} & 
    $Code_w$ & \cellcolor{green!25} \textbf{15,724} & \cellcolor{green!25} \textbf{16,533} & \cellcolor{green!25} \textbf{20,664} \\
    \bottomrule
    \end{tabular}
\end{table*}
\endgroup

\Cref{tab:appendix_zhang_evaluation} presents a detailed, side-by-side comparison of the results from the original methodology by Zhang et al.~\cite{zhang_code_weaknesses} and our re-implementation, which utilized Cppcheck 1.86 for both language and security weakness detection (see \Cref{fig:zhang:methodology_flowchart} for details).
We identified 15,724 C/C++ answers containing security weaknesses.
Of these, 11,142 answers were also flagged as containing weaknesses in the original study's $Answer_w = 11,235$.
This indicates that our approach missed only 93 answers with weaknesses from the original study, demonstrating that our approach captures the same snippets as the original methodology. 
We surmise that the increase in detected snippets and snippet versions in our methodology is rooted in Guesslang's performance in the original methodology, i.e., many C/C++ snippets were misclassified as another language and hence not analyzed with Cppcheck by the authors.
However, since the exact Cppcheck version used by the authors is unknown and we empirically estimated it to be version 1.86, we cannot exclude that our guessed version differs and performs differently from the authors' version.

\begingroup
    \sisetup{detect-weight,
            output-decimal-marker={,},
            group-minimum-digits=2,
            group-separator={.}
            }

\begin{table*}[t]
    \caption{The proportion of $Code_w$ versus the number of code revisions by \hlblue{Zhang et al.}~\cite{zhang_code_weaknesses} and our \hlgreen{replication study} using SOTorrent22 and Cppcheck v2.13. (cf.~Table~2 in \cite{zhang_code_weaknesses})}
    \label{tab:zhang_proportion_comparison}
    \centering
    \small
    \setlength\tabcolsep{4pt}
    \begin{tabular}{|c|rrrr@{}rr@{\hspace{1.0em}}rrrr}\toprule
      & \multicolumn{4}{c}{\textbf{Authors Results based on SOTorrent18}} &&& \multicolumn{4}{c}{\textbf{Results based on SOTorrent22}}\\
    \cmidrule(){2-5}\cmidrule(){8-11}
    \textbf{\#revisions} & \textit{Snippets} & \textit{Unchanged} & \textit{Improved} & \textit{Deteriorated} &&& \textit{Snippets} & \textit{Unchanged} & \textit{Improved} & \textit{Deteriorated} \smallskip \\
    0 & 8,103 & NA & NA & NA                                 &&& 24,388 & NA & NA & NA \smallskip \\
    $\geq 1$ & 3,645 & 1,886 (51.7\%) & 1218 (33.4\%) & 541 (14.8\%)           &&& 5,866 & 5,511 (93.9\%) & 221 (3.8\%) & 134 (2.3\%) \smallskip \\
    1 & 2,369 & 1,340 (56.6\%) & \cellcolor{blue!25}\textbf{714 (30.1\%)} & 315 (13.3\%) &&& 4,391 & 4,179 (95.2\%) & \cellcolor{green!25}\textbf{136 (3.1\%)} & 76 (1.7\%) \smallskip \\
    2 & 774 & 349 (45.1\%) & \cellcolor{blue!25}\textbf{294 (38.0\%)} & 131 (16.9\%)                       &&& 1,058 & 969 (91.6\%) & \cellcolor{green!25}\textbf{54 (5.1\%)} & 35 (3.3\%) \smallskip \\
    $ \geq 3$ & 502 & 197 (39.2\%) & \cellcolor{blue!25}\textbf{210 (41.8\%)} & 95 (18.9\%)      &&& 417 & 363 (87.1\%) & \cellcolor{green!25}\textbf{31 (7.4\%)} & 23 (5.5\%) \smallskip \\
    \bottomrule
    \end{tabular}

\end{table*}
\endgroup

\Cref{tab:zhang_proportion_comparison} compares the proportion of $Code_w$ versus the number of code revisions that were detected with the original methodology by Zhang et al.~\cite{zhang_code_weaknesses} for SOTorrent18 and our re-implementation based on SOTorrent22.

\begingroup
    \sisetup{detect-weight,
            output-decimal-marker={,},
            group-minimum-digits=2,
            group-separator={.}
            }
\begin{table*}[t]
    \caption{The results of Cppcheck v1.86 and Cppcheck v2.13 on different versions of the SOTorrent dataset.}
    \label{tab:appendix:code_snippets_dataset_distribution}
    \centering
    \small
    \begin{tabular}{l|ccc|ccc}\toprule
     & \multicolumn{3}{c|}{\textbf{SOTorrent18}} & \multicolumn{3}{c}{\textbf{SOTorrent22}}\\
    \midrule
     & \textit{Answer \#} & \textit{Code Snippet \#} & \textit{Code Version \#} & \textit{Answer \#} & \textit{Code Snippet \#} & \textit{Code Version \#}\\\midrule
    \textbf{Cppcheck $v1.86$} &  15,724 &  16,533 &  20,664 &  19,485 &  20,450 &  25,832 \\
    \textbf{Cppcheck $v2.13$} &  23,253 &  24,699 &  30,923 &  28,521 &  30,254 &  38,248 \\
    \bottomrule
    \end{tabular}
    \label{tab:appendix:zhang}
\end{table*}
\endgroup

\Cref{tab:appendix:zhang} lists the vulnerable answers, snippets, and versions detected with different versions of Cppcheck on different versions of SOTorrent.
We examined with a paired t-test whether the effect of changing the SOTorrent dataset version was stable across different versions of Cppcheck.
Shapiro-Wilk tests confirmed the normality of the distribution of differences between paired observations for the SOTorrent versions for Cppcheck v1.86 ($0.832, p=0.193$) and v2.13 ($0.852, p=0.247$), respectively.
The t-test indicates that the effect of upgrading the SOTorrent dataset is not consistent across Cppcheck versions ($t=-8.90, p<0.05$).
The change from SOTorrent18 to SOTorrent22 significantly impacted the results when using Cppcheck v2.13 compared to v1.86, implying that changes in the dataset can have a different impact depending on the Cppcheck version.
Although using Cppcheck v1.86 on SOTorrent22 would better isolate the effect of code evolution, we still decided to report the results of Cppcheck v2.13 in our replication study.
We reason that if the authors had conducted their experiment later, they would have used the newer tool version.

\subsection{Details about Replicating RQ3}\label[secinapp]{sec:appendix:casestudy_1_rq3}

Our data set based on SOTorrent22 and Cppcheck v2.13 comprises 495,330 C/C++ code snippets from 75,779 users.
This number is smaller than the original data (85k users), which we surmise is due to a larger fraction of non-C/C++ snippets falsely identified as such by Guesslang in the original methodology. See \Cref{fig:zhang:methodology_flowchart} for a comparison of the original methodology with the approach used in the replication study.

\begin{figure}[h]
    \centering
    \includegraphics[width=\linewidth]{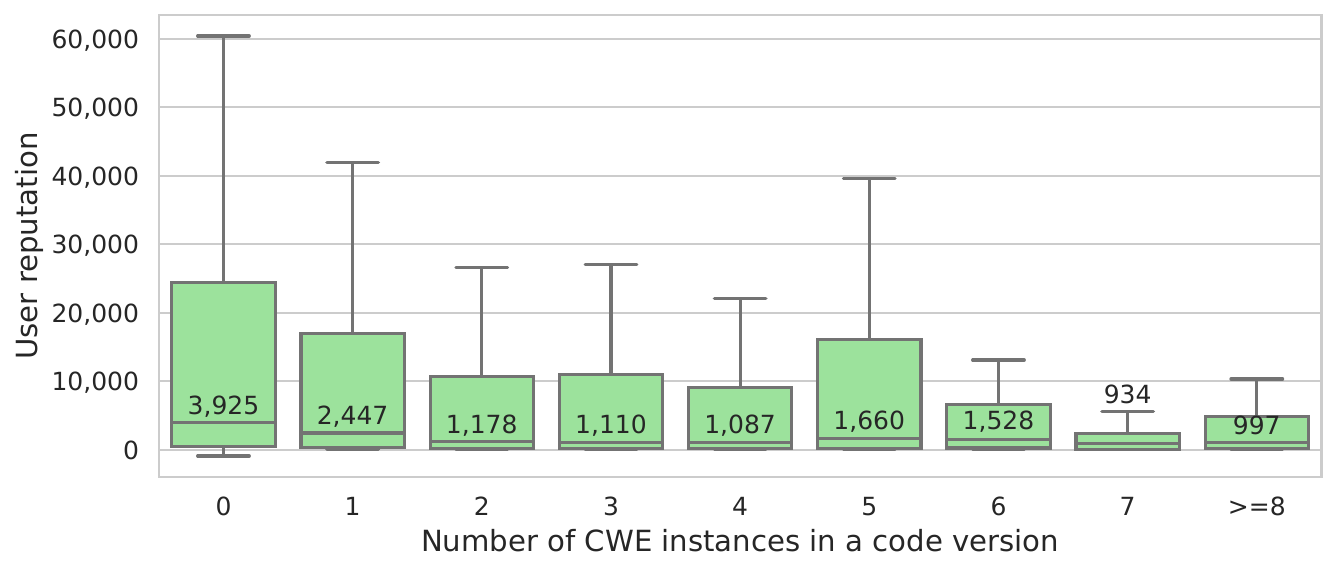}
    \caption{The distribution of user reputation points for users who contribute code versions both without and with different numbers of CWE instances. (cf.~Figure~11 in \cite{zhang_code_weaknesses})}
    \label{fig:zhang_rq3_cwe_vs_reputation}
\end{figure}

Zhang et al.~\cite{zhang_code_weaknesses} compared the reputation of the contributor for a code version with different numbers of CWE instances in the code.
Their results showed \textit{``that users with higher reputation tend to introduce fewer CWE instances in their contributed code versions.''}
We also explored the relationship between the users' reputation and the number of CWE instances.
\Cref{fig:zhang_rq3_cwe_vs_reputation} depicts the relation between user's reputation and the number of CWE instances.
Both linear regression and the Pearson correlation analysis suggest that while a statistically significant relationship ($p < 0.001$) exists between reputation and the number of CWE instances by a user, the effect size is very small.
The linear regression had a practically negligible effect on the number of CWE, with an $R^2 = 0.001$.
The Pearson correlation coefficient is $-0.0227$, indicating a very weak inverse relationship.
This implies that reputation alone is not a strong predictor of the number of CWEs. Further research with additional variables or different analytical approaches might be needed to understand better the factors influencing the number of CWEs.

\begin{figure}[h]
    \centering
    \includegraphics[width=\linewidth]{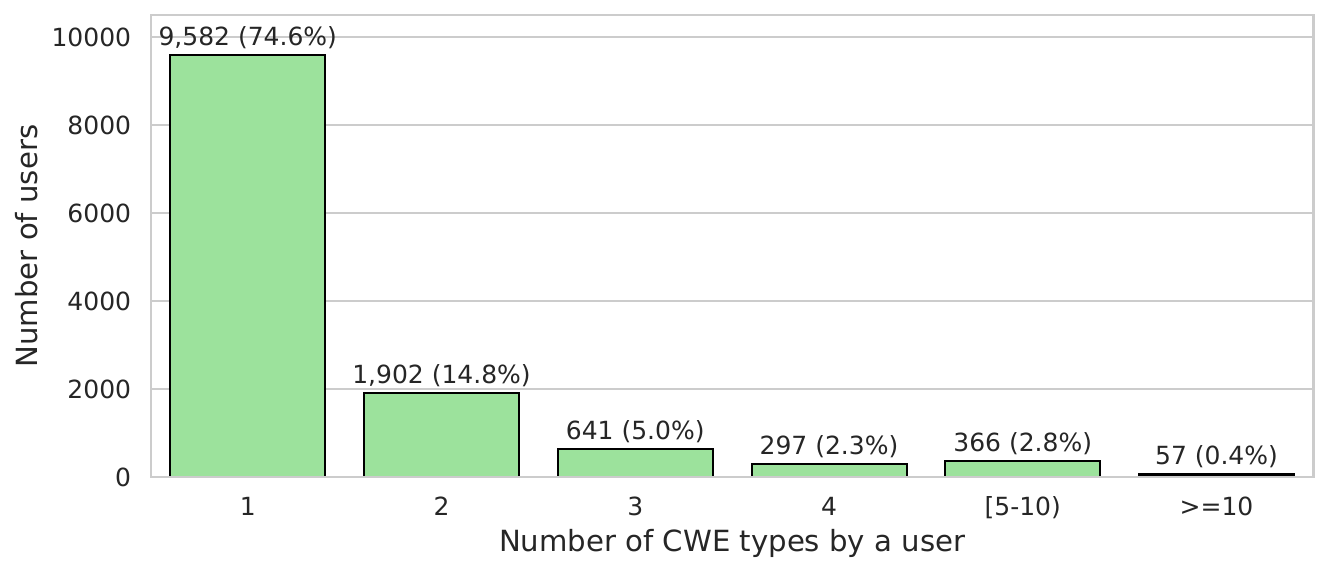}
    \caption{Number of distinct CWE types introduced by users.}
    \label{fig:zhang_rq3_cwe_types_per-user}
\end{figure}

Next, Zhang et al.~found that \textit{``78.0 percent of users contribute code with only one CWE type.''}
\Cref{fig:zhang_rq3_cwe_types_per-user} depicts the number of users vs.~number of distinct CWE types by the user in our replication study.
We found that 74.6 percent (i.e., 9,582) of users contribute only one CWE type.

\begin{figure}[h]
    \centering
    \includegraphics[width=\linewidth]{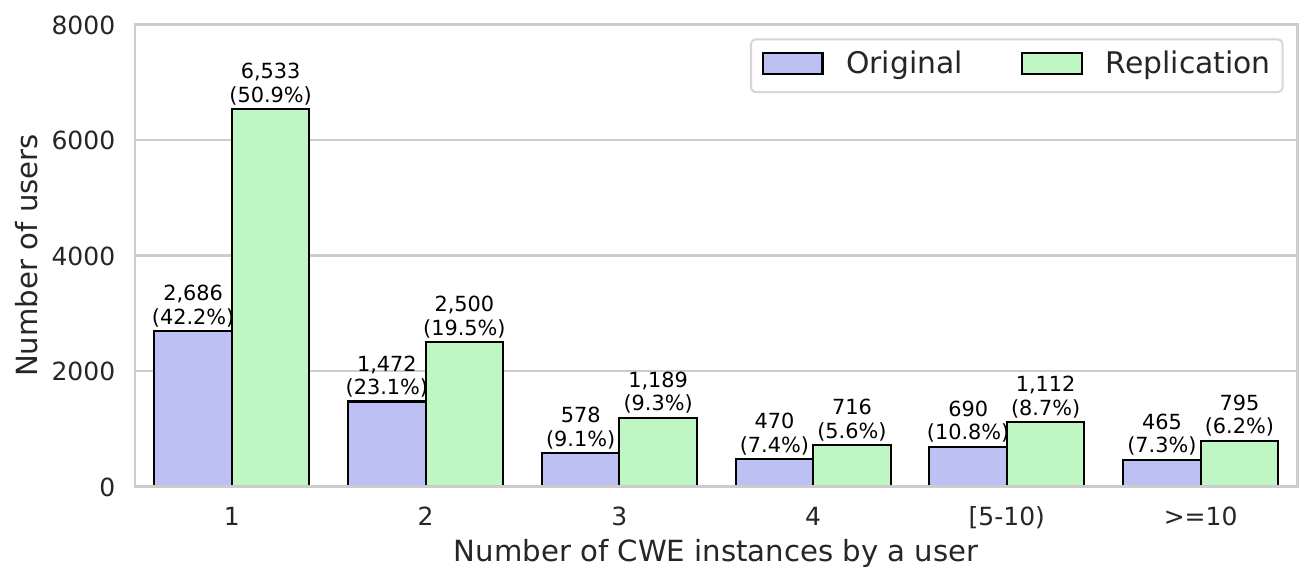}
    \caption{The number/proportion of users contributing a different number of CWE instances. (cf.~Figure~12 in \cite{zhang_code_weaknesses})}
    \label{fig:zhang_rq3_cwe_instances_per-user}
\end{figure}

The next result reported by Zhang et al. is that \textit{``42.2 percent (i.e., 2,686) of the users contribute only one CWE instance in all their $Version_w$.''} and that \textit{``81.8 percent (i.e., 5,206) of the users contribute less than five CWE instances in all their $Version_w$.''}
In \Cref{fig:zhang_rq3_cwe_instances_per-user}, we compare the authors' results with ours.
We find that 50.9 percent (i.e., 6,533) of the users contribute only one CWE instance in all their $Version_w$ and that 96.7 percent of the users contribute less than five CWE instances in all their $Version_w$
We compare the two distributions in \Cref{fig:zhang_rq3_cwe_instances_per-user} with a $\chi^2$ test and find that there is a statistically significant difference with moderate effect ($\chi^2=146.623, p < 0.001$, Cramér's $V=0.086$).

\begin{figure}[h]
    \centering
    \includegraphics[width=\linewidth]{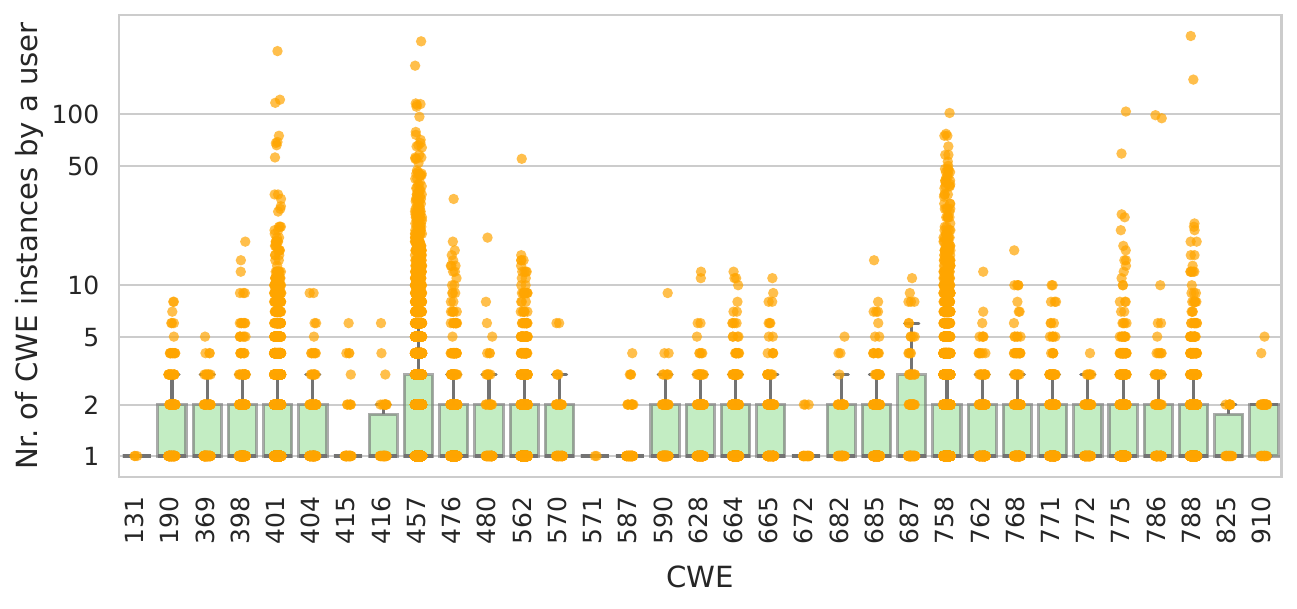}
    \caption{The distribution of contributed CWE instances by different users for different CWE types. (cf.~Figure~13 in \cite{zhang_code_weaknesses})}
    \label{fig:zhang_rq3_cwe_type_frequency}
\end{figure}

Zhang et al.~further found that ``users tend to commit the same types of CWE instances repeatedly'' by studying the distribution of the number of contributed CWE instances by different users.
We depict this distribution for our replication study in \Cref{fig:zhang_rq3_cwe_type_frequency}.
We also find that certain CWE types have a higher frequency than others---however, the types shifted between the original study and the replication study.
Zhang et al.~found that users highly frequently contributed CWE-401/775/908, while we found that CWE-457/476/758/788 are significantly more frequent, while CWE 908 disappeared.

\begin{figure}[h]
    \centering
    \includegraphics[width=\linewidth]{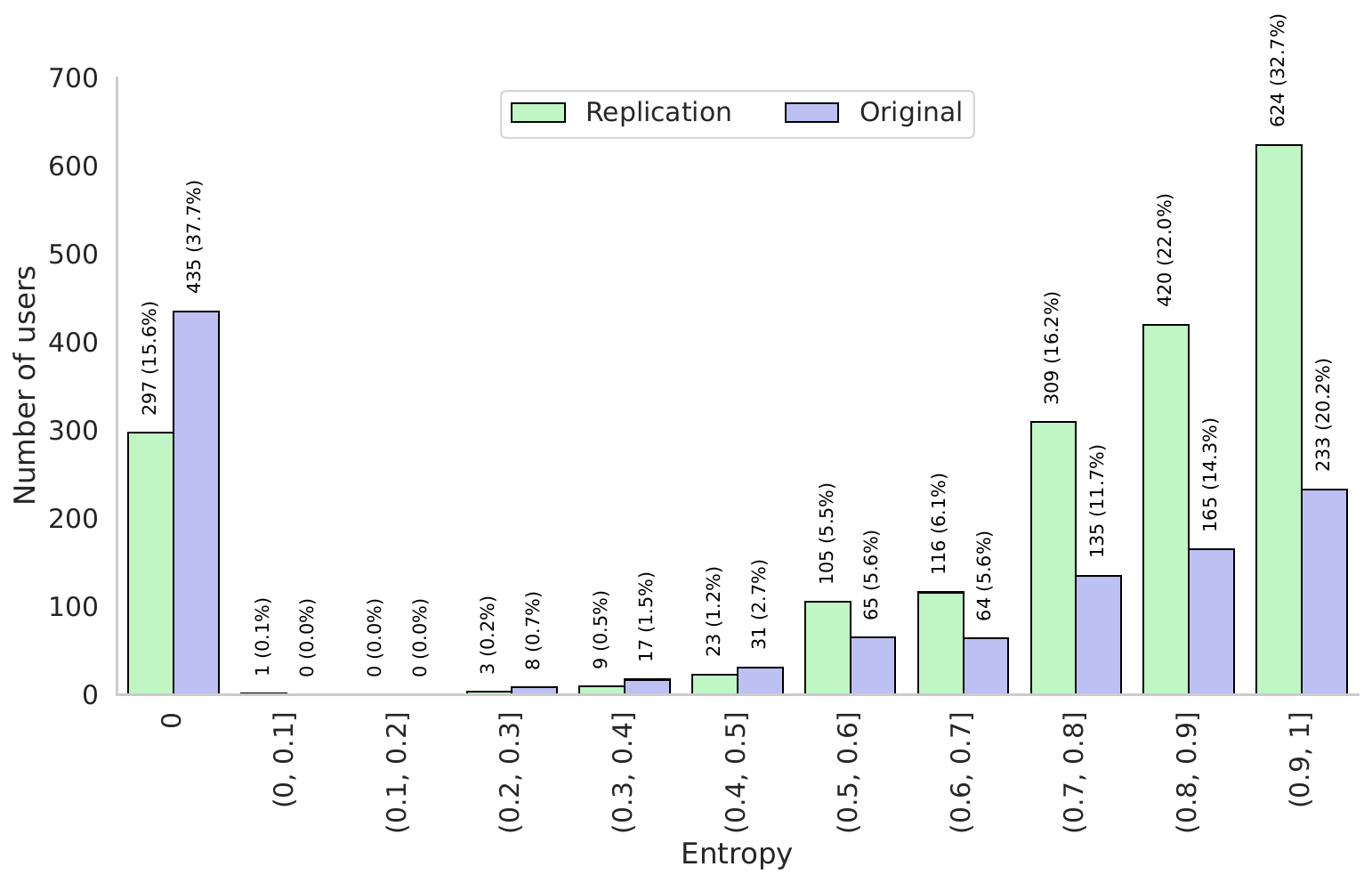}
    \caption{The distribution of entropy for CWE instances of different types. (cf.~Figure~14 in \cite{zhang_code_weaknesses})}
    \label{fig:zhang_rq3_entropy}
\end{figure}

Lastly, Zhang et al.~calculated the normalized entropy of the CWE types each user posted in $Code_w$ to better understand how users contribute different CWE types.
Here, an entropy value of 0 indicates that the user only contributed a single type of CWE in their code snippets.
\Cref{fig:zhang_rq3_entropy} compares the distribution of the normalized entropy for the original results and our replication study.
The authors observed that \textit{``37.7 percent of users are likely to introduce a single type of CWE instances in their posted code versions.''}
In contrast, we found that most users (82.5 percent) had a normalized entropy $>.5$, and only 15.6 percent had an entropy value of 0.
A $\chi^2$ test for homogeneity between the entropy in the original paper and our replication shows that there is a statistically significant difference between the distributions ($\chi^2(8, N=N=3,059) = 243.501, p < 0.001$, Cramér's $V=0.278$).

\subsection{A remark on the methodology for RQ3}\label[secinapp]{sec:appendix:casestudy_1_rq3_mistake}

While replicating the results for Zhang et al.'s RQ3, we noticed a potential mistake in how CWE instances are counted for users. Concretely, the authors stated in their paper:

\begin{displayquote}
\textit{one user has contributed CWE-775, i.e., missing release of ﬁle descriptor or handle after effective lifetime, for 79 times.}\footnote{\url{https://stackoverﬂow.com/users/3422102/}}
\end{displayquote}

Our data set based on SOTorrent22 found that this user has 104 code revisions with this particular CWE. However, only in 59 (57\%) of these revisions did this user \textit{add} the CWE-755 to the code; in the other cases, the author posted a revision to a snippet that already contained CWE-755 and the revision did not fix this CWE. This particular way of counting ``inherited CWE instances'' bears the risk of over-reporting CWE instances for users, thus inflating the observed dangerous behavior of users by ``punishing'' them for not fixing CWEs during revisions.

\begin{figure}[h]
    \centering
    \includegraphics[width=\linewidth]{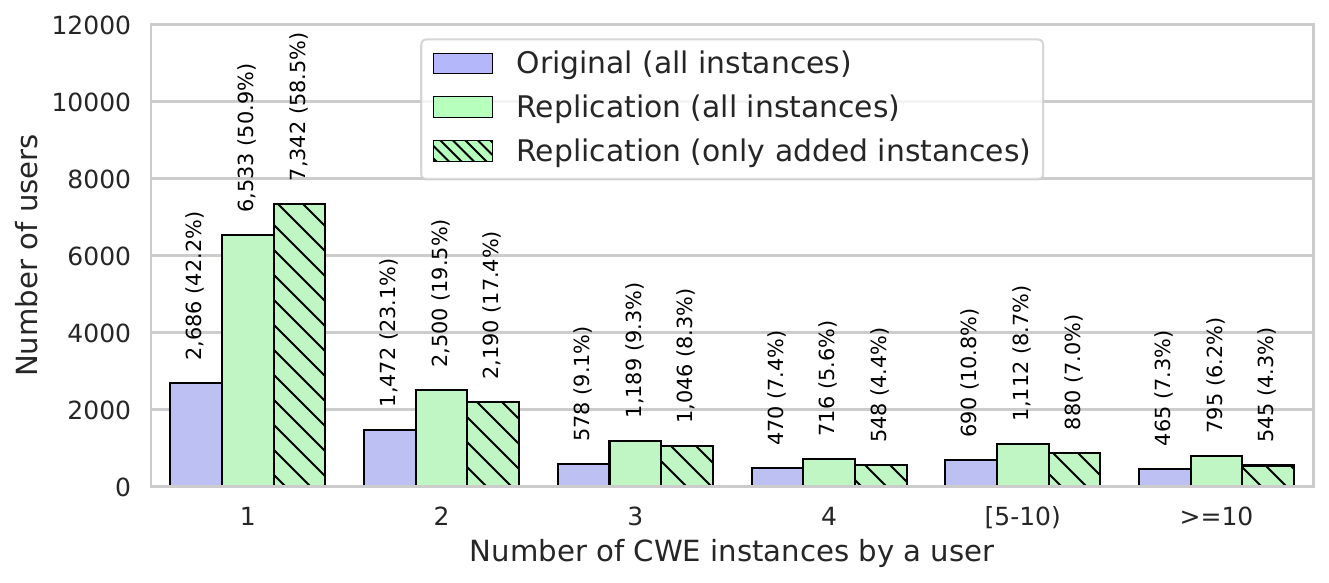}
    \caption{The number/proportion of users contributing a different number of CWE instances, including a replication \textit{that only considers newly added CWE instances for code revisions}.}
    \label{fig:zhang_new_rq3_cwe_instances_per-user_comparison}
\end{figure}

\begin{figure}[h]
    \centering
    \includegraphics[width=\linewidth]{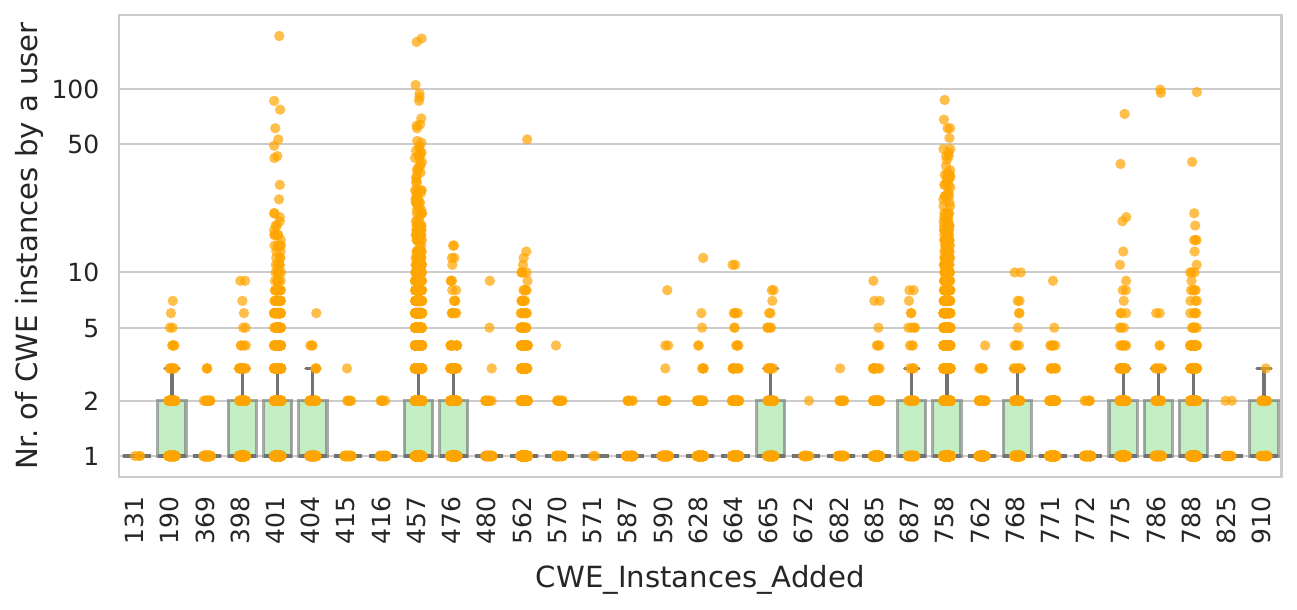}
    \caption{The distribution of contributed CWE instances by different users for different CWE types \textit{when only considering newly added CWE instances for code revisions}.}
    \label{fig:zhang_new_rq3_cwe_type_frequency}
\end{figure}

\begin{figure}[h]
    \centering
    \includegraphics[width=\linewidth]{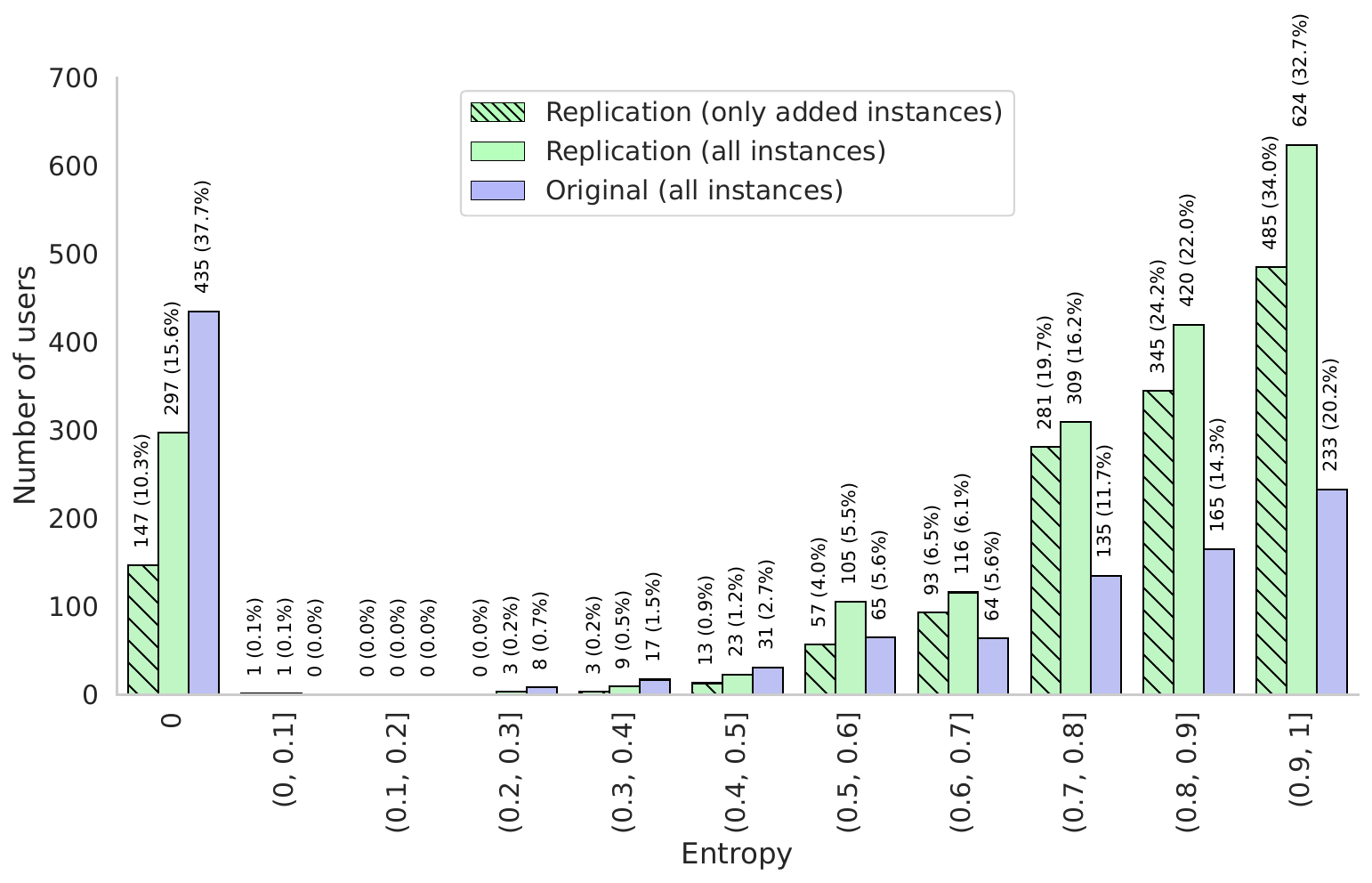}
    \caption{The distribution of entropy for CWE instances of different types, including a replication \textit{that only considers newly added CWE instances for code revisions}.}
    \label{fig:zhang_new_rq3_entropy}
\end{figure}

We briefly explored the impact of this approach by re-running our replication but only counting \textit{added} CWE instances.
While we did not find a significant drop in the number of users who contributed CWEs---indicating that this ``punishment'' affected users who contributed CWEs anyway---we found that the distribution of CWEs and their entropy changed.
\Cref{fig:zhang_new_rq3_cwe_instances_per-user_comparison} depicts the number of users contributing different numbers of CWE instances.
When only considering \textit{added} CWE instances, the number of users with only one CWE instance is higher, while the number of users with multiple CWE instances drops.
A $\chi^2$ test shows this difference between the two flavors of replication studies to be significant ($\chi^2(5, N=25,396) = 169.420, p <0.001$, Cramér's $V=0.080$).
\Cref{fig:zhang_new_rq3_cwe_type_frequency} shows the CWE instance distribution based on this alternative methodology.
Compared to \Cref{fig:zhang_rq3_cwe_type_frequency}, several CWEs have a lower instance count.
Further, \Cref{fig:zhang_new_rq3_entropy} compares the entropy per user between the original study, our replication with the same methodology as the original, and our replication considering only added CWEs.
When considering only \textit{added} CWE instances, the entropy per user is higher, indicating that users add different CWE types with similar probabilities in their code revisions.
The difference in entropy between our replication studies has been confirmed as statistically significant with a $\chi^2$ test for homogeneity ($\chi^2(8, N=3,329) = 31.031, p < 0.001$, Cramér's $V=0.083$).

\section{Case Study 2: \DC: Discovering Insecure code snippets}\label[secinapp]{sec:appendix:casestudy_2}

We replicated the accuracy measurements for C/C++ and Android posts as presented in Table~3 and Table~4 of the original paper~\cite{dicos}.
Our comparative findings for C/C++ posts are shown side-by-side in \Cref{tab:appendix_dicos_c_cpp_accuracy_measurement} and for Android in \Cref{tab:appendix_dicos_android_accuracy_measurement}.
For C/C++ posts, we observed an \hlgreen{accuracy of \textbf{11\%}} (\hlblue{authors reported 93\%}), a \hlgreen{recall of 92\%} (\hlblue{vs.~94\%}), and a \hlgreen{precision of \textbf{27\%}} (\hlblue{vs.~90\%}). Similarly, for Android posts, we found a significant drop in precision (\hlgreen{12\%} vs.~\hlblue{86\%}) and accuracy (\hlgreen{41\%} vs.~\hlblue{86\%}) while recall remained high (\hlgreen{78\%} vs.~\hlblue{89\%}).

\begin{figure}[h]
    \centering
    \includegraphics[width=\linewidth]{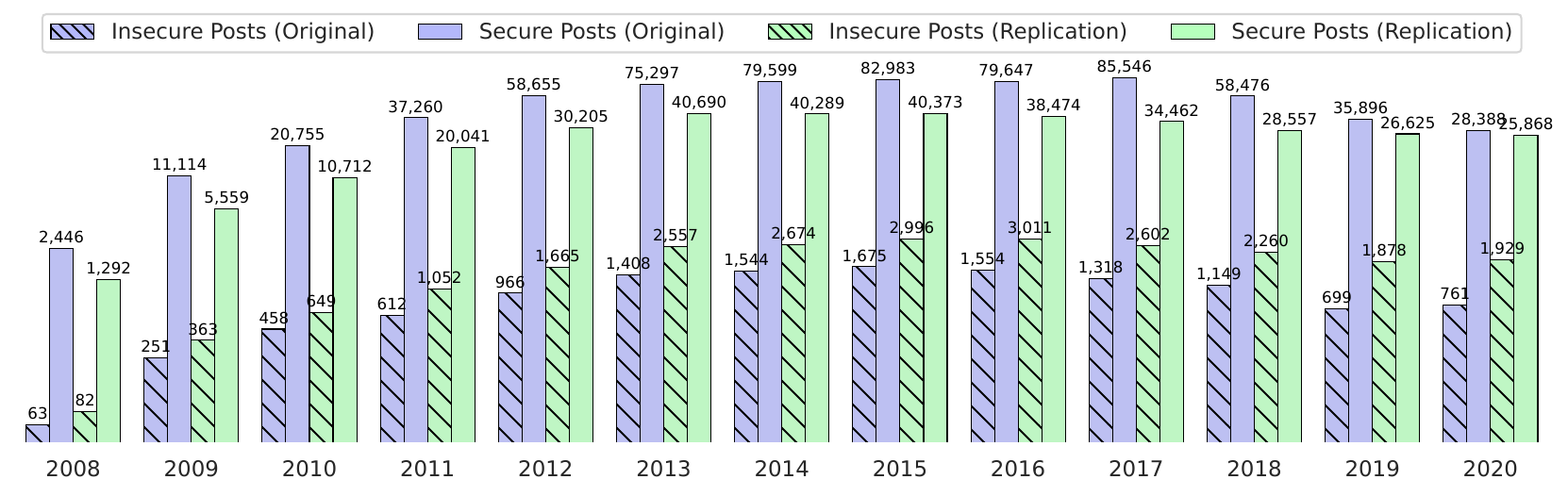}
\caption{Comparison of yearly distribution of secure and insecure posts discovered by \DC (logarithmic scale) as reported by the \hlblue{authors} (Figure 6 in \cite{dicos}) and found in our \hlgreen{replication study}.}
\label{fig:appendix:dicos_rq1_comparison}
\end{figure}

As stated in \Cref{sec:dicos}, we reproduced the authors' results for \textbf{RQ1} using the same dataset version. Our comparative findings are shown in \Cref{fig:appendix:dicos_rq1_comparison}.
The authors concluded that older posts are less likely to introduce insecure posts than newer ones. While we observed similar trends, we found different yearly numbers of secure/insecure posts from those reported by the authors. For instance, in 2008, the authors observed \hlblue{63 insecure and 2,446 secure posts} while we observed \hlgreen{82 insecure and 1,292 secure posts}.

A two-sample z-test for the \textit{overall} proportions of the insecure to secure posts ratio between the original and our replication results shows a significant difference ($Z=-121.962$, $p<0.001$).
To evaluate whether the proportion of insecure posts differed significantly between the originally reported results and our replication study over the years, we conducted two-proportion z-tests for each year from 2008 to 2020.
The Holm–Bonferroni correction~\cite{holm1979simple} was applied to account for multiple comparisons across the 13 years.
Our results indicate that for all years, the differences in the proportion of insecure posts between the two studies are statistically significant after the Holm–Bonferroni correction (adjusted $\alpha$ levels ranged from $0.0038$ to $0.05$, all p-values < 0.05).
This suggests a consistent and significant disparity in the proportion of insecure posts across the entire study period.

\Cref{fig:appendix:dicos_rq2_comparison} compares the original results (refer to Figure 7 in \cite{dicos}) and the replication findings. As explained in \Cref{sec:dicos}, we found a significantly higher ratio of insecure posts in both accepted (\hlblue{1.67\%} $\nearrow$ \hlgreen{92.83\%}) and non-accepted answers (\hlblue{1.99\%} $\searrow$ \hlgreen{93.7\%}). Nevertheless, the original conclusion that secure posts outnumber insecure posts in both categories still remains valid.
A two-sample z-test for proportions of the insecure to secure posts ratio between the original and our replication results shows a significant difference ($Z=-126.888$, $p<0.001$).

\Cref{fig:appendix:dicos_rq3_comparison} presents a comparison between the authors' original results (refer to Figure 8 in \cite{dicos}) and the replication findings for the types of insecure code snippets with all three types of changes (i.e., changes in security-sensitive APIs, security-related keywords, and control flows).
Overall, there is a significant decrease in the number of snippets with all three change types, for instance, \textit{Undefined behavior} (\hlblue{367} $\searrow$ \hlgreen{16}), \textit{null-terminated string issue} (\hlblue{175} $\searrow$ \hlgreen{10}) and out-of-bounds error (\hlblue{17} $\searrow$ \hlgreen{0}).

\begin{table*}[t]
\caption{Comparison of \DC Discovery Accuracy for C/C++ posts: Precision, Recall, and Accuracy metrics \hlblue{measured by authors} (Table 3 in \cite{dicos}) vs. our \hlgreen{Replicated Metrics (SOTorrent22)}
}
\label{tab:appendix_dicos_c_cpp_accuracy_measurement}
\centering
\small
\begin{tabular}{N N N N N N N N N N N }\toprule
    \multicolumn{1}{N }{\textbf{}} & \multicolumn{5}{c }{\textbf{Accuracy measurement result of \DC for C/C++}} & \multicolumn{5}{c }{\textbf{Accuracy measurement results of \DC for C/C++}} \\  
    \multicolumn{1}{c }{\textbf{}} & \multicolumn{5}{c }{\textbf{\textit{SOTorrent20 (\hlblue{Original})}}} & \multicolumn{5}{c }{\textbf{\textit{SOTorrent22 (\hlgreen{Replication})}}} \\ 
    \cmidrule(lr){1-1}
    \cmidrule(lr){2-6}
    \cmidrule(ll){7-11}
    \multicolumn{1}{ N }{\textbf{ID}} & \textbf{\#Posts} & \textbf{\#TP} & \textbf{\#FP} & \textbf{\#TN} & \textbf{\#FN} & \textbf{\#Posts} & \textbf{\#TP} & \textbf{\#FP} & \textbf{\#TN} & \textbf{\#FN} \\ 
    \cmidrule(lr){1-1}
    \cmidrule(lr){2-6}
    \cmidrule(ll){7-11}
    
    \multicolumn{1}{ N }{\textbf{G1}} & 731 & 704 & 27 & \textcolor{gray!25}{N/A} & \textcolor{gray!25}{N/A} & 746 & 92 & 654 & \textcolor{gray!25}{N/A} & \textcolor{gray!25}{N/A} \\
    
    \multicolumn{1}{ c }{\textbf{G2}} & 200 & 171 & 29 & \textcolor{gray!25}{N/A} & \textcolor{gray!25}{N/A} & 200 & 13 & 187 & \textcolor{gray!22}{N/A} & \textcolor{gray!25}{N/A} \\
    
    \multicolumn{1}{ c }{\textbf{G3}} & 100 & 82 & 18 & \textcolor{gray!25}{N/A} & \textcolor{gray!25}{N/A} & 300 & 26 & 274 & \textcolor{gray!25}{N/A} & \textcolor{gray!25}{N/A} \\
    
    \multicolumn{1}{ c }{\textbf{G4}} & 200 & \textcolor{gray!25}{N/A} & \textcolor{gray!25}{N/A} & 151 & 49 & 200 & \textcolor{gray!25}{N/A} & \textcolor{gray!25}{N/A} & 191 & 9 \\
    
    \multicolumn{1}{ c }{\textbf{G5}} & 100 & \textcolor{gray!25}{N/A} & \textcolor{gray!25}{N/A} & 92 & 8 & 100 & \textcolor{gray!25}{N/A} & \textcolor{gray!25}{N/A} & 98 & 2 \\ \midrule
    
    \multicolumn{1}{ c }{\textbf{Total}} & 1,331 & 957 & 74 & 243 & 57 & 1,546 & 131 & 1115 & 289 & 11 \\ \midrule
    \multicolumn{2}{ c }{\textbf{Precision}}  &  &  &  & \hlblue{\textbf{0.91}} &  &  &  &  & \hlgreen{\textbf{0.11}} \\
    \multicolumn{2}{ c }{\textbf{Recall}}  &  &  &  & \hlblue{\textbf{0.93}} &  &  &  &  & \hlgreen{\textbf{0.92}} \\
    \multicolumn{2}{ c }{\textbf{Accuracy}}  &  &  &  & \hlblue{\textbf{0.89}} &  &  &  &  & \hlgreen{\textbf{0.27}} \\
    \bottomrule
\end{tabular}

\end{table*}

\begin{table*}[t]
\caption{ Comparison of \DC Discovery Accuracy for Android posts: Precision, Recall, and Accuracy metrics \hlblue{measured by authors} (Table 4 in \cite{dicos}) vs. our \hlgreen{Replicated Metrics (SOTorrent22)}
}
\label{tab:appendix_dicos_android_accuracy_measurement}
\centering
\small
\begin{tabular}{N N N N N N N N N N N }\toprule
    \multicolumn{1}{N }{\textbf{}} & \multicolumn{5}{c }{\textbf{Accuracy measurement result of \DC for Android}} & \multicolumn{5}{c }{\textbf{Accuracy measurement results of \DC for Android}} \\  
    \multicolumn{1}{c }{\textbf{}} & \multicolumn{5}{c }{\textbf{\textit{SOTorrent20 (\hlblue{Original})}}} & \multicolumn{5}{c }{\textbf{\textit{SOTorrent22 (\hlgreen{Replication})}}} \\ 
    \cmidrule(lr){1-1}
    \cmidrule(lr){2-6}
    \cmidrule(ll){7-11}
    \multicolumn{1}{ N }{\textbf{ID}} & \textbf{\#Posts} & \textbf{\#TP} & \textbf{\#FP} & \textbf{\#TN} & \textbf{\#FN} & \textbf{\#Posts} & \textbf{\#TP} & \textbf{\#FP} & \textbf{\#TN} & \textbf{\#FN} \\ 
    \cmidrule(lr){1-1}
    \cmidrule(lr){2-6}
    \cmidrule(ll){7-11}
    
    \multicolumn{1}{ N }{\textbf{G1}} & 57 & 53 & 4 & \textcolor{gray!25}{N/A} & \textcolor{gray!25}{N/A} & 42 & 3 & 39 & \textcolor{gray!25}{N/A} & \textcolor{gray!25}{N/A} \\
    
    \multicolumn{1}{ c }{\textbf{G2}} & 200 & 175 & 25 & \textcolor{gray!25}{N/A} & \textcolor{gray!25}{N/A} & 200 & 20 & 180 & \textcolor{gray!22}{N/A} & \textcolor{gray!25}{N/A} \\
    
    \multicolumn{1}{ c }{\textbf{G3}} & 100 & 80 & 20 & \textcolor{gray!25}{N/A} & \textcolor{gray!25}{N/A} & 300 & 40 & 260 & \textcolor{gray!25}{N/A} & \textcolor{gray!25}{N/A} \\
    
    \multicolumn{1}{ c }{\textbf{G4}} & 200 & \textcolor{gray!25}{N/A} & \textcolor{gray!25}{N/A} & 167 & 33 & 200 & \textcolor{gray!25}{N/A} & \textcolor{gray!25}{N/A} & 188 & 8 \\
    
    \multicolumn{1}{ c }{\textbf{G5}} & 100 & \textcolor{gray!25}{N/A} & \textcolor{gray!25}{N/A} & 93 & 7 & 100 & \textcolor{gray!25}{N/A} & \textcolor{gray!25}{N/A} & 90 & 10 \\ \midrule
    
    \multicolumn{1}{ c }{\textbf{Total}} & 657 & 308 & 49 & 260 & 40 & 842 & 63 & 479 & 278 & 18 \\ \midrule
    \multicolumn{2}{ c }{\textbf{Precision}}  &  &  &  & \hlblue{\textbf{0.86}} &  &  &  &  & \hlgreen{\textbf{0.12}} \\
    \multicolumn{2}{ c }{\textbf{Recall}}  &  &  &  & \hlblue{\textbf{0.89}} &  &  &  &  & \hlgreen{\textbf{0.78}} \\
    \multicolumn{2}{ c }{\textbf{Accuracy}}  &  &  &  & \hlblue{\textbf{0.86}} &  &  &  &  & \hlgreen{\textbf{0.41}} \\
    \bottomrule
\end{tabular}
\end{table*}

\section{Programming Language Detection}\label[secinapp]{sec:appendix:languagedetection}

\begin{figure}[!htbp]
    \centering
    \includegraphics[width=\linewidth]{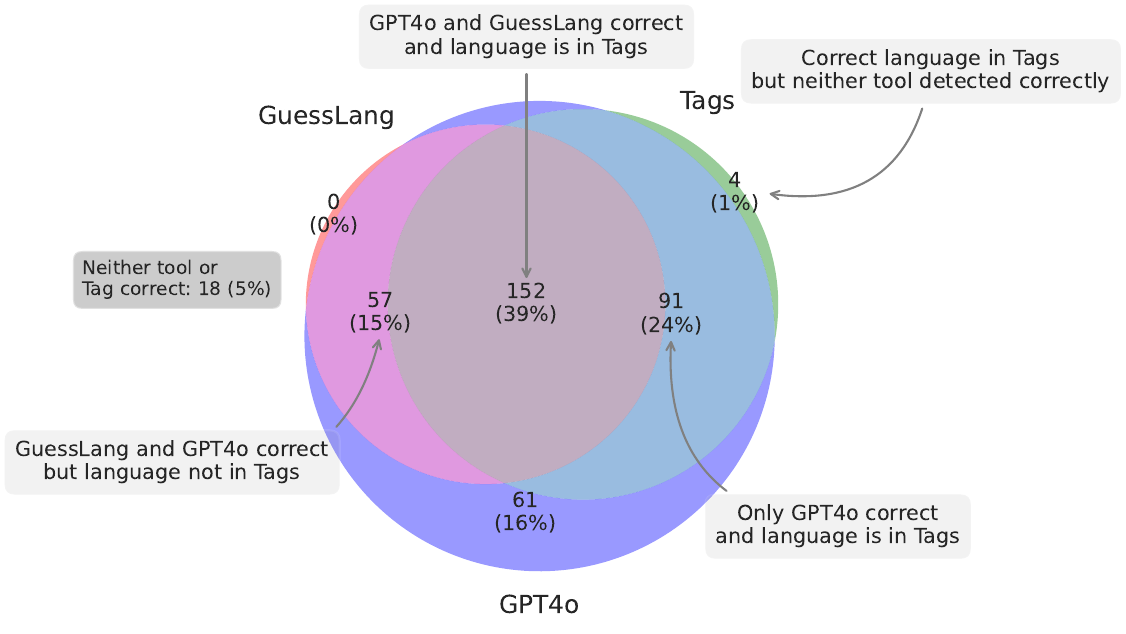}
    \caption{Intersection between the manually verified language predictions by GuessLang, GPT4o, and post Tags.}
    \label{fig:appendix:venn_verification}
\end{figure}

During our replication of the work by Zhang et al.~\cite{zhang_code_weaknesses}, we recognized the issue of detecting the correct programming language of code snippets.
Considering that various works focus on snippets in a specific language (see \Cref{tab:comparision_table}) and rely on \textit{Tags} or machine learning, e.g., Guesslang~\cite{guesslang}, we were interested in their efficacy.
We randomly sampled 385 code snippets from the 20,158,096 snippets on Stack Overflow.
This forms a representative sample size with a 95\% confidence interval and 5\% margin of error.
We used Guesslang v2.0.1 and GPT4o (via the batch API of OpenAI) to determine the programming language of the snippet.
GPT4o was prompted to determine the most likely programming language in a given code snippet.
Two researchers manually verified the detected languages and marked the true positives (correct language detected) and negatives (wrong language detected).
The researchers also checked if the post's tags contained the correct language.
We found that GPT4o achieved an accuracy of $0.93$ and Guesslang of $0.55$.
Only 247 (64.16\%) posts listed the programming language in their tags.
We observed that the tags often refer to concepts or frameworks, e.g., code snippets in Java were tagged with \textit{Android} but not \textit{Java}.
However, since Android encompasses more than Java code, one cannot reliably deduce Java snippets from Android tags.
Further, tags are applied exclusively to question posts, each of which can include one or more answers. A single post may contain multiple snippets written in various programming languages..
The correct language was only in three (0.78\%) cases in the tags when neither of the two tools could correctly detect it.
In 117 (30.39\%) cases, GPT4o detected the correct language while the tags did not contain it.
However, in 21 (5.45\%) cases, neither the two tools nor the tags detected the correct language, indicating that even with a combination of these three approaches, there is a residual risk of missing the correct language.
We only found two cases in which Guesslang (and Tags) were correct while GPT4o was wrong, underlining GTP4o's better reliability.
\Cref{fig:appendix:venn_verification} illustrates the intersection between the three tools.

Since the works studied in this paper focus on snippets in Java, JavaScript, Python, and C/C++, we further considered how well GuessLang and GPT4o detected these selected languages.
To this end, we computed their precision and recall on our sample, where a positive case is a snippet written in either of these languages (see \Cref{tab:appendix:landdetectperformance}).
Our random sample contained 156 positive cases and 229 negative cases.
GPT4o achieved the best performance, with a precision of $0.93$, recall of $0.99$, and F1 score of $0.96$.
Unfortunately, GPT4o is also the least scaling and economically reasonable of these three approaches, creating a call for action to improve the community's Guesslang tool (e.g., with new fine-tuned LLM).

\begin{table}[!htbp]
    \caption{Performance of different approaches in detecting Python, JavaScript, Java, and C/C++ code snippets.}
    \label{tab:appendix:landdetectperformance}
    \centering
    \small
    \begin{tabular}{l|lll}
        \textbf{Tool}    & \textbf{Precision} & \textbf{Recall} & \textbf{F1} \\\midrule
        GPT4o   & 0.93 & 0.99 & 0.96 \\
        Guesslang & 0.89 & 0.69 & 0.78 \\
        Tags & 0.43 & 0.67 & 0.52
    \end{tabular}
\end{table}

\label{sec:appendix}

\end{document}